\documentclass[preprint,12pt]{elsarticle}
\usepackage{caption}
\usepackage{amsmath}
\usepackage{amssymb}
\usepackage{amsthm}
\usepackage{mathrsfs}

\usepackage{epstopdf}

\usepackage{subfig}
\usepackage{booktabs} %The package enhances the quality of tables in LATEX, providing extra commands as well as behind-the-scenes optimisation
\usepackage{multirow} %Create tabular cells spanning multiple rows

\usepackage{url}
\usepackage{hyperref}
\usepackage{xcolor}
\usepackage{cleveref} %Intelligent cross-referencing

\usepackage{lineno}

\usepackage{threeparttable}
\usepackage{color}
\usepackage{epstopdf}
\usepackage{float}
\usepackage[top=2.5cm, bottom=2.5cm, left=2cm, right=2cm]{geometry}
\renewcommand{\vec}[1]{\boldsymbol{#1}}

\biboptions{numbers,sort&compress}
\journal{Journal of Computational Physics}

\begin{document}

% \begin{CJK*}{UTF8}{gbsn}y

\begin{frontmatter}



% \title{Unified gas-kinetic scheme for chemical nonequilibrium in multi-scale flow simulations}
\title{Unified gas-kinetic scheme for reactive flow with multi-scale transport and chemical non-equilibrium}


\author[a]{Yufeng Wei}
\author[a]{Junzhe Cao}
\author[a,b,c]{Kun Xu\corref{cor1}}

\cortext[cor1] {Corresponding author.}
\ead{makxu@ust.hk}

\address[a]{Department of Mathematics, Hong Kong University of Science and Technology, Hong Kong, China}
\address[b]{Department of Mechanical and Aerospace Engineering, Hong Kong University of Science and Technology, Hong Kong, China}
\address[c]{HKUST Shenzhen Research Institute, Shenzhen, 518057, China}
\begin{abstract}
Reactive flows for rarefied gas mixtures involve a multi-scale transport characterized by particle collisions and free streaming, and non-equilibrium physics containing multi-species interactions, and chemical non-equilibrium. These flows are pivotal in aerospace engineering and semiconductor manufacturing, impacting spacecraft control and thermal protection in near-space flight, and plasma etching for chip processing. Therefore, the simulation of these multi-scale and non-equilibrium flow is of scientific and industrial significance. Following the methodology of direct modeling, the unified gas-kinetic scheme (UGKS) is constructed to describe the multi-scale transport of gas molecules across all flow regimes. This study extends the UGKS to the reactive flows with chemical non-equilibrium for the capture of more complex flow physics. Test cases, including shock structures, hypersonic flows around a two-dimensional cylinder and the three-dimensional re-entry and space vehicle, and nozzle plume flow into a vacuum, are used to validate the UGKS through the comparison with the direct simulation Monte Carlo method. The study shows that the methodology of direct modeling and the extended UGKS have great potential for simulating multi-scale flows with complex non-equilibrium physics. 
\end{abstract}


\end{frontmatter}



\section{Introduction}
Flows with multi-scale transport and chemical non-equilibrium are encountered in diverse engineering applications ranging from aerospace engineering to semiconductor manufacturing \cite{boyd2017nonequilibrium,Kushner2009}. In aerospace engineering, hypersonic vehicles operate in rarefied environments where the flow around the vehicle's leading edge experiences intense compression and the flow over the trailing edge rapidly expands at extremely high Mach numbers \cite{xu2015}. The variation in the mean free path spans more than four orders of magnitude \cite{jiang2016}. Additionally, the high temperature in the compression region leads to real gas effects \cite{anderson2006}, and the resulting thermal and chemical non-equilibrium have significant impacts on aerodynamics and aerothermal properties. In semiconductor manufacturing, plasma etching requires neutral gas ionization to generate charged particles. The chemical reactions between electrons, ions, and neutral gases directly affect the mass and energy fluxes of ions on plates. However, efficiently and accurately solving these multi-scale non-equilibrium flows still faces challenges. The main difficulty lies in accurately capturing the multi-scale issues brought by molecular free transport and collisions and precisely characterizing the non-equilibrium physics resulting from mechanical collisions and chemical reactions of gas mixtures.

There are two main numerical methods, i.e., the stochastic particle methods and deterministic methods for the simulation of highly non-equilibrium flow.
For stochastic particle methods, the most representative approach is the direct simulation Monte Carlo (DSMC) method \cite{bird1970direct}. It captures the nonequilibrium physics through particles within local velocity space and therefore achieves high computational efficiency for hypersonic rarefied flow \cite{wagner1992convergence,bird1990application,blanchard1997aerodynamic,fang2020dsmc}. During gas evolution, the free streaming and collision of particles strictly follow conservation law, which guarantees high robustness. 
% Particle
For extremely complex molecular collision models, appropriately simplified models are employed in the DSMC method to enhance computational efficiency. 
Additionally, as its name has emphasized, the DSMC method directly incorporates the molecular collision model for solving, thereby shortening the modeling period \cite{bird1970direct,boyd2017nonequilibrium,zhang2014consistent,haas1993models,schwartzentruber2015progress,kim2014monte,Huang2024,adamovich1998vibrational,bondar2007dsmc,wysong2016comparison,bird2011qk,baikov2016inverse,luo2017ab,borges2017dsmc,zakeri2016new,borges2017dsmc}. 
In the DSMC method, the widely used models are phenomenological and based on the Larsen--Borgnakke collision rule \cite{borgnakke_statistical_1975}, which describes the probability of energy exchange during internal energy excitation using non-elastic collision numbers. The Larsen--Borgnakke model can accurately obtain bulk viscosity by restoring the energy exchange rate. However, the current DSMC method does not include corresponding parameters to independently adjust thermal conductivity. Moreover, there is no research evidence to prove that the DSMC method can correctly recover thermal conductivity in simulating molecular gas flow, in reality \cite{jianan2022kinetic}. 
Furthermore, because of its stochastic nature and the splitting treatment, scaling up the time step and mesh size or applying implicit acceleration techniques proves challenging, thereby constraining its efficiency in multi-scale non-equilibrium simulations. Moreover, its statistical noise hampers its application in low-speed microflow scenarios.
% DVM
The discrete velocity method (DVM) \cite{chu1965kinetic} is a popular numerical method among deterministic methods. It evolves the discrete distribution functions by solving the kinetic model equations. Consequently, the solution of non-equilibrium flow \cite{aimi_numerical_2007,groppi_shock_2008,wu2017computable,peng_ao-ping_validation_2017,li2022kinetic,wu2021derivation,lizhihui20140302,jun-lin_wu_utility_2022,zeng2023general} closely depends on the kinetic models \cite{BGK1954,morse1964kinetic,shakhov1968generalization,rykov1978macroscopic,brull_ellipsoidal_2009,hanson1967kinetic,wu2015kinetic,andries2002consistent,bisi_kinetic_2005,bisi2010kinetic,groppi_bhatnagargrosskrook-type_2004,bobylev2018general,borsoni2022general,haack2021consistent,li2023kinetic}. In the deterministic method, the gas evolution flux on cell interface is often required to be constructed based on the same velocity space. The easiest way is to employ a global velocity space for the whole computational domain. Therefore, accurate solutions without statistical noise can be obtained. Additionally, its deterministic nature enables the adoptions of numerical acceleration techniques to achieve high efficiency and the efficient simulations of low-speed flow. However, similar to the DSMC method, the splitting solution of the traditional DVM also restricts the grid and time step.

% UGKS
In recent years, with the adaptation of the direct modeling methodology \cite{xu2015}, several numerical methods have been proposed. The DVM-based unified gas-kinetic scheme (UGKS) \cite{ugks2010}, the particle-based unified wave-particle (UGKWP) method \cite{liu2020unified,zhu2019unified}, discrete UGKS (DUGKS) \cite{guo2013discrete} and discrete UGKWP method \cite{yang2023discrete} are developed. With the coupled particle collisions and free transport in gas evolution, these methods release the constraints on the mesh size and time step when collisions are intensive. At the same time, the multi-scale particle methods have been constructed as well \cite{fei2018particle}. All these multi-scale methods have the unified preserving (UP) property in capturing the Navier--Stokes solution in the continuum regime \cite{guo2023unified} without the limitations on the cell size and time step being less than mean free path and collision time. 
The direct modeling methodology models the particle evolution from the combined free transport and collision within a time step. 
The schemes have been constructed for multi-scale simulations 
with the inclusion of rotation and vibration excitations \cite{xu2021modeling,zhang2015vib}. The corresponding schemes for plasma, radiation transport, and particle flow have been developed as well \cite{wang2015unified,liu2014unified,liu2016unified,liu2017unified,sun2017implicit,liu2019unified,tan2020time,ziyang_xin_discrete_2023,quan2024radiative}. Additionally, the techniques, such as using implicit acceleration for deterministic methods \cite{chen2017,zhu2016implicit,zhu2017implicit,zhu2018implicit,jiang2019implicit,xiao2020velocity,yuan2020conservative,zhang2024conservative,wei2024,long2024} or introducing multiple sets of particles in stochastic methods \cite{xu2021rot,liu2021unified,wei2022unified,yang2023unified,li2020unified,pu2024}, have been used to accelerate convergence and reduce memory usage. 

% This paper
This paper, following the methodology of direct modeling, extends the UGKS to the gas mixture flow with chemical non-equilibrium. Starting from the Boltzmann equation, the BGK-type models for gas mixture \cite{andries2002consistent} and chemical reaction \cite{groppi_bhatnagargrosskrook-type_2004} are chosen as the relaxation model, along with the hard-sphere model and the Arrhenius equations for calculating collision cross sections \cite{jun-lin_wu_utility_2022}. 
The current UGKS aims to give a design paradigm for a multi-scale solution approach to non-equilibrium chemical reactions. Different types of kinetic models for multi-species and chemical reactions as well as molecule interaction models can be employed to adapt different flows \cite{li2024kinetic,bisi2010kinetic,bobylev2018general}. Also, the scheme as a first attempt, provides a basic scheme for other direct modeling methods, such as the UGKS with implicit and adaptive acceleration and the UGKWP method beyond single relaxation time kinetic models \cite{xu2021modeling}.

\section{Kinetic model for chemical non-equilibrium}
A simple reversible bimolecular gas reaction characterized by the chemical law
\begin{equation*}
    \mathcal{A} + \mathcal{B}  \rightleftharpoons \mathcal{C} + \mathcal{D},
\end{equation*}
is considered in this paper. The stoichiometric coefficients of the reactants and products are
$
\Lambda_\mathcal{A} = \Lambda_\mathcal{B} = -\Lambda_\mathcal{C} = -\Lambda_\mathcal{D} = 1.
$
The mass conservation law for the chemical reaction satisfies 
$
    M = m_{\mathcal{A}} + m_{\mathcal{B}} = m_{\mathcal{C}} + m_{\mathcal{D}}. 
$
where $m_\alpha$ denotes the molecular mass of the species $\alpha$. Here, the subscription $\alpha$ belongs to a set of species $N = \left\{\mathcal{A, B, C, D}\right\}$. The energy threshold of the reaction 
$\Delta E = - \sum_{\alpha \in N} \Lambda_\alpha \mathcal{E}_\alpha$ is represented by the internal energy of chemical link $\mathcal{E}_\alpha$ of the species. The endothermic reaction from the left to the right in the reaction is represented by $\Delta E > 0$. 
\subsection{Boltzmann-Type Equations}
The extended kinetic equations of the Boltzmann type for the reaction read
\begin{equation}
    \frac{\partial f_\alpha}{\partial t}
    +\vec{u}_\alpha \cdot \frac{\partial f_\alpha}{\partial \vec{x}}
    = I_\alpha + J_\alpha,
    \quad \alpha\in N,
\end{equation}
where $f_\alpha=f_\alpha(\vec{x},\vec{u}_\alpha,t)$ is the distribution function for gas molecules $\alpha$ at physical space location $\vec{r}$ with the microscopic translational velocity $\vec{u}_\alpha = \left( u_\alpha, v_\alpha, w_\alpha \right)^T$. 
By taking the moment of the distribution function $f_\alpha$ within the velocity space, one can obtain the macroscopic variables $\vec{W}_\alpha = \left( \rho_\alpha, \rho_\alpha \vec{U}_\alpha, \rho_\alpha E_\alpha \right)$, i.e., density of mass, momentum, and energy of the species $\alpha$
\begin{equation}
    \label{eq:conFromPDF}
    \vec{W}_\alpha = \int \vec{\psi}_\alpha f_\alpha {\rm d} {\vec{u}_\alpha},
\end{equation}
where $\vec{\psi}_\alpha = \left(1, \vec{u}_\alpha, \frac12 \vec{u}_\alpha^2 \right)^T$ and $\int (\cdot){\rm d}\vec{u}_\alpha = \int_{-\infty}^{\infty} \int_{-\infty}^{\infty} \int_{-\infty}^{\infty}(\cdot){\rm d} u_\alpha {\rm d} v_\alpha {\rm d} w_\alpha$. 
Note that the equations here only consider translational kinetic energy. Equations accounting for internal state excitations would be more complex. Discussions on this topic can be found in the references \cite{groppi_kinetic_1999}.
The mechanical collision term is denoted by $I_\alpha = \sum^{N}_\beta I(f_\alpha,f_\beta)$ with the usual elastic scattering collision operator for the binary $(\alpha, \beta)$ interaction. 
\begin{equation*}
    I_\alpha =\sum_{\beta \in N } \int \int_0^{2\pi} \int_0^{ \pi}
        \left(f_\alpha^{\prime} f_\beta^{\prime}-f_\alpha f_\beta\right)
        c_{\beta\alpha} \sigma_{\beta \alpha} 
        {\rm d} \varepsilon {\rm d} \chi {\rm d} \vec{u}_\beta
\end{equation*}
where $f_\alpha^\prime = f_\alpha(\vec{x}, \vec{u}_\alpha^\prime, t)$ is the distribution function of the post-collisional velocity $\vec{u}_\alpha^\prime$, and the differential cross section $\sigma_{\alpha \beta}$ depends on the relative speed $c_{\beta\alpha} = \lvert \vec{u}_\beta - \vec{u}_\alpha\rvert$ and the deflection angle $\chi$. 
For the chemical collision term, the requirement for the occurrence of the forward reaction in terms of relative velocity, i.e., $c_{\mathcal{AB}}^2 > 2\Delta E / m_{\mathcal{AB}}$ should be considered. Combined with the principle of microscopic reversibility, the collision term for reactants can be expressed
\begin{equation*}
    J_\mathcal{A(B)}=  \int  \int_0^{4\pi} 
        H\left(c_{\mathcal{AB}}^2-\frac{2 \Delta E}{m_{\mathcal{AB}}}\right) 
        c_{\mathcal{AB}} \sigma_{\mathcal{AB}}^{\mathcal{CD}} 
     \left[
        \left(\frac{m_{\mathcal{AB}}}{m_{\mathcal{CD}}}\right)^3 
        f_{\mathcal{C}}^\prime f_{\mathcal{D}}^\prime-f_{\mathcal{A}} f_\mathcal{B}\right] 
     {\rm d} \vec{u}_{\mathcal{B(A)}} {\rm d}{\Omega}^{\prime},
\end{equation*}
where $H(x)$ is the Heaviside step function accounting for the existence of the energy threshold, and $\Omega$ is solid angle. For the collision term of the products, the key distinction from the reactants lies in the fact that particles $\mathcal{C}$ and $\mathcal{D}$ can collide for the loss term whatever their relative speed, and are produced with unrestricted relative speed in the gain term
\begin{equation*}
    J_\mathcal{C(D)}=  \int  \int_0^{4\pi} 
        c_{\mathcal{CD}} \sigma_{\mathcal{CD}}^{\mathcal{AB}}
     \left[
        \left(\frac{m_{\mathcal{AB}}}{m_{\mathcal{CD}}}\right)^3 
        f_{\mathcal{A}}^\prime f_{\mathcal{B}}^\prime-f_{\mathcal{C}} f_\mathcal{D}\right] 
     {\rm d} \vec{u}_{\mathcal{D(C)}} {\rm d}{\Omega}^{\prime},
\end{equation*}
where $m_{\alpha \beta} = m_\alpha m_\beta / M$ is the reduced mass, 
and the $\sigma_{\mathcal{CD}}^{\mathcal{AB}}$ can be expressed in terms of 
$\sigma_{\mathcal{AB}}^\mathcal{CD}$
according to the principle of microscopic reversibility \cite{light1969rate}. 
Subsequently, only $\sigma_{\mathcal{AB}}^\mathcal{CD}$ will be used to represent the cross-section and frequency of chemical reactions.

\subsection{Bhatnagar--Gross--Krook-Type Model}
The chemical reaction for gas mixtures, the concept in the reference \cite{andries2002consistent} can be applied by expressing the exchange rates of mass, momentum, and energy in terms of collision frequencies. Through the cross sections, the elastic collisions between components can be expressed as
\begin{equation*}
    \nu_k^{\alpha \beta} = \nu_k^{\beta \alpha} 
    = 2 \pi c_{\alpha \beta} 
    \int_{0}^{\pi} \sigma_{\alpha \beta} \left( 1- \cos\chi \right)^k \sin \chi {\rm d} \chi, \quad k \in \left\{0,1\right\},
\end{equation*}
and the chemical microscopic collision frequency is
\begin{equation*}
    \nu_{\mathcal{AB}}^{\mathcal{CD}} = 2 \pi c_{\mathcal{AB}} \int_{0}^{\pi} \sigma_{\mathcal{AB}}^{\mathcal{CD}} \sin \chi {\rm d}\chi.
\end{equation*}
In the construction of collision equilibria, several  kinetic models for gas mixtures \cite{andries2002consistent,brull2014bgk} can be used. This paper uses the mechanical collision model given in Ref.~\cite{andries2002consistent}
\begin{equation}
    \label{eq:I}
    \begin{aligned}
        & \int I_\alpha {\rm d} \vec{u_\alpha}=0, \\
        & \int \vec{u_\alpha} I_\alpha {\rm d} \vec{u_\alpha} 
          = \sum_{\beta \in N} 
          \frac{2 \rho_\alpha \rho_\beta}{m_\alpha + m_\beta} 
         \nu_1^{\alpha \beta}
         \left(\vec{U}_\beta-\vec{U}_\alpha\right), \\
        & \int \frac12 \vec{u_\alpha}^2 I_\alpha {\rm d} \vec{u_\alpha}= 
           \sum_{\beta \in N} 
     \frac{4 \rho_\alpha \rho_\beta}
     {\left(m_\alpha+m_\beta\right)^2} \nu_1^{\alpha \beta}
     \left[m_\beta E_\beta-m_\alpha E_\alpha
     +\frac{m_\alpha-m_\beta}{2} \vec{U}_\beta \cdot \vec{U}_\alpha\right].
    \end{aligned}
\end{equation}
For the chemical collision term $J_\alpha$, the collision equilibria can be analytically obtained if the distribution function $f_s$ is replaced with the Maxwell distribution in certain integrals \cite{bisi2002grad,groppi_bhatnagargrosskrook-type_2004}. It should be noted that the approximation is reasonable in slow chemical reactions, where the elastic relaxation time is much smaller than the chemical relaxation time. Additionally, it is also valid in the bulk time domain after the initial layer where a fast transient pushes distribution functions towards local mechanical equilibrium. Under these constraints, the mass, momentum, and energy exchange of chemical reactions can be expressed as 
\begin{equation}
    \label{eq:J}
    \begin{aligned}
        & \int J_\alpha {\rm d} \vec{u_\alpha}
        =\Lambda_\alpha m_\alpha \mathcal{S}, \\
        & \int \vec{u_\alpha} J_\alpha {\rm d} \vec{u_\alpha} 
        = \Lambda_\alpha m_\alpha \vec{U} \mathcal{S}, \\
        & \int \frac12 \vec{u_\alpha}^2 J_\alpha {\rm d} \vec{u_\alpha}= 
           \Lambda_\alpha \mathcal{S}
            \left[
                \frac{1}{2} m_\alpha \vec{U}^2+\frac{3}{2} k_B T
            + \frac{M-m_\alpha}{M} \left(
                \frac{\eta^{3 / 2} \exp(-\eta)}
                {\Gamma\left(3/2, \eta\right)}k_B T 
                -\frac{1-\Lambda_\alpha}{2} \Delta E 
                \right)
            \right],
    \end{aligned}
\end{equation}
where $\Gamma$ is upper incomplete Gamma function, $\eta = {\Delta E}/{ (k_B T)}$, and $\mathcal{S}$ is reaction velocity
\begin{equation}
    \label{eq:chem-rate}
    \mathcal{S} = \nu_{\mathcal{AB}}^{\mathcal{CD}}
    \frac{2}{\sqrt{\pi}} \Gamma\left(\frac{3}{2}, \eta\right)
    \left[
        n_{\mathcal{C}} n_{\mathcal{D}} 
    \left(\frac{m_{\mathcal{AB}}}{m_{\mathcal{CD}}}\right)^{3/2}
     \exp (\eta)-n_{\mathcal{A}} n_{\mathcal{B}}
    \right].
\end{equation}
The velocity $\vec{U}$ and temperature $T$ of the system is derived from the global variables, i.e., number density $n$, mass density $\rho$, momentum $\rho \vec{U}$, and energy $\rho E$
\begin{equation*}
    \begin{gathered}
    n = \sum_{\alpha \in N} n_\alpha, \quad
    \rho=\sum_{\alpha\in N} \rho_\alpha, \quad
    \rho \vec{U}=\sum_{\alpha\in N} \rho_\alpha \vec{U}_\alpha, \quad
    \rho E = \sum_{\alpha \in N} \rho_\alpha E_\alpha,
    \end{gathered} 
\end{equation*}
with the expressions of the energy in terms of velocity and temperature for the global system
\begin{equation*}
    \rho E  = \frac12 \rho \vec{U}\cdot \vec{U} + \frac{3}{2} n k_B T, \quad
\end{equation*}
and for the species $\alpha$
\begin{equation}
    \label{eq:TtoRhoE}
    \rho_{\alpha} E_{\alpha}  = \frac12 \rho_{\alpha} \vec{U}_{\alpha} \cdot \vec{U}_\alpha + \frac{3}{2} n_{\alpha} k_B T_{\alpha}, \quad
\end{equation}
where $k_B$ is the Boltzmann constant, and $n_\alpha = \rho_\alpha / m_\alpha$ represents the number density of the species $\alpha$. 
Then, the Bhatnagar--Gross--Krook approximation can be applied to the Boltzmann-type description, 
\begin{equation}
    \label{eq:bgk-model}
    \frac{\partial f_\alpha}{\partial t}
    +\vec{u_\alpha} \cdot \frac{\partial f_\alpha}{\partial \vec{x}}
    =\nu_\alpha\left(g^c_\alpha-f_\alpha\right), 
    \quad \alpha\in N,
\end{equation}
with the Maxwellian distribution function
\begin{equation}\label{eq:g-c}
    g_\alpha^c=\rho_\alpha^c
    \left(
        \frac{m_\alpha}{2\pi k_B T_\alpha^c }
    \right)^{3 / 2} 
    \exp 
    \left(
        - \frac{m_\alpha}{2 k_B T_\alpha^c }
        \left(\vec{u}-\vec{U}_\alpha^c\right)^2
    \right),
\end{equation}
where the density $\rho_\alpha^c$, velocity $\vec{U}^c_\alpha$, and temperature $T_\alpha^c$ can be determined by the macroscopic variables $\vec{W}^c = \left( \rho_\alpha^c, \rho_\alpha^c\vec{U}_\alpha^c, \rho_\alpha^c E_\alpha^c \right)^T$, i.e., densities of mass, momentum, and energy, after mechanical and chemical collisions obtained by Eqs.~\eqref{eq:I} - \eqref{eq:J}
\begin{equation}
    \label{eq:Wc}
    \vec{W}^c_\alpha - \vec{W}_\alpha = 
    \frac{1}{\nu_\alpha} \int \vec{\psi}_\alpha 
    \left(I_\alpha + J_\alpha\right) 
    {\rm d} {\vec{u}_\alpha},
\end{equation}
with the consideration of
\begin{equation}
    \label{eq:TcToEc}
    \rho_{\alpha}^c E_{\alpha}^c  = \frac12 \rho_{\alpha}^c \vec{U}_{\alpha}^c \cdot \vec{U}_\alpha^c + \frac{3}{2} \rho_{\alpha}^c \frac{k_B}{m_\alpha} T_{\alpha}^c.
\end{equation}

Another important parameter in the BGK-type model, the relaxation frequency $\nu_\alpha$, is introduced 
\begin{equation}
    \label{eq:nu}
    \nu_\alpha = \left\{
    \begin{aligned}
        &\sum_{ \beta \in N} \nu_0^{\alpha \beta} n_\beta
        +\frac{2}{\sqrt{\pi}} \Gamma\left(\frac{3}{2}, \eta \right) 
        \nu_{\mathcal{AB}}^{\mathcal{CD}}
        \frac{n_\mathcal{A} n_\mathcal{B}}{n_\alpha}, 
        & \alpha\in \left\{
            \mathcal{A, B}
            \right\},  \\
        &\sum_{\beta \in N} \nu_0^{\alpha \beta} n_\beta 
        +\frac{2}{\sqrt{\pi}} \Gamma\left(\frac{3}{2}, \eta \right) 
        \nu_{\mathcal{AB}}^{\mathcal{CD}}
        \frac{n_\mathcal{C} n_\mathcal{D}}{n_\alpha} \left(
            \frac{m_{\mathcal{AB}}}{m_\mathcal{CD}}
        \right)^{3/2} \exp\left(\eta\right), 
        & \alpha \in \left\{
                \mathcal{C, D}
            \right\},
        \end{aligned}
    \right.
\end{equation}
where the first term in Eq.~\eqref{eq:nu} denotes the mechanical frequency calculated by the summation of a relevant number of collisions per unit time for all species. The chemical frequencies for reactants and products are reflected in the second term, where the principle of microscopic reversibility is applied for a single collisional frequency. The factor $2/\sqrt{\pi} \Gamma(3/2, \eta) < 1$ accounts for the presence of the threshold. The mechanical frequency can be approximated by different types of molecules \cite{morse1963energy}. In this paper, the hard-sphere model is used for molecular interaction
\begin{equation}
    \label{eq:hard-sphere}
    \nu_0^{\alpha \beta}=\frac{4 \sqrt{\pi}}{3}\left(\frac{2 k_B T_\alpha}{m_\alpha}+\frac{2 k_B T_\beta}{m_\beta}\right)^{1 / 2}\left(\frac{d_\alpha+d_\beta}{2}\right)^2,
\end{equation}
where $d_\alpha$ is the molecular diameter of the species $\alpha$. 
The frequency of chemical collision in the Eq.~\eqref{eq:chem-rate} 
\begin{equation*}
    \nu_{\mathcal{AB}}^{\mathcal{CD}}=\mathcal{S}\left[\frac{2}{\sqrt{\pi}} \Gamma\left(\frac{3}{2}, \eta\right)\left(n_{\mathcal{C}} n_{\mathcal{D}} 
    \left(\frac{m_{\mathcal{AB}}}{m_{\mathcal{CD}}}\right)^{3/2}
    \exp (\eta)-n_{\mathcal{A}} n_{\mathcal{B}}\right)\right]^{-1}
\end{equation*}
can be further defined in a computable way \cite{jun-lin_wu_utility_2022}, with the consideration of the gas-kinetic theory and hydrodynamics \cite{lo2015modelling,boyd2017nonequilibrium} 
\begin{equation*}
    \frac{{\rm d} n_{\mathcal{A}}}{{\rm d}t} = \mathcal{S}=-K_f n_\mathcal{A} n_\mathcal{B}+K_b n_\mathcal{C} n_\mathcal{D},
\end{equation*}
where $K_f$ and $K_b$ are forward and backward rate constants, respectively, described by Arrhenius equations
\begin{equation}
    \label{eq:arrhenius-f}
    K_{f(b)} = A_{f(b)} T^B 
    \exp\left(
        - \frac{E_{a,f(b)}}{k_B T}
    \right),
\end{equation}
where the subscript $f$ denotes variables related to the forward reaction and $b$ denotes variables related to the backward reaction. The variables $A_{f(b)}$ represent pre-exponential factors, $B_{f(b)}$ are temperature exponents, and $E_{a,f(b)}$ are activation energies for forward and backward reactions, respectively.

\section{Numerical Method}
For the numerical solution of the kinetic model of chemical reaction Eq.~\eqref{eq:bgk-model}, 
due to the fact that the collision operator cannot preserve mass, momentum, and energy conservation within the species, i.e., $\int \vec{\psi}_\alpha(g^c_\alpha - f_\alpha){\rm d}\vec{u}_\alpha \neq \vec{0}$, the standard unified gas-kinetic scheme (UGKS) cannot be directly applied to solve the current kinetic model of chemical reaction. Therefore, the numerical scheme consists of two steps. First, the UGKS is used to solve the relaxation process of free transport and conservative collision within the species $\alpha$, which satisfies the compatibility conditions
\begin{equation}
    \label{eq:react-g-f}
    \begin{aligned}
        &\frac{\partial {\tilde f}_\alpha}{\partial t}
        +\vec{u} \cdot \frac{\partial {\tilde f}_\alpha}{\partial \vec{x}}
        =\nu_\alpha\left(g_\alpha-{\tilde f}_\alpha\right),   \\
        & {\tilde f}_\alpha\left(\vec{x}, 0\right) = f_\alpha\left(\vec{x}, 0\right),
        \quad t \in [0, \Delta t],
    \end{aligned}
\end{equation}
where $g_\alpha$ is the Maxwellian distribution function that satisfies the compatibility condition $\int \vec{\psi}(g_\alpha - f_\alpha){\rm d}\vec{u} = \vec{0}$, and can be written as
\begin{equation*}
    g_\alpha=\rho_\alpha
    \left(
        \frac{m_\alpha}{2\pi k_B T_\alpha }
    \right)^{3 / 2} 
    \exp 
    \left(
        - \frac{m_\alpha}{2 k_B T_\alpha }
        \left(\vec{u}-\vec{U}_\alpha\right)^2
    \right),
\end{equation*}
where the mass density $\rho_\alpha$, velocity $\vec{U}_\alpha$, and temperature $T_\alpha$ in the equilibrium state are obtained from Eq.~\eqref{eq:conFromPDF} and Eq.~\eqref{eq:TtoRhoE}.
The second step is to solve the relaxation process between species that do not satisfy the conservation law for a single species, with the initial condition given by Eq.~\eqref{eq:react-g-f}
\begin{equation}
    \label{eq:react-gc-g}
    \begin{aligned}
        & \frac{\partial {\tilde{\tilde{f}}}_\alpha}{\partial t} 
          = \nu_\alpha \left(g_\alpha^c - g_\alpha\right), \\
        & {\tilde{\tilde{f}}}_\alpha\left(\vec{x}, 0\right) = {\tilde f}_\alpha\left(\vec{x}, \Delta t\right),
        \quad t \in [0, \Delta t].
    \end{aligned}
\end{equation}

Under the framework of the finite volume method, the governing equation for solving chemical kinetic model Eq.~\eqref{eq:react-g-f} within a discrete finite volume cell $i$ and time scale $\Delta t = t^{n+1} - t^n$ at $k$-th discrete particle velocity point is given by
\begin{equation}
    \label{eq:fvm-pdf}
    {\tilde f}_{\alpha, i,k}^{n+1} = 
		{f}_{\alpha, i,k}^n
		- \frac{\Delta t}{V_i}\sum_{j \in Q(i)} \mathcal{F}^\alpha_{ij,k} \mathcal{A}_{ij}
		+ \int_{0}^{\Delta t} \nu_{\alpha,i} \left(g_{\alpha,i,k} - {\tilde f}_{\alpha,i,k}\right) {\rm d} t,
\end{equation}
where the initial time $t^n$ is considered as $t=0$ for the sake of brevity, $V_i$ denotes the volume of cell $i$, $Q(i)$ is the set of all interface-adjacent neighboring cells of cell $i$, and $j$ is one of the neighboring cells of $i$. The interface between them is labeled as $ij$, having an area of $\mathcal{A}_{ij}$. $\mathcal{F}^\alpha_{ij,k}$ is the time-averaged microscopic flux of distribution function crossing the interface $ij$ of the species $\alpha$ for the $k$-th discrete velocity point, which can be constructed by the evolution of the gas distribution function on the cell interface
\begin{equation}
    \label{eq:micro-flux}
	\mathcal{F}^\alpha_{ij,k} = \frac{1}{\Delta t} 
	\int_0^{\Delta t} \vec{u}_{\alpha,k} \cdot 
	\vec{n}_{ij} f^\alpha_{ij,k}(t) {\rm d}t,
\end{equation}
where $\vec{u}_{\alpha,k}$ is $k$-th discrete particle velocity of the species $\alpha$ and $\vec{n}_{ij}$ is the normal direction of the cell interface, and $f^\alpha_{ij,k}(t)$ is the time-dependent distribution function on the cell interface of the species $\alpha$, which is derived from the internal solution of the Eq.~\eqref{eq:react-g-f} along the characteristic line $\vec{x_\alpha}^\prime = \vec{u_\alpha} t^\prime - \vec{u_\alpha}t$ at the center of the cell interface $\vec{x}_{ij}$
\begin{equation}
    \label{eq:bgk-solution}
    f_\alpha(\vec{x}_{ij},t) =
    \nu_\alpha \int_{0}^t e^{-\nu_\alpha(t-t')}
    g_\alpha(\vec{x}_\alpha^\prime,t^\prime)
    {\rm d} t^\prime
    + e^{-\nu_\alpha t}f_\alpha^0(\vec{x}_{ij}-\vec{u_\alpha}t),
\end{equation}
where $f^0_\alpha(\vec{x})$ represents the initial distribution function at the time $t^n$, and $g(\vec{x}_\alpha^\prime, t^\prime)$ describes the spatial distribution and temporal variation of the equilibrium distribution function near the position $\vec{x}_{ij}$ at the initial state $t_n$. Note that during obtaining the integral solution, an approximation is introduced, assuming that the collisional frequency $\nu_\alpha$ is constant within the time interval $t$. Therefore, the integral solution describes the physical evolution process in a local spatiotemporal domain, corresponding to the discrete scales of the physical time step $\Delta t$ and the control volume $\Omega_i$.
Additionally, the integral solution encompasses both the free transport and collision of particles within the time interval $t$, with the proportion determined by $e^{-\nu_\alpha t}$. From a numerical perspective, when constructing flux for gas evolution, this coupling between particle transport and collision is able to remove the limitations imposed by the splitting method on time steps and physical grid sizes. It lays the theoretical foundation for accurately restoring multi-scale transport physics. 

To achieve second-order accuracy, the initial distribution function $f^0_\alpha(\vec{x})$ and the equilibrium state $g_\alpha(\vec{x},t)$ in the integral solution of the Eq.~\eqref{eq:bgk-solution} are expanded with discretized forms
\begin{equation}
	\label{eq:ugks-taylor}
	\begin{aligned}
		g_{\alpha,k}(\vec{x}, t) &= g_{\alpha,k}^0 
		+ \vec{x} \cdot \frac{\partial g^0_{\alpha,k}}{\partial \vec{x}} 
		+ \frac{\partial g^0_{\alpha,k}}{\partial t} t, \\
		f^0_{\alpha,k}(\vec{x}) &= f_{\alpha,k}^{l,r} 
		+ \vec{x} \cdot \frac{\partial f_{\alpha,k}^{l,r}}{\partial \vec{x}},
	\end{aligned}	
\end{equation}
where $f_{\alpha,k}^{l,r}$ is the distribution function of $k$-th discrete particle velocity $\vec{u}_{\alpha,k}$ constructed by distribution functions $f_{\alpha,k}^l$ and $f_{\alpha,k}^r$ interpolated from cell centers to the left and right sides of the interface
\begin{equation*}
	f_{\alpha,k}^{l,r}= f_{\alpha,k}^l H\left(\bar{u}_{\alpha,k}^{ij}\right)+f_{\alpha,k}^r\left[1-H\left(\bar{u}_{\alpha,k}^{ij}\right)\right],
\end{equation*}
where $\bar{u}_{\alpha,k}^{ij}=\vec{u_{\alpha,k}}\cdot\vec{n}_{ij}$ represents the projection of the discrete velocity point $\vec{u}_k$ onto the interface normal vector $\vec{n}_{ij}$. The macroscopic variables $\vec{W}^0_\alpha$ corresponding to the equilibrium state at the interface are also obtained through discrete summation, i.e. taking moment discretely, of particle collisions from the left and right sides
\begin{equation*}
	\vec{W}^0_\alpha = \sum g^0_{\alpha,k} \vec{\psi}_{\alpha,k} \mathcal{V}_{\alpha,k} 
		      = \sum f_{\alpha,k}^{l,r} \vec{\psi}_{\alpha,k} \mathcal{V}_{\alpha,k},
\end{equation*}
where $\vec{\psi}_{\alpha,k} = \left(1, \vec{u}_{\alpha,k}, \frac12 \vec{u}_{\alpha,k}^2\right)$ and  $\mathcal{V}_{\alpha,k}$ denotes the represents the volume of velocity space unit $k$ or the integral weight at the discrete velocity point $\vec{u}_{\alpha,k}$. 
The spatial derivative term $\partial_{\vec{x}} g^0_{\alpha,k}$ for equilibrium state can be obtained from the gradients of macroscopic variables $\partial \vec{W}_\alpha^0 / \partial \vec{x}$ and compatibility conditions. The time derivative term in the equilibrium state is determined by compatibility conditions
\begin{equation*}
	\int (g_\alpha-f_\alpha) \vec{\psi}_\alpha {\rm d} \vec{u}_\alpha = \vec{0},
\end{equation*}
then 
\begin{equation*}
	\frac{\partial \vec{W}_\alpha^{0}}{\partial t} = 
	- \int \vec{u}_\alpha \cdot \frac{\partial g_\alpha^{0}}{\partial \vec{x}} \vec{\psi}_\alpha
	{\rm d}\vec{u}_\alpha.
\end{equation*}
Correspondingly, the temporal gradient of equilibrium state $\partial_t g_{\alpha,k}^0$ can be evaluated from the above $\partial \vec{W}_\alpha^0 / \partial t$.
The spatial gradient $\partial_{\vec{x}} f^{l,r}_{\alpha,k }$ of the distribution function is directly obtained through the reconstruction of the distribution function on the cell. 
With the completion of introducing all the terms involved in UGKS construction Eq.~\eqref{eq:ugks-taylor}, by substituting them into the integral solution Eq.~\eqref{eq:bgk-solution} and the microscale flux expression Eq.~\eqref{eq:micro-flux}, we can obtain the evolution flux of the distribution function of the species $\alpha$
\begin{equation} 
	\label{eq:ugks-micro-flux}
	\begin{aligned}
	\mathcal{F}_{ij,k}^\alpha
	&= \vec{u}_{\alpha,k} \cdot \vec{n}_{ij}
	\left(
      C_1^\alpha g^0_{\alpha,k}
	+ C_2^\alpha \vec{u}_{\alpha,k} \cdot \frac{\partial g^0_{\alpha,k}}{\partial \vec{x}}
	+ C_3^\alpha \frac{\partial g^0_{\alpha,k}}{\partial t} \right) +
	\vec{u}_{\alpha,k} \cdot \vec{n}_{ij} 
	\left(
	  C_4^\alpha f^{l,r}_{\alpha,k}
	+ C_5^\alpha \vec{u}_{\alpha,k} \cdot \frac{\partial f^{l,r}_{\alpha,k}}{\partial \vec{x}} \right) \\
	&= {\mathcal{F}}_{ij,k}^{\alpha, eq} + {\mathcal{F}}_{ij,k}^{\alpha, fr},
	\end{aligned}
\end{equation}
where $\mathcal{F}^{\alpha,eq}_{ij,k}$ and $\mathcal{F}^{\alpha,fr}_{ij,k}$  correspond to the fluxes of the equilibrium state and initial distribution function at the discrete velocity point $k$, respectively. $C_1^\alpha$ to $C_5^\alpha$ are time integral coefficients determined by time step and collisional frequency of the species $\alpha$
\begin{equation}
	\label{eq:ugks-tCoef}
	\begin{aligned}
	C_1^\alpha &= 1 - \frac{1}{\nu_\alpha \Delta t} \left( 1 - e^{-\nu_\alpha\Delta t } \right) , \\
	C_2^\alpha &= -\frac{1}{\nu_\alpha} + \frac{2}{ \nu_\alpha^2 \Delta t} - e^{-\nu_\alpha \Delta t } \left( \frac{2}{\nu_\alpha^2 \Delta t } + \frac{1}{\nu_\alpha}\right) ,\\
	C_3^\alpha &=  \frac12 \Delta t - \frac{1}{\nu_\alpha} + \frac{1}{\nu_\alpha^2 \Delta t} \left( 1 - e^{- \nu_\alpha \Delta t} \right) , \\
	C_4^\alpha &= \frac{1}{ \nu_\alpha \Delta t} \left(1 - e^{- \nu_\alpha \Delta t }\right), \\
	C_5^\alpha & = \frac{1}{\nu_\alpha} e^{- \nu_\alpha \Delta t } 
                 - \frac{1}{ \nu_\alpha^2 \Delta t}(1 -  e^{- \nu_\alpha \Delta t }) . 
	\end{aligned}
\end{equation}

With the completion of the construction of both microscopic, the distribution functions in the first step Eq.~\eqref{eq:react-g-f} can be updated. The trapezoidal rule is used for the time integration of the collision term in Eq.~\eqref{eq:fvm-pdf} to reduce the stiffness
\begin{equation}
    \label{eq:update-micro}
	{\tilde f}_{\alpha, i,k}^{n+1} = \left( 1 + \frac12 {\Delta t}{ \nu_{\alpha,i}^{n+1}} \right)^{-1}
	\left\{
		{f}_{\alpha, i,k}^n
		- \frac{\Delta t}{V_i}\sum_{j \in Q(i)} \mathcal{F}^\alpha_{ij,k} \mathcal{A}_{ij}
		+ \frac{\Delta t}{2}
		\left[
			{g_{\alpha,i,k}^{n+1}}{\nu_{\alpha, i}^{n+1}}
          + \left(
            {g_{\alpha, i,k}^n-{f}_{\alpha, i,k}^n}
          \right)
          {\nu_{\alpha, i}^n}
		\right]
	\right\},
\end{equation}
where $g_{\alpha,i,k}^{n+1}$ represents the Maxwell distribution corresponding to the macroscopic variables $\vec{\tilde W}^{n+1}_{\alpha,i}$ at the $n+1$ step, which can be calculated by the discrete governing equation for the macroscopic variables through taking moments of Eq.~\eqref{eq:react-g-f}
\begin{equation*}
\vec{\tilde W}_{\alpha, i}^{n + 1}
=
\vec{W}_{\alpha,i}^n
- \frac{\Delta t}{\Omega_i}
\sum\limits_{j \in N(i)} {\vec{F}^\alpha_{ij}{\mathcal A}_{ij}},
\end{equation*}
where the macroscopic fluxes $\vec{F}^\alpha_{ij}$ is the moments of the microscopic ones. It is also known that the flux of particle collision $\mathcal{F}_{ij,k}^{\alpha,eq}$ denoted by the Maxwellian distribution function can be integrated analytically through macroscopic variables analytically. Therefore, the macroscopic fluxes read as
\begin{equation} \label{eq:ugks-macro-flux}
	\vec{F}_{ij}^\alpha
	= \int \mathcal{F}_{ij}^{\alpha, eq} {\vec{\psi}_\alpha} {\rm d}\vec{u}_\alpha +
	   \sum_k \mathcal{F}_{ij,k}^{\alpha, fr} {\vec{\psi}}_{\alpha,k} \mathcal{V}_{\alpha,k}.
\end{equation}

Once we obtain the distribution function and macroscopic variables after the free transport and conservative collisions, the next step is to solve Eq.~\eqref{eq:react-gc-g}, which does not satisfy the compatibility condition within the species
\begin{equation*}
    {\tilde{\tilde{f}}}_{\alpha, i, k}^{n+1} = {\tilde f}^{n+1}_{\alpha, i, k} 
    + {\Delta t}{\nu_{\alpha, i}^{n+1}}\left(g^{c,n+1}_{\alpha,i,k}-g^{n+1}_{\alpha,i,k}\right),
\end{equation*}
where $g^{c,n+1}_{\alpha,i}$ is the Maxwellian equilibrium state Eq.~\eqref{eq:g-c} in the chemical reaction model, which needs to be obtained from its corresponding macroscopic variables $\vec{\dot W}^{c,n+1}_{\alpha,i}$ through Eq.~\eqref{eq:Wc}, along with the integration of the collision term in Eq.~\eqref{eq:I} and Eq.~\eqref{eq:J}. Finally, by taking moments of Eq.~\eqref{eq:react-gc-g}, the final macroscopic variables can be updated
\begin{equation*}
    \vec{\tilde{\tilde{W}}}_{\alpha, i}^{n + 1}
    =
    \vec{\tilde W}_{\alpha}^{n+1}
    + \Delta t \nu^{n+1}_\alpha  \left(
         \vec{\tilde W}_{\alpha, i}^{c,n + 1}
       - \vec{\tilde W}_{\alpha, i}^{n + 1}
    \right).
\end{equation*}
Additionally, this study also introduces numerical methods for thermal flux correction
\begin{equation*}
      ({F}_{\rho E}^\alpha)^{\prime} 
    = {F}_{\rho E}^\alpha 
    + \left(\frac{1}{{\rm Pr}} - 1\right) q_{ij}^\alpha,
\end{equation*}
where the heat flux can be denoted as 
\begin{equation*}
    q_{ij}^\alpha = \int_0^{\Delta t} \sum_k 
                    \frac{1}{2}
                    \left(
                         \vec{u}_k - \vec{U}_{\alpha,ij}
                    \right)
                         \cdot \vec{n}_{ij}
                    \left(
                         \vec{u}_k - \vec{U}_{\alpha,ij}
                    \right)^2
                    f_{ij,k}^\alpha \mathcal{V}_k {\rm d} t.
\end{equation*}

Thus far, the numerical method for solving the chemical kinetic model has been fully introduced. In summary, the core of UGKS to solve the kinetic model of chemical reaction lies in employing the integral solution of the kinetic model. It couples the particle free transport and conservative collisions, thereby reducing the numerical dissipation introduced by splitting solutions at temporal and spatial discrete scales. Moreover, the microscopic and macroscopic evolution fluxes constructed by the integral solution update the distribution function and macroscopic variables in one step. It allows the distribution function to be updated in a semi-implicit way, which effectively eliminates the stiffness introduced by conservative collision terms. The numerical validation will be discussed in the next section.

\section{Numerical Validation}
In this section, a series of test cases are conducted to verify the current UGKS and to study multi-scale non-equilibrium physics in chemical reaction flows. The capacity of the scheme to capture chemical non-equilibrium will be demonstrated by simulating shock structures. The distinct differences in shock structures between gas mixtures with and without chemical reactions will be clearly illustrated. The shock structures under different upstream Mach numbers and chemical reaction rates are also validated in comparison with the results of the DSMC method. The computational accuracy of the current UGKS in hypersonic flow involving multi-scale physics will be quantitatively evidenced through the flow around a circular cylinder at different inflow Knudsen numbers. The influences of different energy release types on the hypersonic flows, such as endothermic reactions, exothermic reactions, and inert scenarios are discussed in detail under the same inflow conditions. The UGKS's characteristics for multi-scale flow simulations will be further showcased through the simulation of nozzle plume flow into a background vacuum. Additionally, the current scheme will undergo validation for three-dimensional scenarios. Hypersonic flows around an Apollo re-entry at an altitude of 100 km will be simulated, with results closely aligning with those derived from the DSMC method. Also, simulations of hypersonic flows around X38-like space vehicles at altitudes of 100 km and 80 km will be conducted, accompanied by detailed comparisons of flow fields. These simulations will incorporate chemical reactions for real gas mixtures without additional notation. 
\begin{equation*}\label{eq:NO}
{\rm O_2 + N \rightleftharpoons NO + O}.
\end{equation*}
The properties of the gas mixtures in calculation of mechanical collision frequencies in Eq.~\eqref{eq:hard-sphere} and chemical collision frequencies in Eq.~\eqref{eq:arrhenius-f} are given in the Table.~\ref{tab:gas-mix} and the Table.~\ref{tab:chem-rate}, respectively. 
\begin{table}[H]
	\caption{Molecular mass $m_\alpha$ and the hard-sphere diameter $d_\alpha$ for mechanical collision under 272 K of ${\rm O_2}$, ${\rm N}$, ${\rm NO}$, and ${\rm O}$ used in collisional frequency and equilibrium distribution functions.}
	\label{tab:gas-mix}
	\centering
	\begin{tabular}{*{4}{c}}
		\toprule
		Mass &Value & Diameter &Value  \\
		\midrule
		$m_{\mathcal{A}}$ &$5.3156\times 10^{-26}$ kg &$d_{\mathcal{A}}$ &$4.07\times 10^{-10}$ m    \\
		$m_{\mathcal{B}}$ &$2.3256\times 10^{-26}$ kg &$d_{\mathcal{B}}$ &$3.00\times 10^{-10}$ m    \\
		$m_{\mathcal{C}}$ &$4.9834\times 10^{-26}$ kg &$d_{\mathcal{C}}$ &$4.20\times 10^{-10}$ m    \\
		$m_{\mathcal{D}}$ &$2.6578\times 10^{-26}$ kg &$d_{\mathcal{D}}$ &$3.00\times 10^{-10}$ m    \\
		\bottomrule
	\end{tabular}
\end{table}
\begin{table}[H]
    \caption{Pre-exponential factor $A$, temperature exponent $B$, reacting energy $E_a$ and energy threshold $\Delta E$ for forward and backward chemical reaction in the Arrhenius equation and kinetic model.}
	\label{tab:chem-rate}
    \centering
    \begin{tabular}{*{4}{c}}
    \toprule
        Forward &Value & Backward &Value  \\
    \midrule
        $A_f$        &$5.2\times 10^{-22}$   &$A_b$        &$3.6\times 10^{-22}$   \\
        $B_f$        &$1.29$                 &$B_b$        &$1.29$                 \\
        $E_{a,f}$    &$4.97\times 10^{-20}$ J &$E_{a,b}$    &$2.72\times 10^{-19}$ J \\
        $\Delta E_f$ &$2.72\times 10^{-19}$ J &$\Delta E_b$ &$-2.72\times 10^{-19}$ J \\
    \bottomrule
    \end{tabular}
\end{table}

\subsection{Shock structure under chemical non-equilibrium}\label{sec:shockstructure}
In comparison with single-species flow, calculating the upstream and downstream states of shock structures for gas mixtures undergoing chemical reactions is more complex. Because the internal energy is uncertain and more energy is released or absorbed by the reaction, the Rankine--Hugoniot(RH) condition for upstream (-) and downstream (+) states cannot be given directly, but should be derived by iterations, which is given in \ref{ch:reacting-shock}. 

At first, the differences between gas mixtures with and without chemical reactions are illustrated by fixing the collisional frequency of chemical reaction $\nu_{\mathcal{AB}}^{\mathcal{CD}}$ equal to 0, 0.03, and 0.003 after the non-dimensionalization. The setting indicates the inert scenarios with pure mechanical collisions and slow chemical reactions at two rates. The non-dimensional variables give the mechanical collision frequencies $\nu_0^{\alpha \beta} = \nu_1^{\alpha \beta}= 1.0$ for $\alpha, \beta \in N$ and the molecular mass $m_\mathcal{A} = 1.0000$, $m_\mathcal{B} = 1.4667$, $m_\mathcal{C} = 1.5332$, $m_\mathcal{D} = 1.5332$ with $\Delta E$ and $k_B$ normalized to unity. The total number density is $n^{-} = 1$ with the concentration fractions $\chi_\alpha = n_\alpha / n$, equal to $\chi_{\mathcal{A}}^{-} = 0.25$, $\chi_{\mathcal{B}}^{-} = 0.35$, $\chi_{\mathcal{C}}^{-} = 0.25$, $\chi_{\mathcal{D}}^{-} = 0.15$, which implies the upstream temperature is $T^{-} = 1.2337$. With the determination of changes in the chemical composition $\Delta \chi = 0.03$ , the upstream Mach number is defined ${\rm Ma}^{-} = 2.4698$. The other upstream and downstream variables determined by the RH conditions are $U^{-} = 3.1069$, $U^{+} = 1.1565$, $n^{+} = 2.6864$, and $T^{+} = 3.3614$. The computaional domain $(-200, 600)$, divided by 1500 cells, has a length of 800 times of mean free path which is determined by the mechanical collision frequency. The discretized velocity space $(-20\sqrt{ 2 k_B / m_\mathcal{\alpha} T^{-}}, 20\sqrt{ 2 k_B / m_\mathcal{\alpha} T^{-}})$ is discretized with 300 velocity points based on the midpoint rule. The left and right boundaries are treated as far fields. The Courant--Friedrichs--Lewy (CFL) number is taken as 0.5.

\begin{figure}[H]
	\centering
	\includegraphics[width=0.4\textwidth]{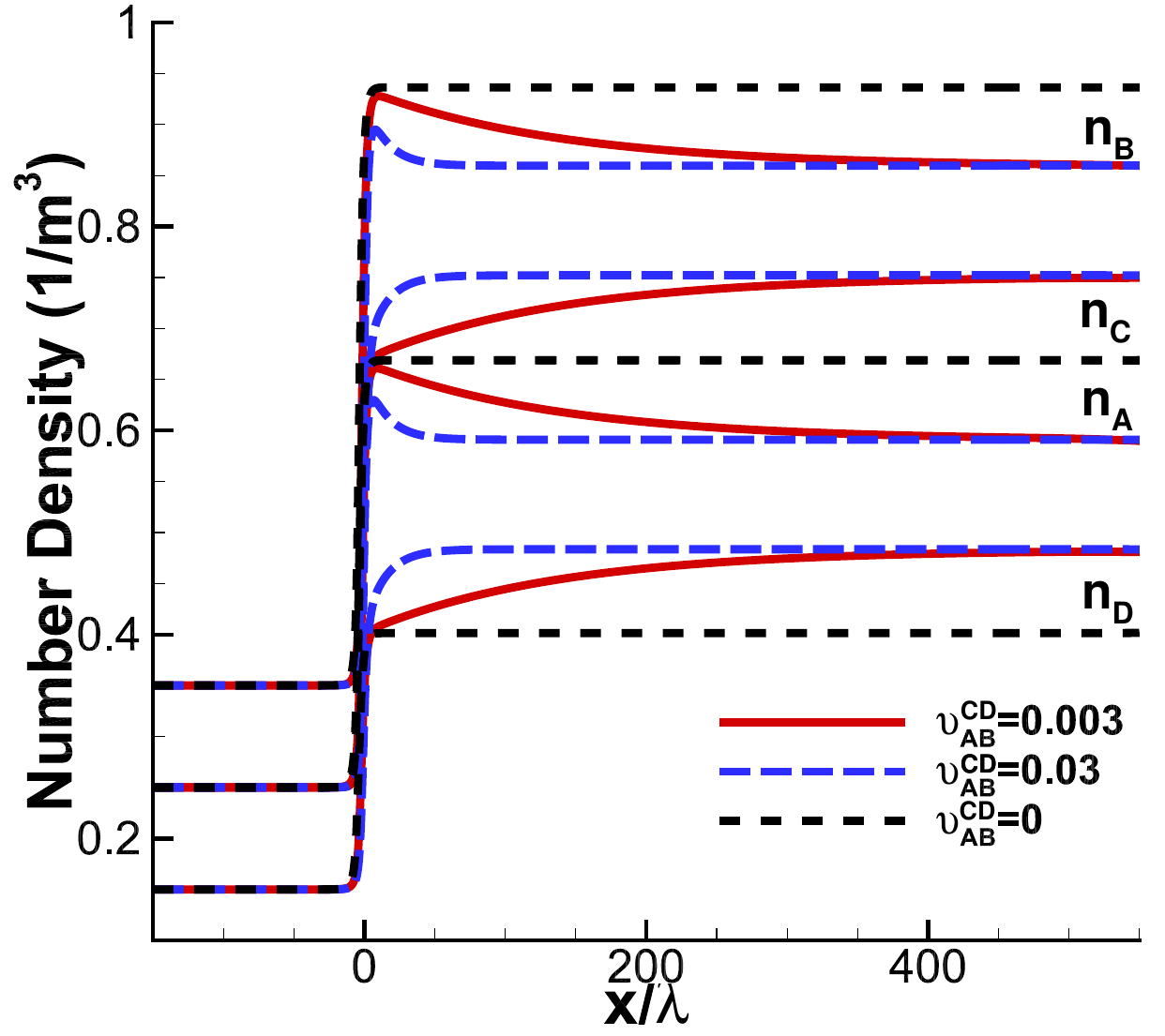}
	\caption{Steady shock structure of the four number densities $n_\alpha$ for three different choices of the chemical collision frequency $\nu_{\mathcal{AB}}^{\mathcal{CD}} = 0.0, 0.03, 0.003$, corresponding to no reaction and slow reactions with two rates.}
	\label{fig:shock-v-const}
\end{figure}

Profiles for number densities are shown in Fig.~\ref{fig:shock-v-const} where the non-equilibrium relaxation processes lie in chemical reactions that can be clearly observed.  Without chemical reactions, the inert scenarios with pure mechanical collisions have a quick relaxation process for the densities of each species, which satisfies the RH conditions for gas mixtures. The existence of chemical reactions leads to a different equilibrium state downstream with lower densities for reactants and higher densities for products, given by RH condition for chemical reactions. From a microscopic perspective, when particles stream from the upstream to the downstream, they first undergo the first equilibrium state given by the RH condition for gas mixtures \cite{wang2015unified}. This is due to the fact that the particles' cross-sections of mechanical collision are much larger than those of the chemical reactions. Subsequently, based on the first equilibrium state, the slow chemical reactants begin to influence the particles' collision downstream, manifested as the longer tails during the relaxation stage with a length around 30 mean free paths under the chemical reactant rate $\nu_{\mathcal{AB}}^{\mathcal{CD}} = 0.03$ and 300 mean free paths under the chemical reaction rate $\nu_{\mathcal{AB}}^{\mathcal{CD}} = 0.003$. Mass transfer occurs in this process, leading to the observation of overshoots in the profiles of reactants. 

\begin{figure}[H]
	\centering
    \subfloat[]
    {
    	\includegraphics[width=0.4 \textwidth]{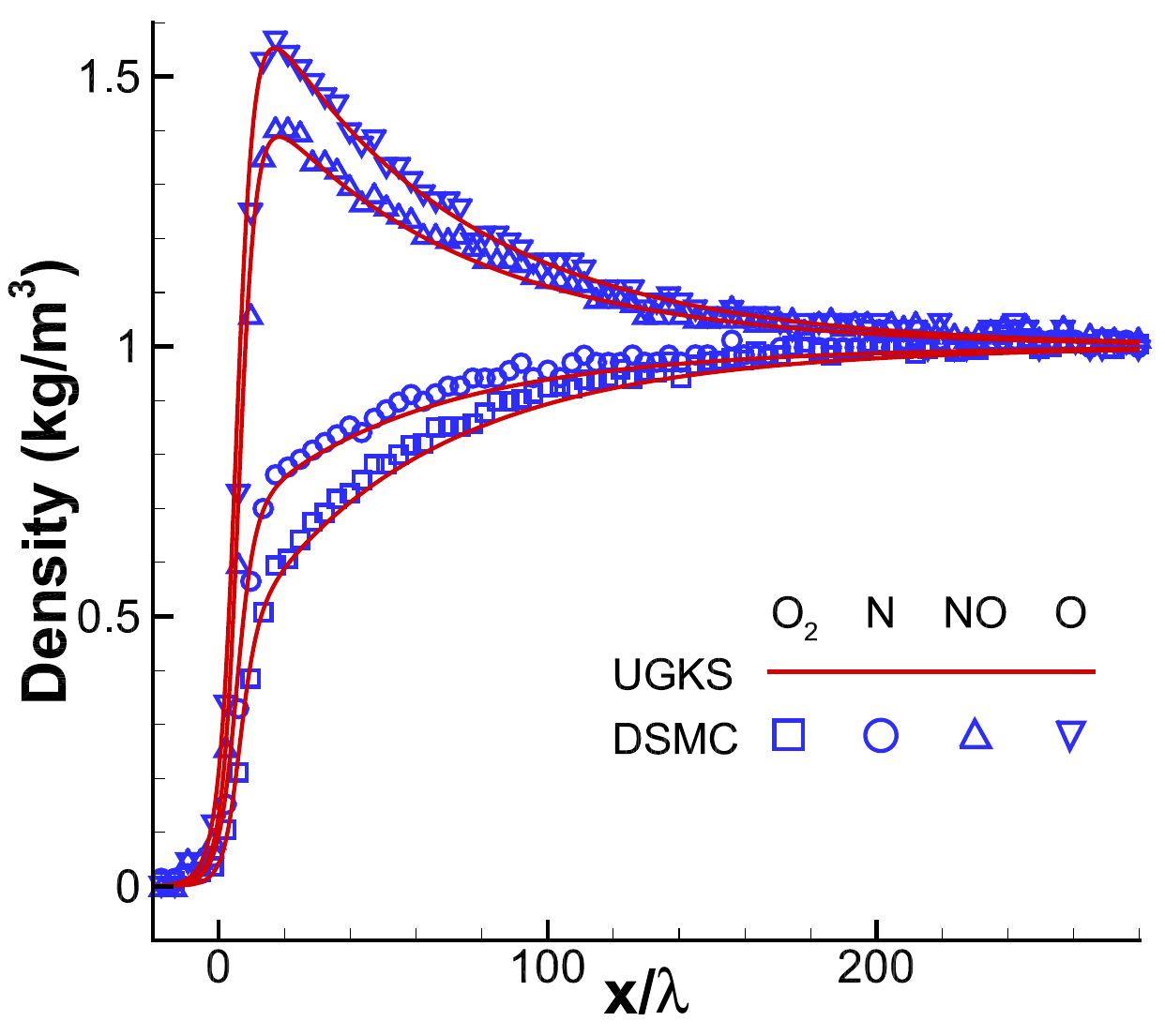}
    }
    \subfloat[]{
		\includegraphics[width=0.4 \textwidth]{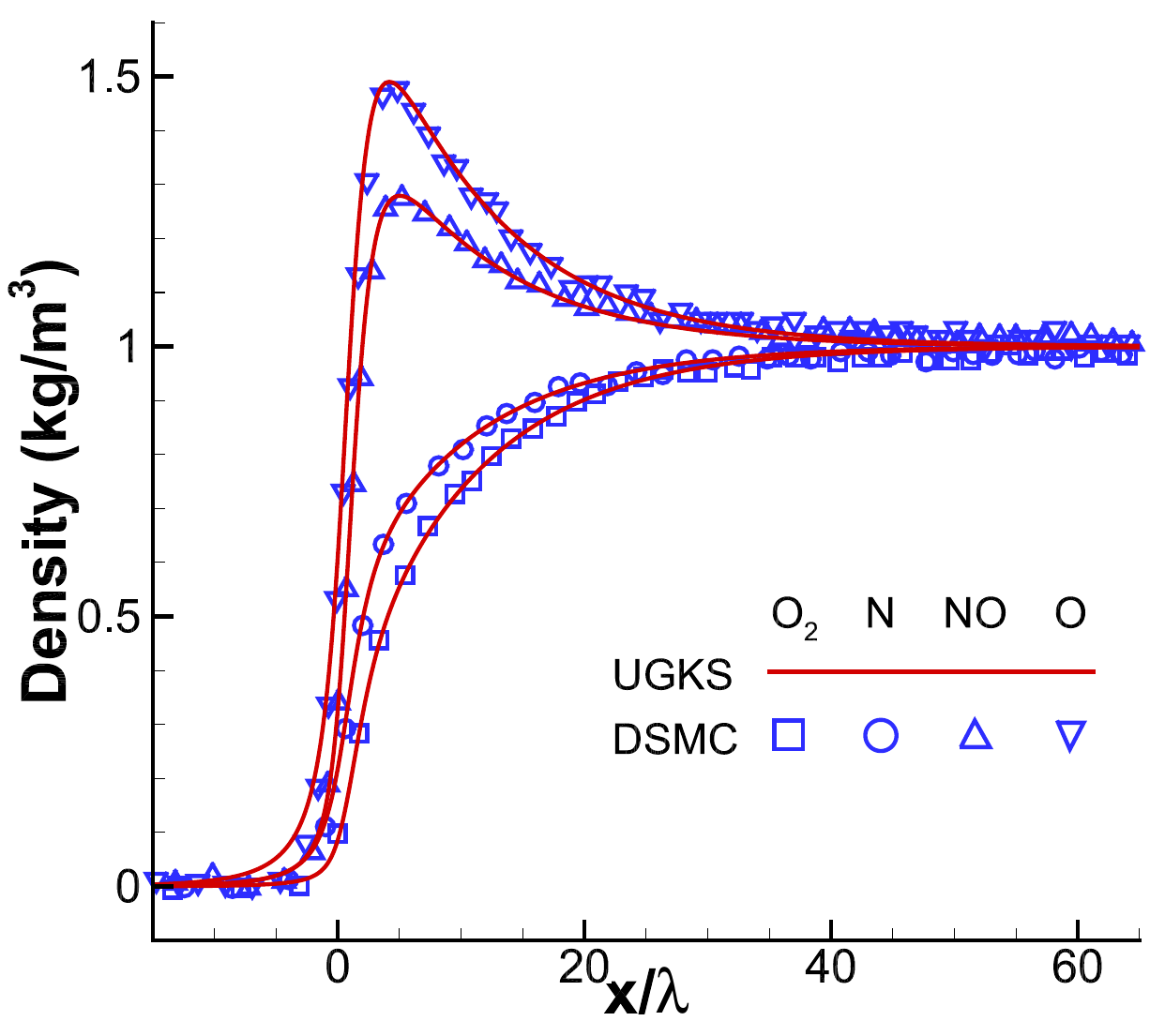}
	} 
	\caption{Steady shock structure of the four densities $\rho_\alpha$ at ${\rm Ma}^{-}=1.5$, corresponding to $\Delta \chi = 0.0317$, and ${\rm Ma}^{-}=2.5$, corresponding to $\Delta \chi = 0.744$ compared with the DSMC method.}
	\label{fig:shock-ma1-5-and-2-5}
\end{figure}

Additionally, the correctness of the current kinetic model and the numerical scheme is validated by comparing the results with the DSMC method at the upstream Mach number ${\rm Ma}^{-} = 1.5$ and ${\rm Ma}^{-} = 2.5$. The real gases with the parameters listed in Table~\ref{tab:gas-mix} and Table~\ref{tab:chem-rate} are applied. The upstream state is set as total number density $n^{-}=10^{20}$ ${\rm m}^{-3}$ with the concentration fractions equal to $\chi_{\mathcal{A}}^{-}=0.0614$, $\chi_\mathcal{B}^{-}=0.1228$, $\chi_{\mathcal{C}}^{-}=0.4913$, and $\chi_{\mathcal{D}}^{-}=0.3245$. The inflow temperature of each species is at equilibrium state with $T^{-}=6000$ K. By iteration, the upstream and downstream parameters at ${\rm Ma}^{-} = 1.5$ and ${\rm Ma}^{-}=2.5$ are obtained, corresponding to $\Delta \chi = 0.0317$ and $\Delta \chi= 0.744$, respectively. The parameters are shown in Table~\ref{tab:reacting-shock-Ma1.5} and Table~\ref{tab:reacting-shock-Ma2.5}. Referring to the molecular diameter $d_\mathcal{A}$, the mean free path is calculated by $l^{-}=0.0135877$ m. The results of UGKS are consistent with those of the DSMC method as shown in Fig.~\ref{fig:shock-ma1-5-and-2-5}.

\begin{table}[h]
	\caption{Upstream and downstream parameters at ${\rm Ma}^{-}=1.5$, corresponding to $\Delta \chi = 0.0317$.}
	\centering
	\begin{tabular}{*{4}{c}}
	\toprule
		Upstream &Value & Downstream &Value    \\
	\midrule
		$n^{-}$  &$1.0000\times 10^{20}~ {\rm m}^{-3}$  
		&$n^{+}$ &$1.5264\times 10^{20}~{\rm m}^{-3}$     \\
		$\rho^{-}$  &$3.9228\times 10^{-6}~{\rm kg/m}^3$  
		&$\rho^{+}$ &$5.9877\times 10^{-6}~{\rm kg/m}^3$  \\
		$U^{-}$  &$2702.2$ m/s & $U^{+}$ &$1770.3$ m/s    \\
		$T^{-}$  &$6000$ K  & $T^{+}$ &$8618$ K  	      \\
		$\chi_{\mathcal{A}}^{-}$  &$0.0614$ & $\chi_{\mathcal{A}}^{+}$ &$0.0931$ \\
		$\chi_{\mathcal{B}}^{-}$  &$0.1228$ & $\chi_{\mathcal{B}}^{+}$ &$0.1545$ \\
		$\chi_{\mathcal{C}}^{-}$  &$0.4913$ & $\chi_{\mathcal{C}}^{+}$ &$0.4596$ \\
		$\chi_{\mathcal{D}}^{-}$  &$0.3245$ & $\chi_{\mathcal{D}}^{+}$ &$0.2928$ \\
	\bottomrule
	\end{tabular}
	\label{tab:reacting-shock-Ma1.5}
\end{table}

\begin{table}[H]
	\caption{Condition parameters at ${\rm Ma}^{-}=2.5$, corresponding to $\Delta \chi = 0.744$.}
	\centering
	\begin{tabular}{*{4}{c}}
	\toprule
		Upstream & Value & Downstream&Value    \\
	\midrule
		$n^{-}$  &$1.0000\times 10^{20}{~ \rm m}^{-3}$  
	  & $n^{+}$ &$2.4898\times 10^{20}{~ \rm m}^{-3}$         \\
		$\rho^{-}$  &$3.9228\times 10^{-6}{~ \rm kg/m}^3$  
		&$\rho^{+}$ &$9.7670\times 10^{-6}{~\rm kg/m}^3$      \\
		$U^{-}$  &$4500.2$ m/s &$U^{+}$ &$1807.4$ m/s         \\
		$T^{-}$  &$6000$ K  & $T^{+}$ &$16238$ K  			  \\
		$\chi_{\mathcal{A}}^{-}$  &$0.0614$   & $\chi_{\mathcal{A}}^{+}$ &$0.1358$  \\
		$\chi_{\mathcal{B}}^{-}$  &$0.1228$   & $\chi_{\mathcal{B}}^{+}$ &$0.1972$  \\
		$\chi_{\mathcal{C}}^{-}$  &$0.4913$   & $\chi_{\mathcal{C}}^{+}$ &$0.4169$  \\
		$\chi_{\mathcal{D}}^{-}$  &$0.3245$   & $\chi_{\mathcal{D}}^{+}$ &$0.2501$  \\
	\bottomrule
	\end{tabular}
	\label{tab:reacting-shock-Ma2.5}
\end{table}

\subsection{Hypersonic Flow around a Circular Cylinder}\label{sec:cylinder}
The hypersonic rarefied flow around a circular cylinder is a typical case containing multi-scale and non-equilibrium physics. At the post-shock region, the flow becomes rather continuum due to the compression, and the temperature rises rapidly, promoting the chemical reaction. In the lee-ward region, the rarefied flow comes about due to the gas expansion. In this classical test case, several simulations are conducted. First, the flow at the ${\rm Kn}_{\infty}=0.0488$ is computed and compared with the DSMC method to validate the current model and scheme. Then, the influence of rarefied effects on chemical reactions is discussed by increasing the inflow Kn number to ${\rm Kn}_{\infty}=0.488$. At last, the influences of energy release type are investigated by setting $\Delta E > 0$, $\Delta E < 0$, and $\Delta E = 0$. The temperature along the stagnation line and the heat flux around the wall show that exothermic and endothermic reactions will result in a twofold difference in aerodynamic heating.

\begin{figure}[H]
	\centering
    \subfloat[]{
    		\includegraphics[width=0.4 \textwidth]{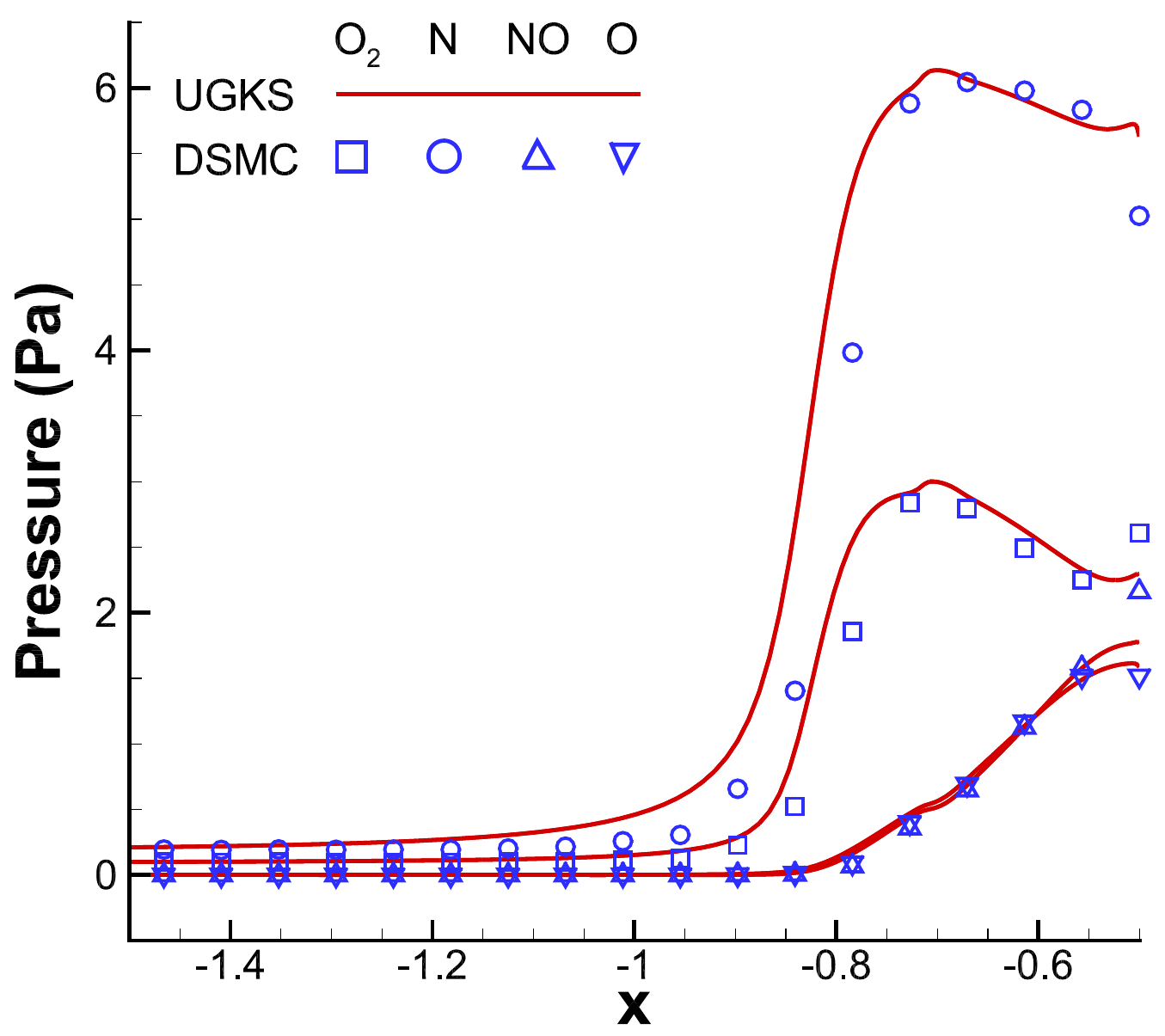}
    	}
    \subfloat[]{\label{che_cylinder_0.0488f}
			\includegraphics[width=0.4 \textwidth]{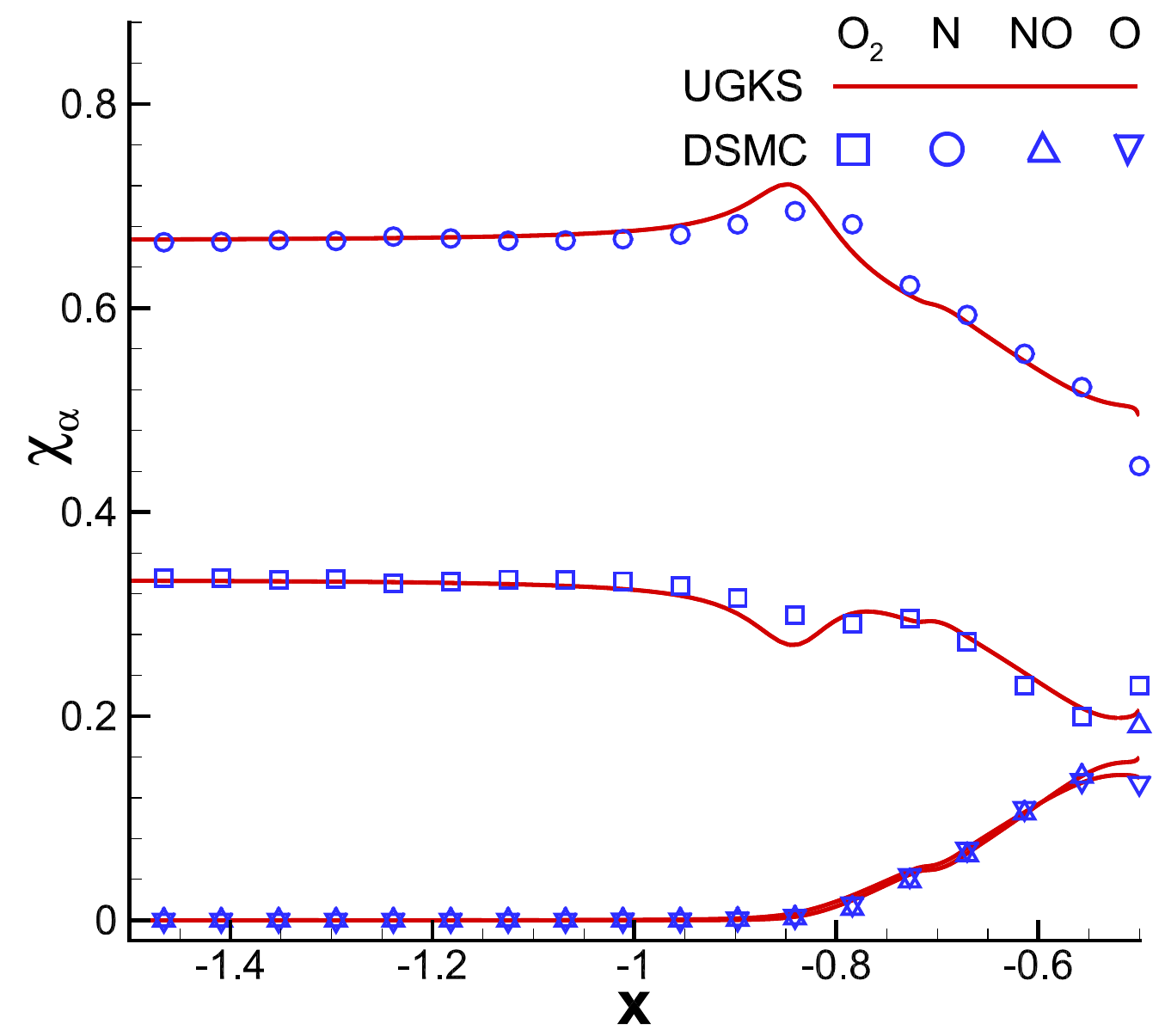}
		}
	\caption{Hypersonic chemical reaction flow around a circular cylinder at ${\rm Ma}_\infty=7.03$, ${\rm Kn}_\infty=0.0488$. Distributions of (a) pressure and (b) concentration fraction of each species along the stagnation line, compared with the DSMC method.}
	\label{fig:cylinder-0-0488-stag}
\end{figure}

\begin{figure}[H]
	\centering
    \subfloat[]{
			\includegraphics[width=0.3 \textwidth]{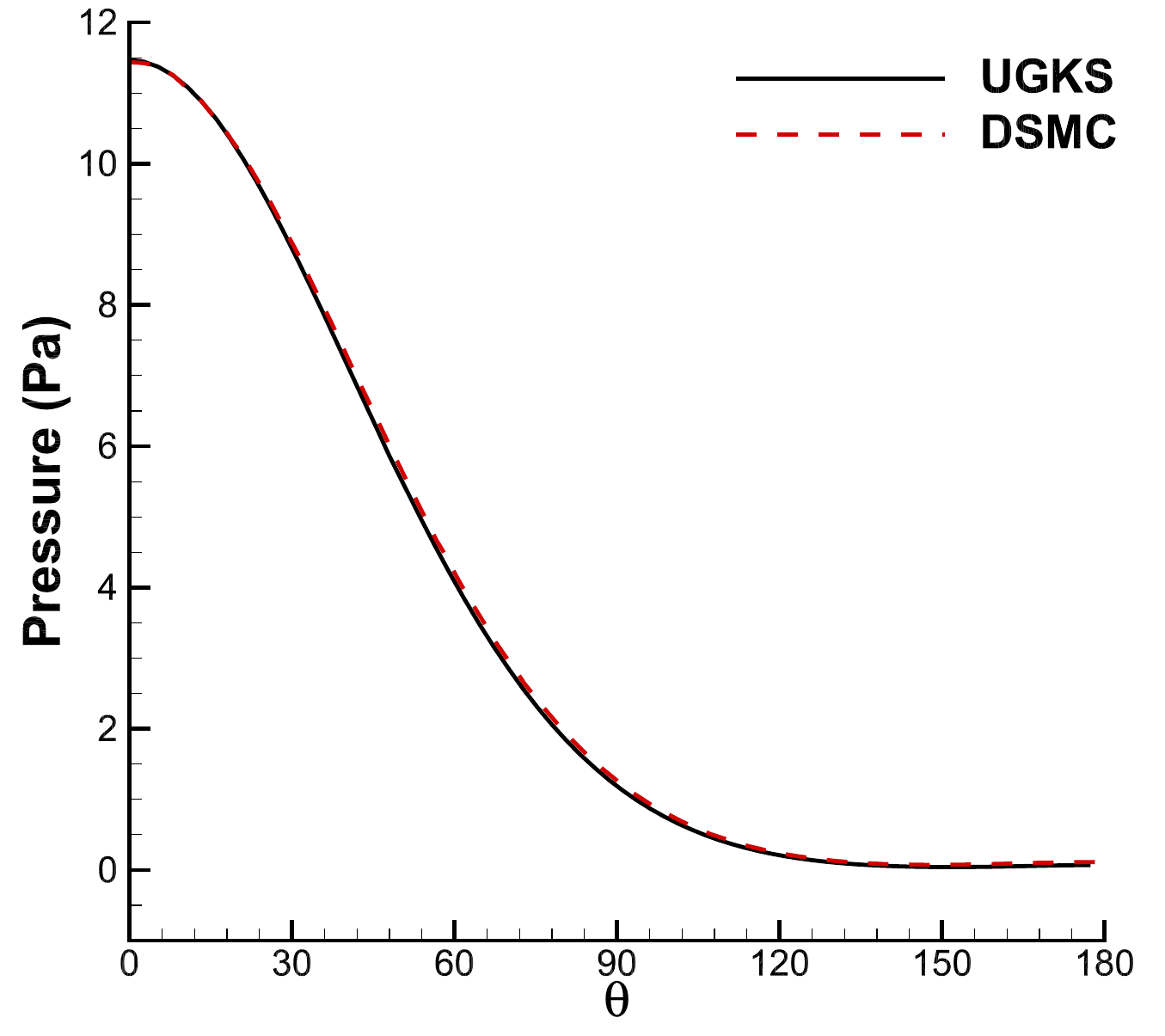}
		}
    \subfloat[]{
    		\includegraphics[width=0.3 \textwidth]{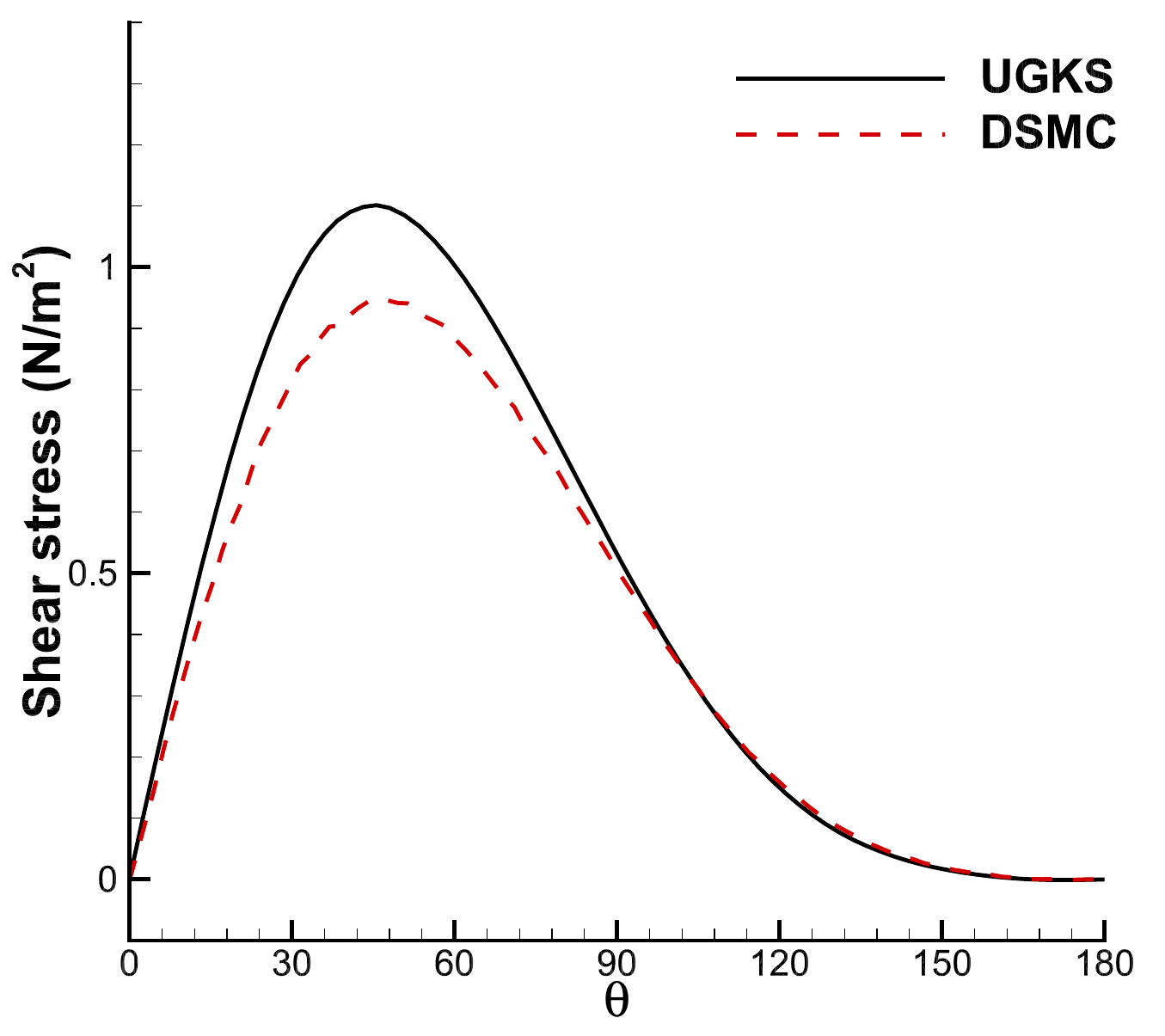}
    	}
    \subfloat[]{
			\includegraphics[width=0.3 \textwidth]{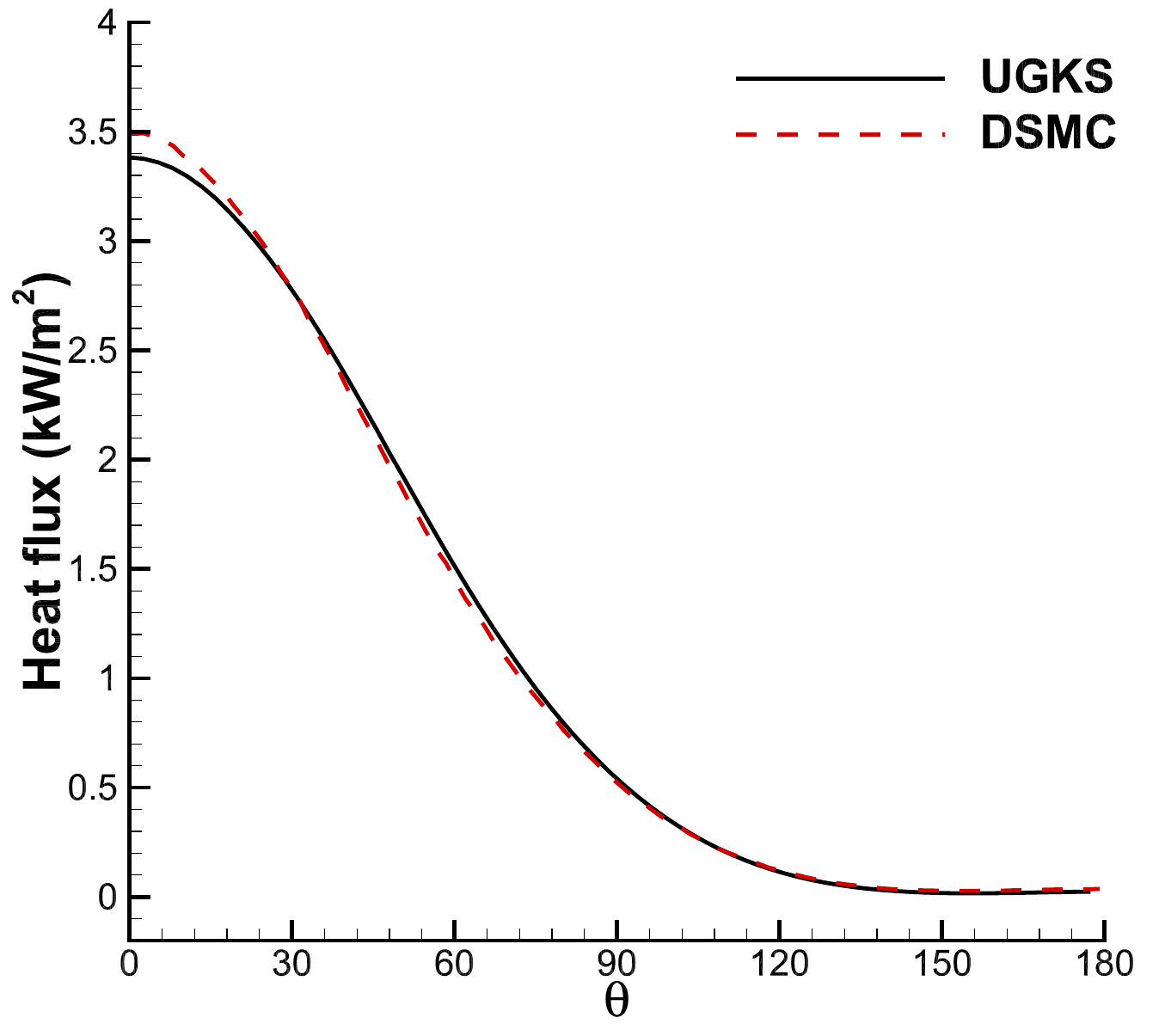}
		}
	\caption{Hypersonic chemical-reaction flow around a circular cylinder at ${\rm Ma}_\infty=7.03$, ${\rm Kn}_\infty=0.0488$. Distributions of (a) pressure, (b) shear stress, and (c) heat flux on the cylinder surface, compared with the DSMC method.}
	\label{fig:cylinder-0-0488-wall}
\end{figure}

\begin{figure}[H]
	\centering
    \subfloat[]
    {
    	\includegraphics[width=0.45 \textwidth]{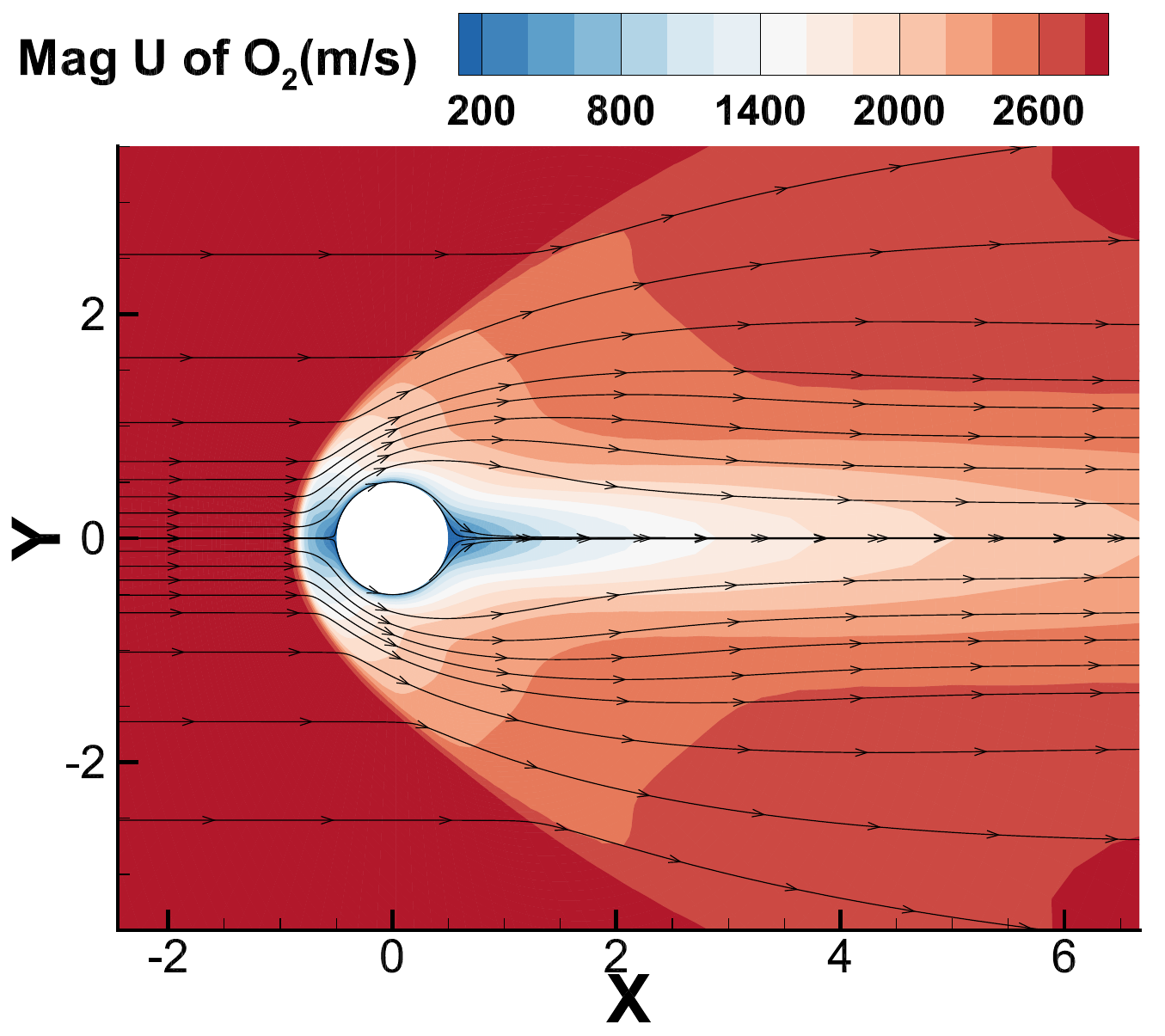}
    }
    \subfloat[]{
			\includegraphics[width=0.45 \textwidth]{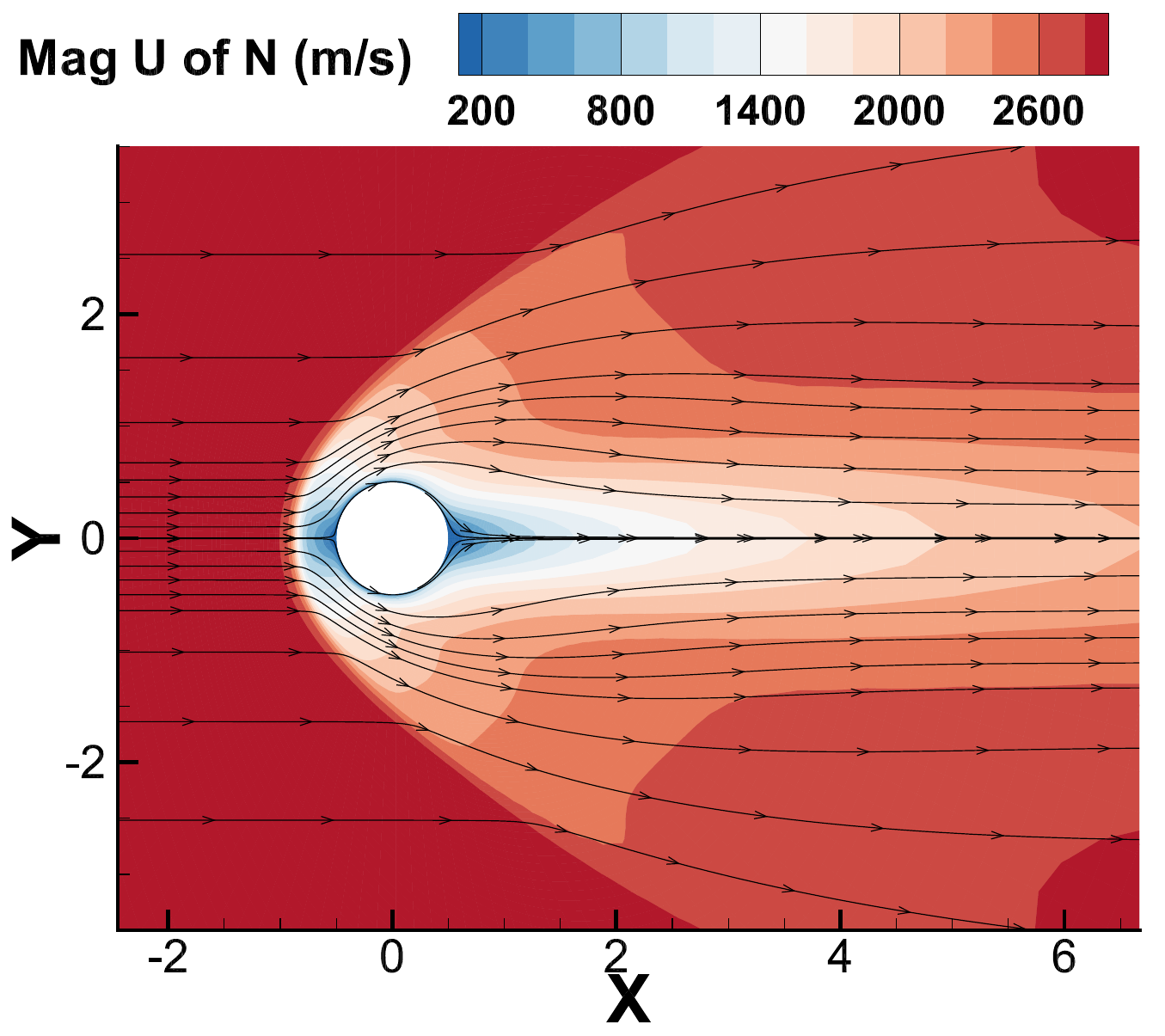}
	} \\
	\vspace{1.5mm}
    \subfloat[]{
    		\includegraphics[width=0.45 \textwidth]{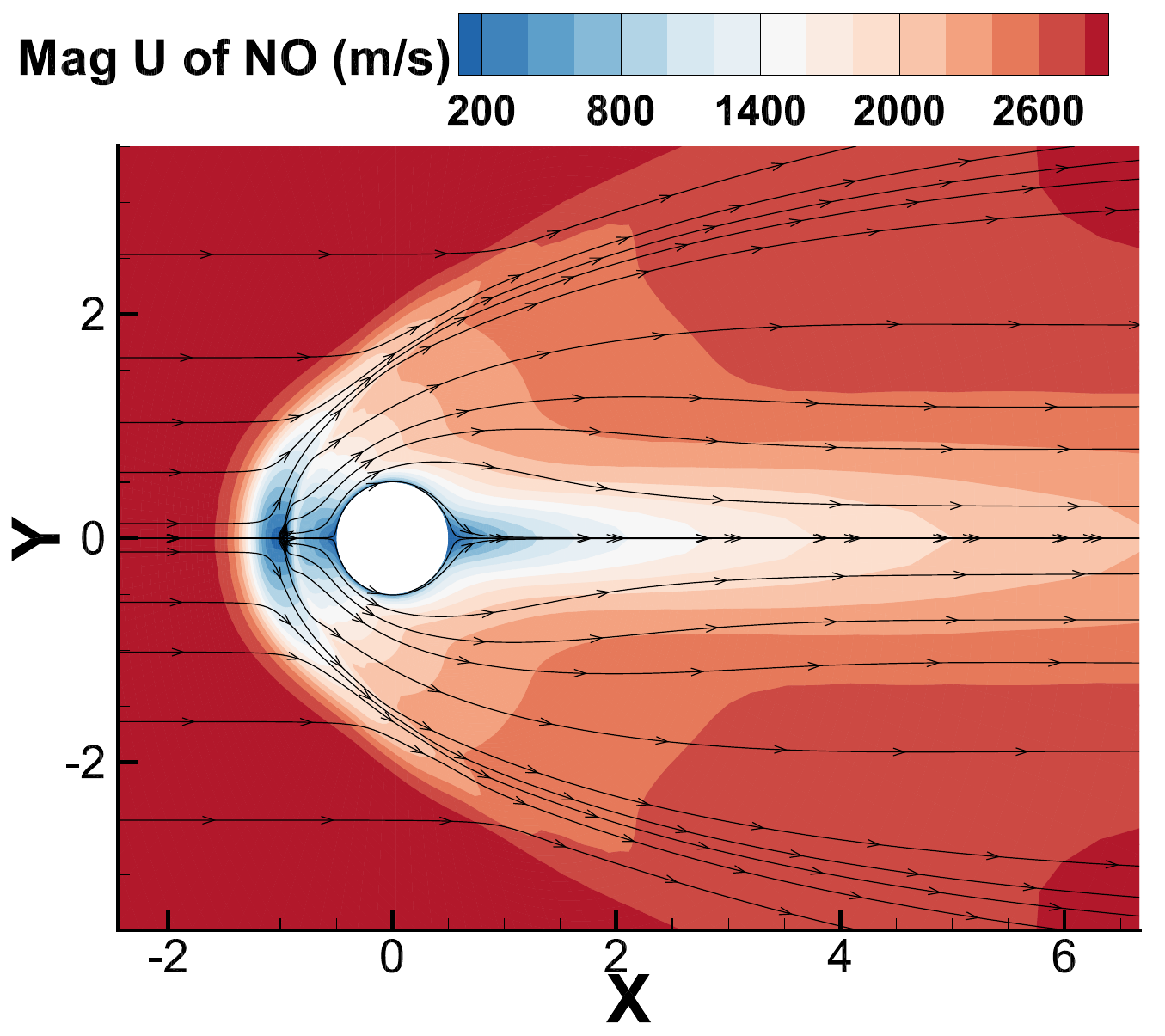}
    } 
    \subfloat[]{
			\includegraphics[width=0.45 \textwidth]{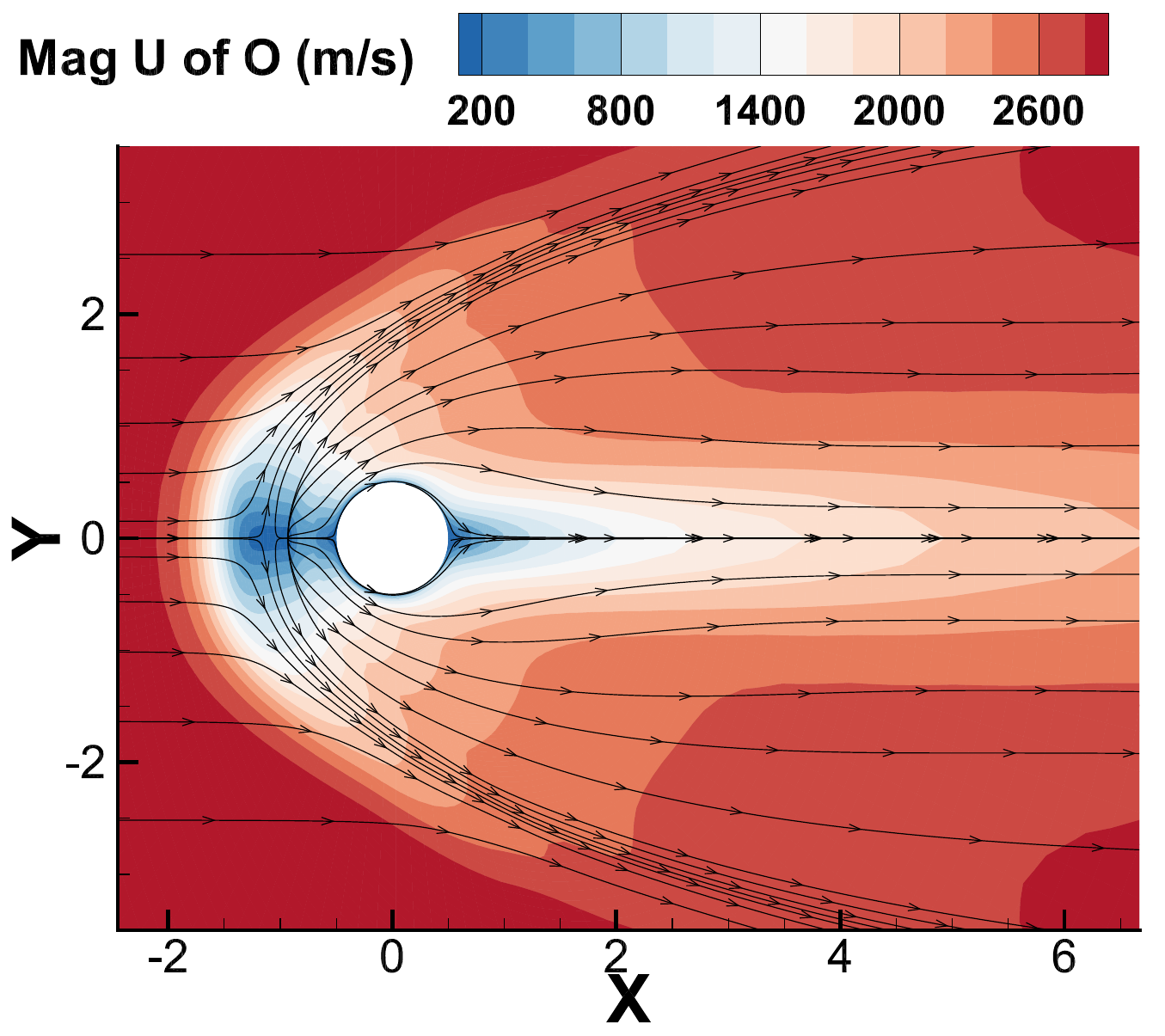}
	}
	\caption{Hypersonic chemical-reaction flow around a circular cylinder at ${\rm Ma}_\infty=7.03$, ${\rm Kn}_\infty=0.0488$. The contours of magnitude of velocity and streamline of the species (a) ${\rm O}_2$, (b) ${\rm N}$, (c) ${\rm NO}$, and (d) ${\rm O}$.}
	\label{fig:cylinder-U-0-0488} 
\end{figure}

\begin{figure}[H]
	\centering
    \subfloat[]{
    		\includegraphics[width=0.3 \textwidth]{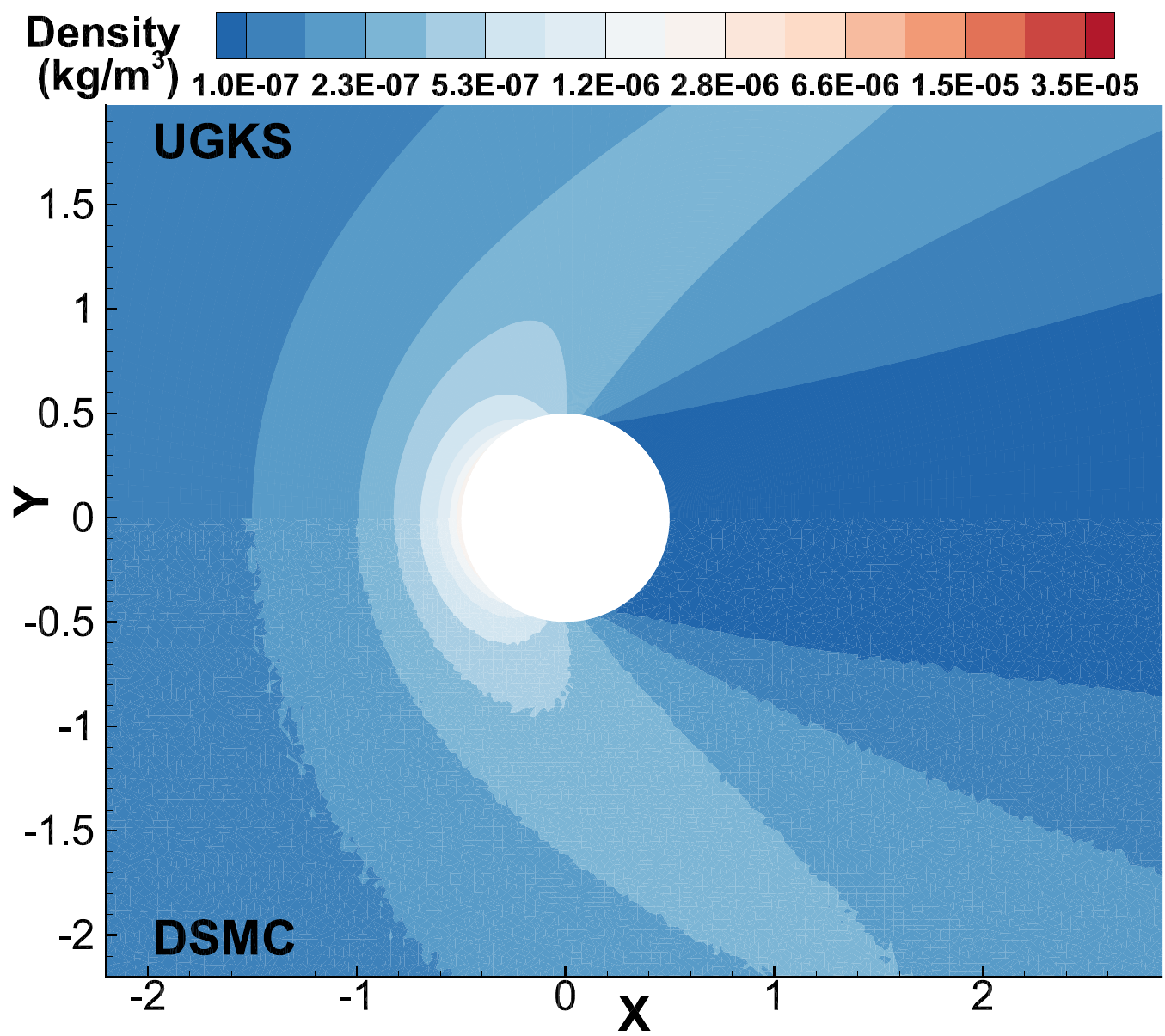}
    	}
    \subfloat[]{
			\includegraphics[width=0.3 \textwidth]{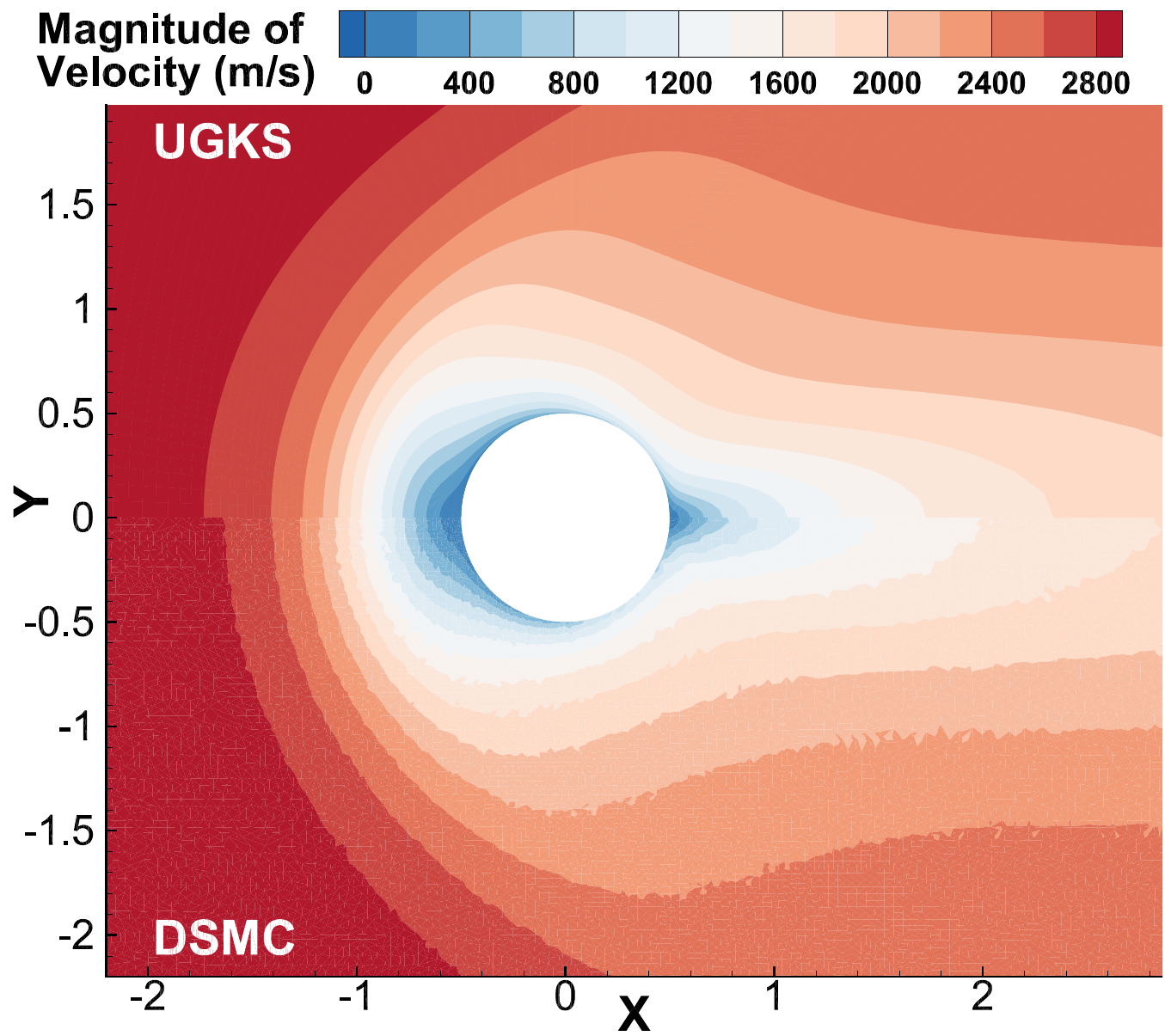}
		}
    \subfloat[]{
    		\includegraphics[width=0.3 \textwidth]{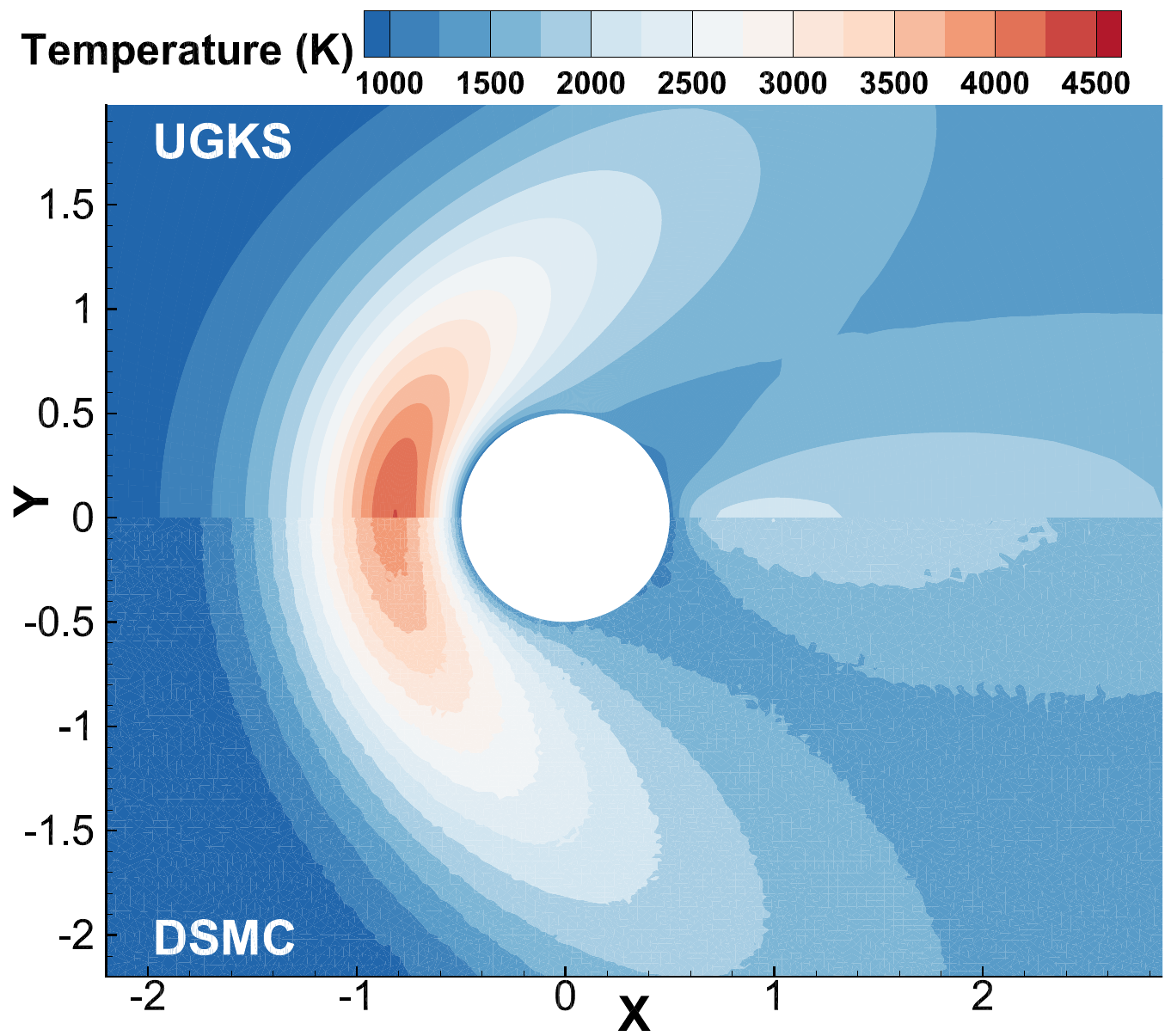}
    	} 
	\caption{Hypersonic chemical-reaction flow around a circular cylinder at ${\rm Ma}_\infty=7.03$, ${\rm Kn}_\infty=0.488$. Contours of (a) global density, (b) magnitude of global velocity, and (c) global temperature, compared with the DSMC method.}
	\label{fig:cylinder-0-488-contour}
\end{figure}

In the first simulation, the state is set as $n_{\rm O_2,\infty}:n_{\rm N,\infty}=1:2$, $n_{\rm O_2,\infty}=1.392\times 10^{19}~{\rm m}^{-3}$, $n_{\rm N,\infty}=2.784\times 10^{19}{~\rm m}^{-3}$, $U_{\infty}=3000$ m/s, $T_{\infty}=500$ K, ${\rm Ma}_{\infty}=7.03$, ${\rm Kn}_{\infty}=0.0488$, and $T_w=600$ K. The reference length is the diameter of the cylinder 1.0 m. The mesh in the physical space is set to be $145\times 160$. The height of the first layer near the wall is 0.001 m. And the mesh in the phase space is set to be $90\times 90$. 
The contours of global density, magnitude of velocity, and temperature are shown in Fig.~\ref{fig:cylinder-0-488-contour}, which are compared with those of DSMC method. It can be seen that the peak values, contours, and the positions of detached shock waves in various flow fields match well with the results of DSMC method. To provide a more detailed display of the flow states of each component, the pressure along stagnation lines and concentration fractions are presented in Fig.~\ref{fig:cylinder-0-0488-stag}. The concentration fractions in Fig.~\ref{fig:cylinder-0-0488-stag}(b) show the chemical reactions become relatively intensive in the post-shock region, caused by the increment of global temperature in this highly compressed domain. Also, by comparing with the results from DSMC method, it shows that the UGKS can effectively capture the pressure and species variation process along stagnation lines. Furthermore, the high-order quantities such as pressure, heat flux, and shear stress on the surface of the cylinder are also compared in Fig.~\ref{fig:cylinder-0-0488-wall}. The pressure and heat flux exhibit good agreement with the DSMC results, while there is some error in the shear. This indicates that there is still room for improvement in the single relaxation model when calculating high-order quantities. 
Furthermore, an interesting observation is that stagnation points away from the wall can be observed in the streamlines of the products, shown in Fig.~\ref{fig:cylinder-U-0-0488}, with the stagnation points more pronounced for component O. This may be due to the fact that behind the wave, chemical reactions become increasingly intense, and newly generated particles only acquire initial velocities behind the wave.

\begin{figure}[H]
	\centering
    \subfloat[]
    {
    	\includegraphics[width=0.3 \textwidth]{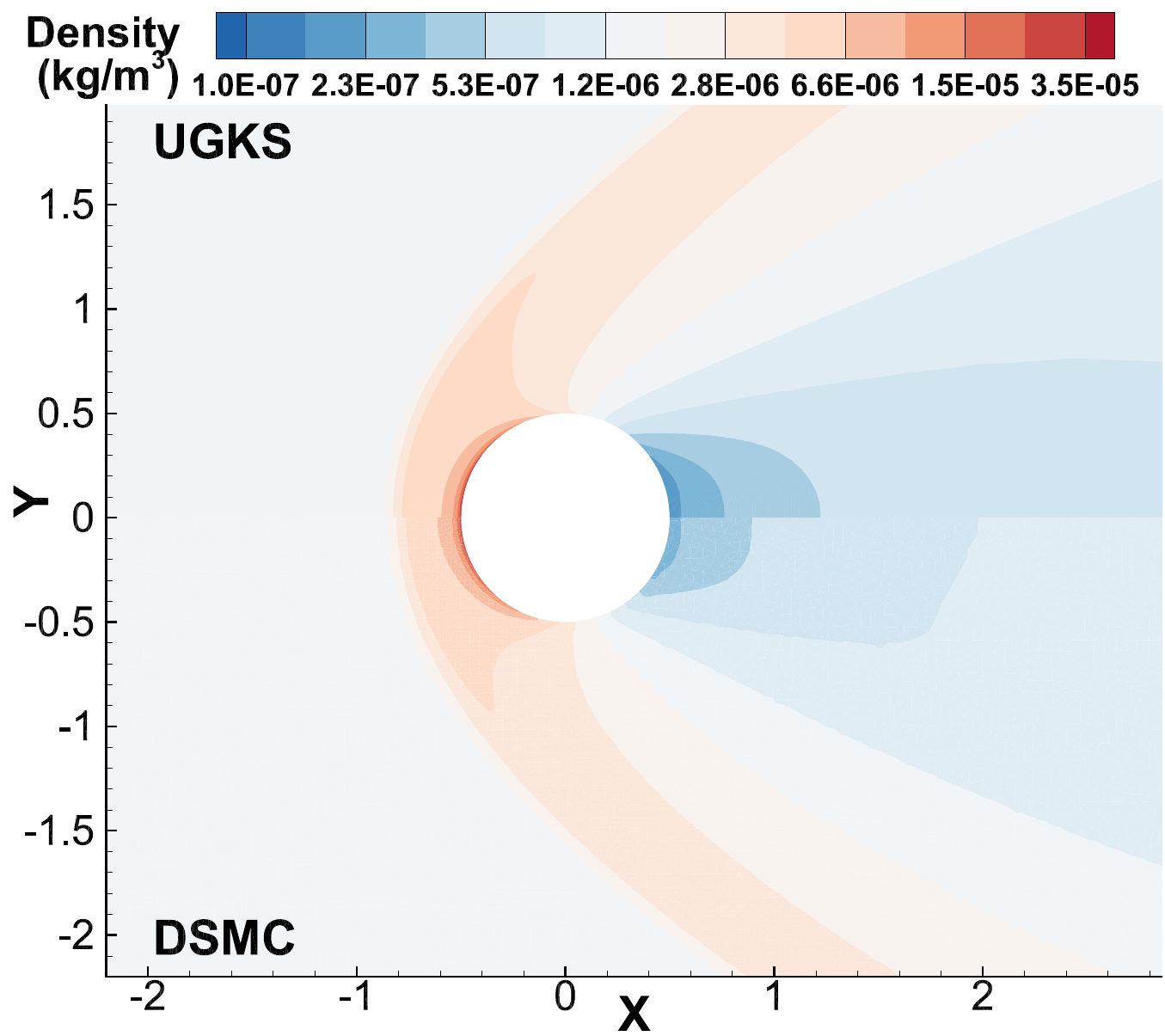}
    }
    \subfloat[]{
			\includegraphics[width=0.3 \textwidth]{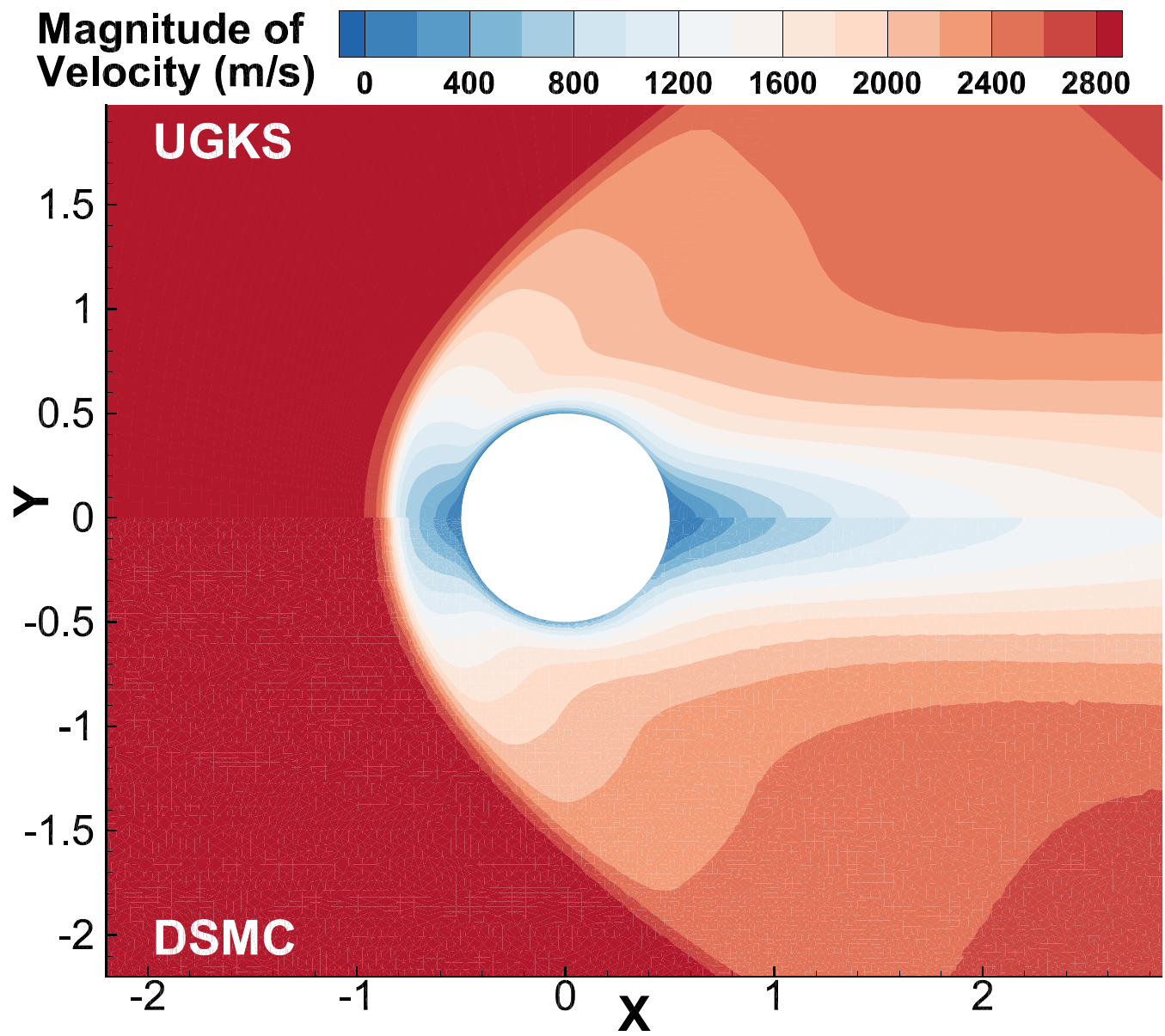}
	}
    \subfloat[]{
    		\includegraphics[width=0.3 \textwidth]{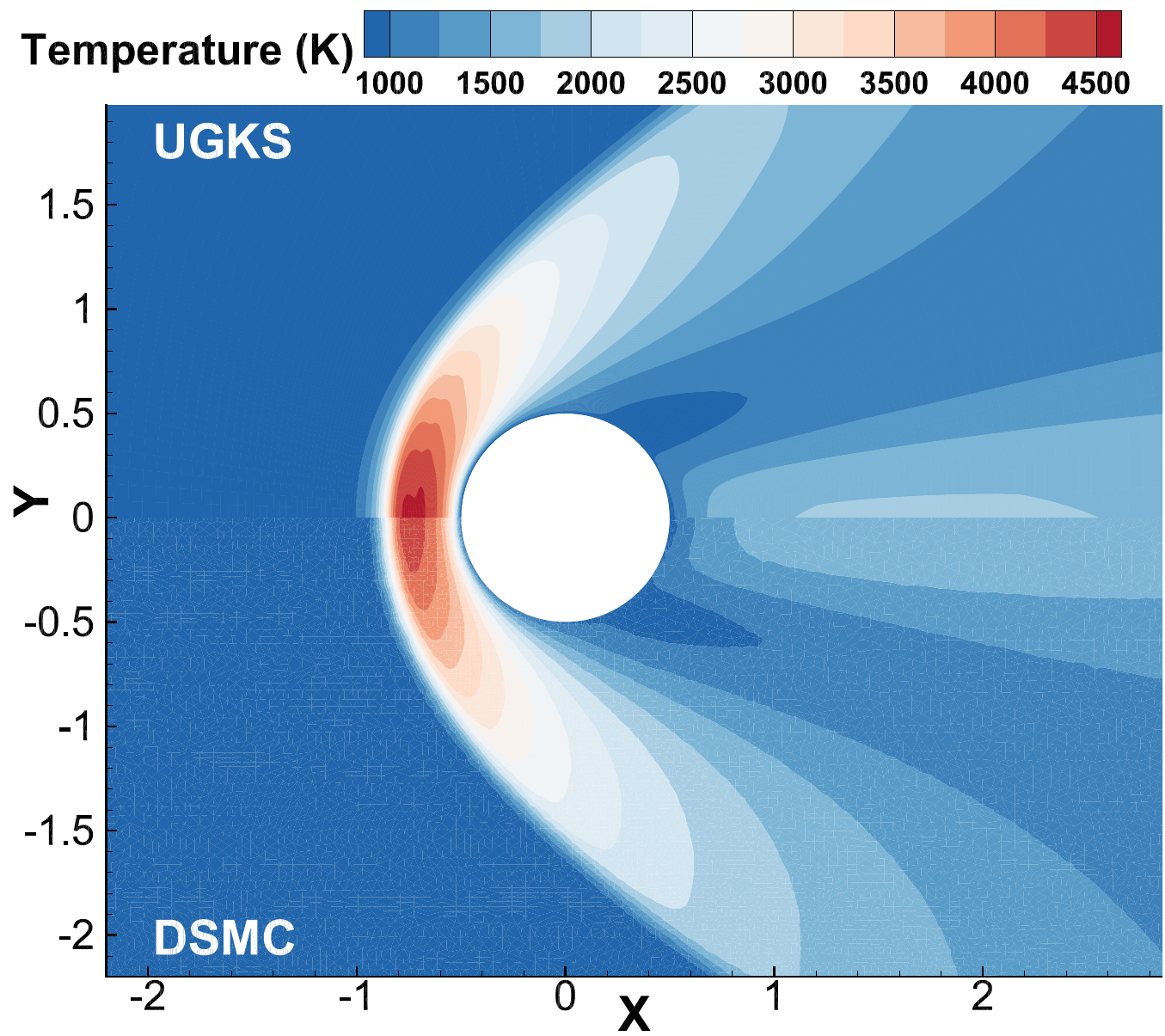}
    	} 
	\caption{Hypersonic chemical-reaction flow around a circular cylinder at ${\rm Ma}_\infty=7.03$, ${\rm Kn}_\infty=0.0488$. Contours of (a) global density, (b) magnitude of global velocity, and (c) global temperature, compared with the DSMC method.}
	\label{fig:cylinder-0-0488-contour}
\end{figure}

\begin{figure}[H]
	\centering
    \subfloat[]{
    		\includegraphics[width=0.4 \textwidth]{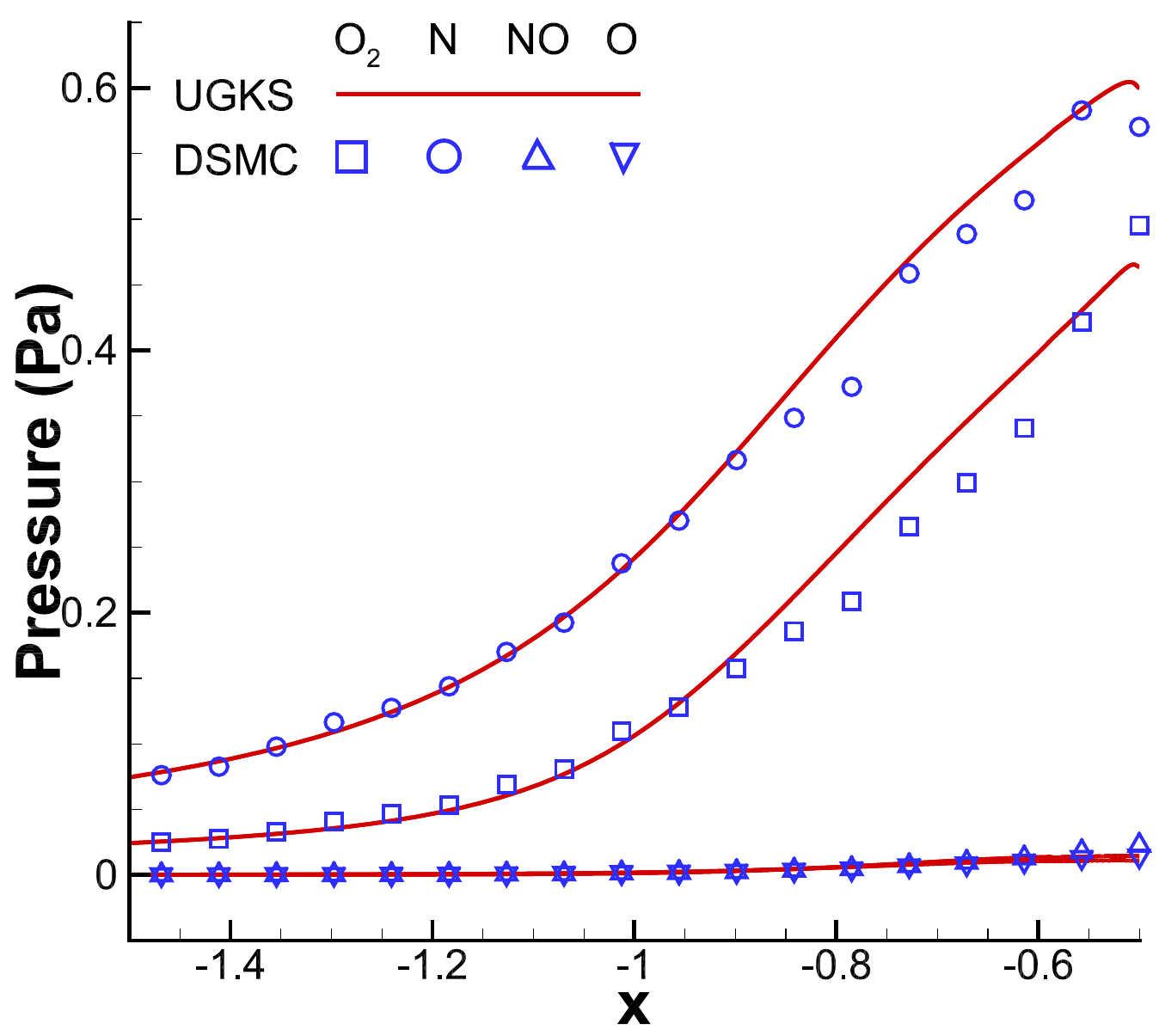}
    	}
    \subfloat[]{
			\includegraphics[width=0.4 \textwidth]{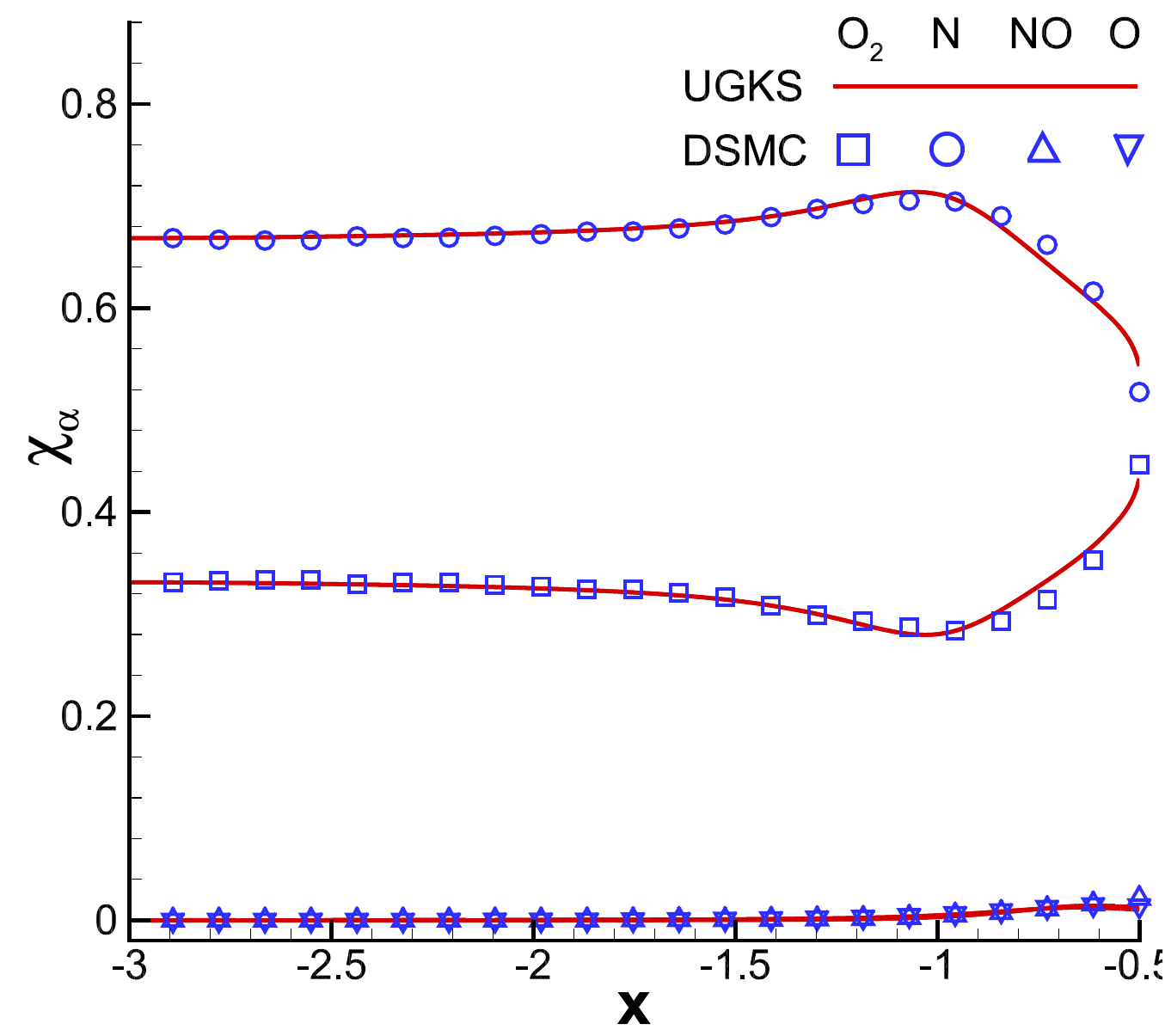}
		}
	\caption
	{Hypersonic chemical reaction flow around a circular cylinder at ${\rm Ma}_\infty=7.03$, ${\rm Kn}_\infty=0.488$. Distributions of (a) pressure and (b) concentration fraction of each species along the stagnation line, compared with the DSMC method.}
	\label{fig:cylinder-0-488-stag}
\end{figure}

\begin{figure}[H]
	\centering
    \subfloat[]{
			\includegraphics[width=0.3 \textwidth]{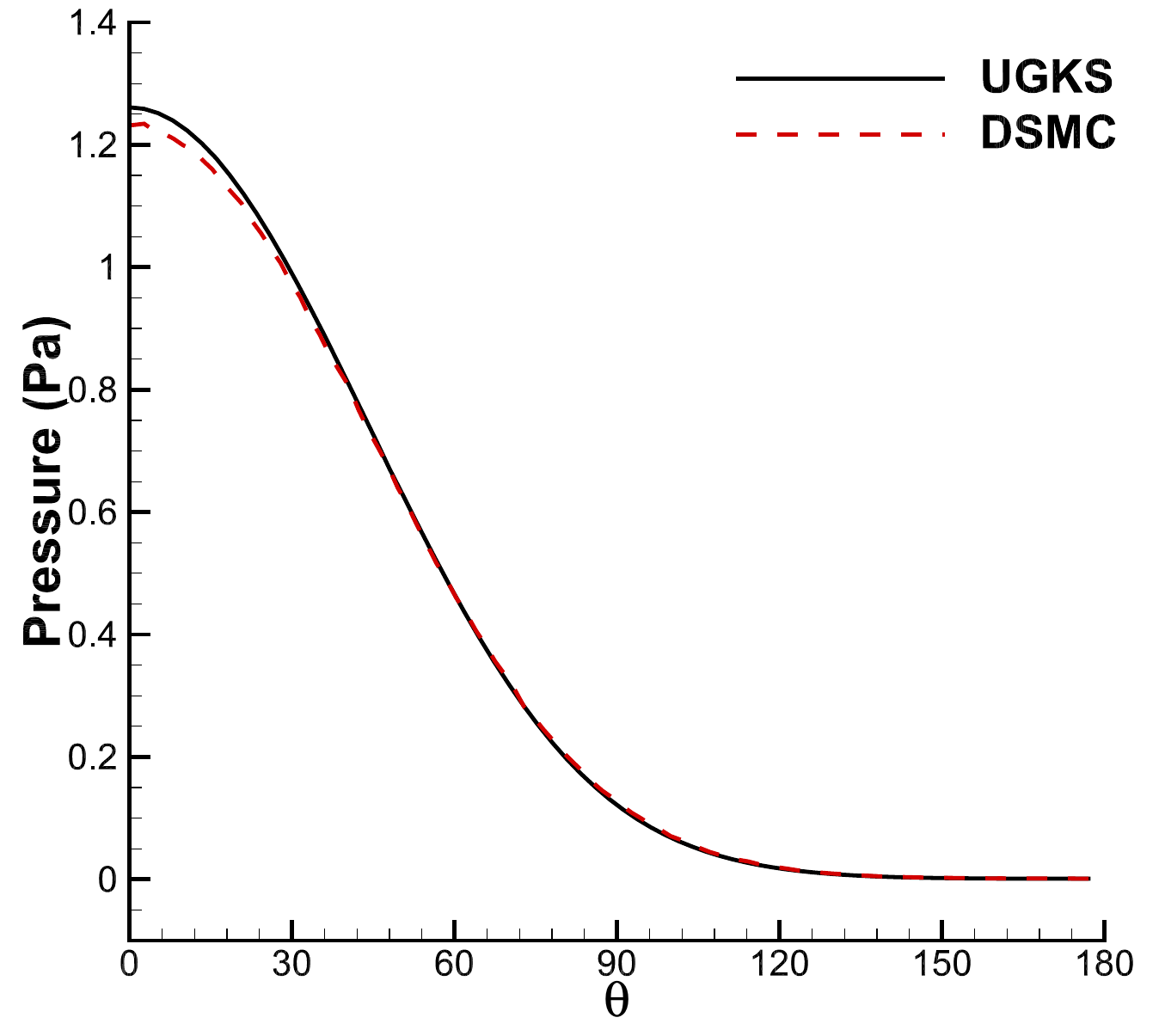}
		}
    \subfloat[]{
    		\includegraphics[width=0.3 \textwidth]{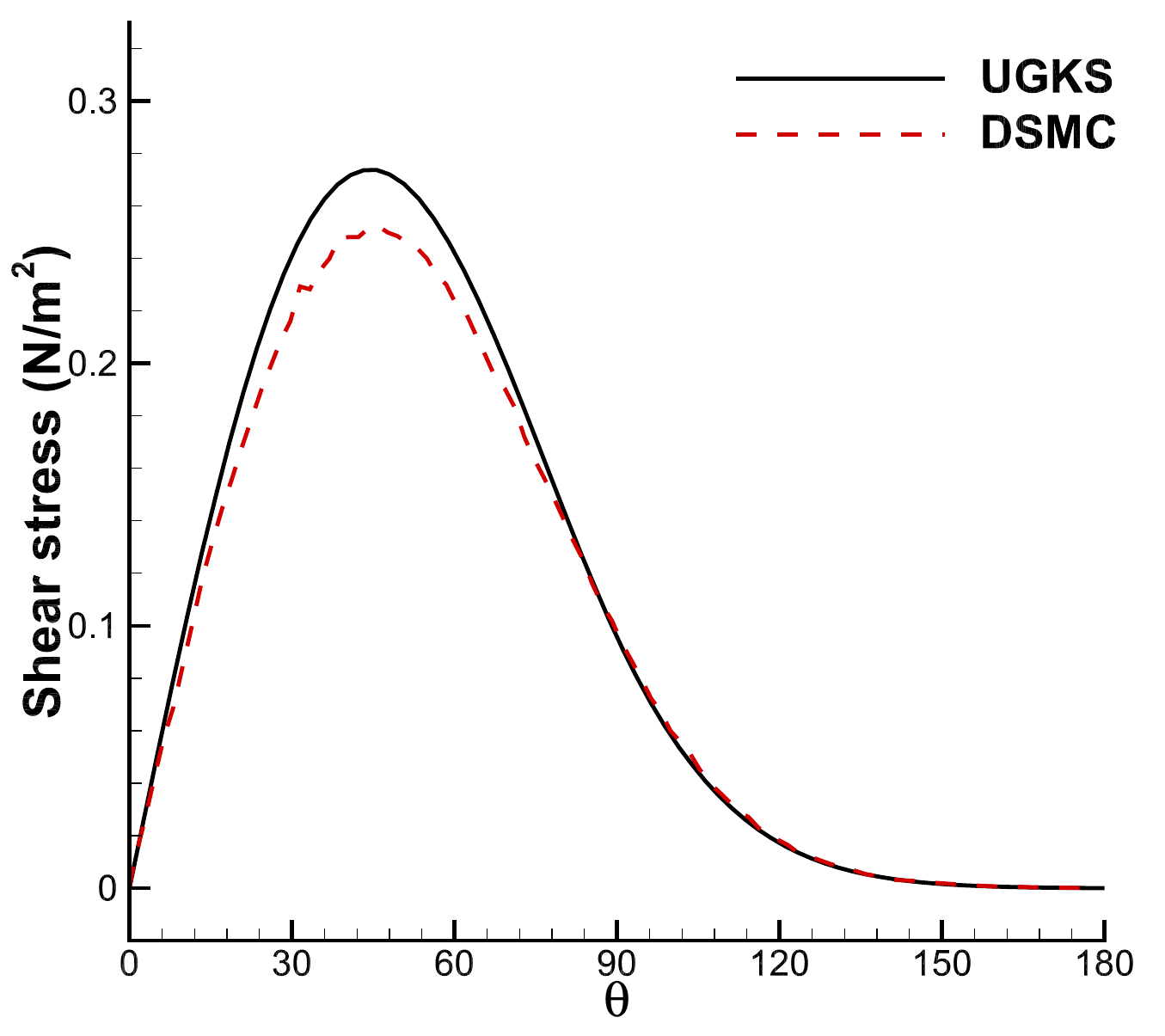}
    	}
    \subfloat[]{
			\includegraphics[width=0.3 \textwidth]{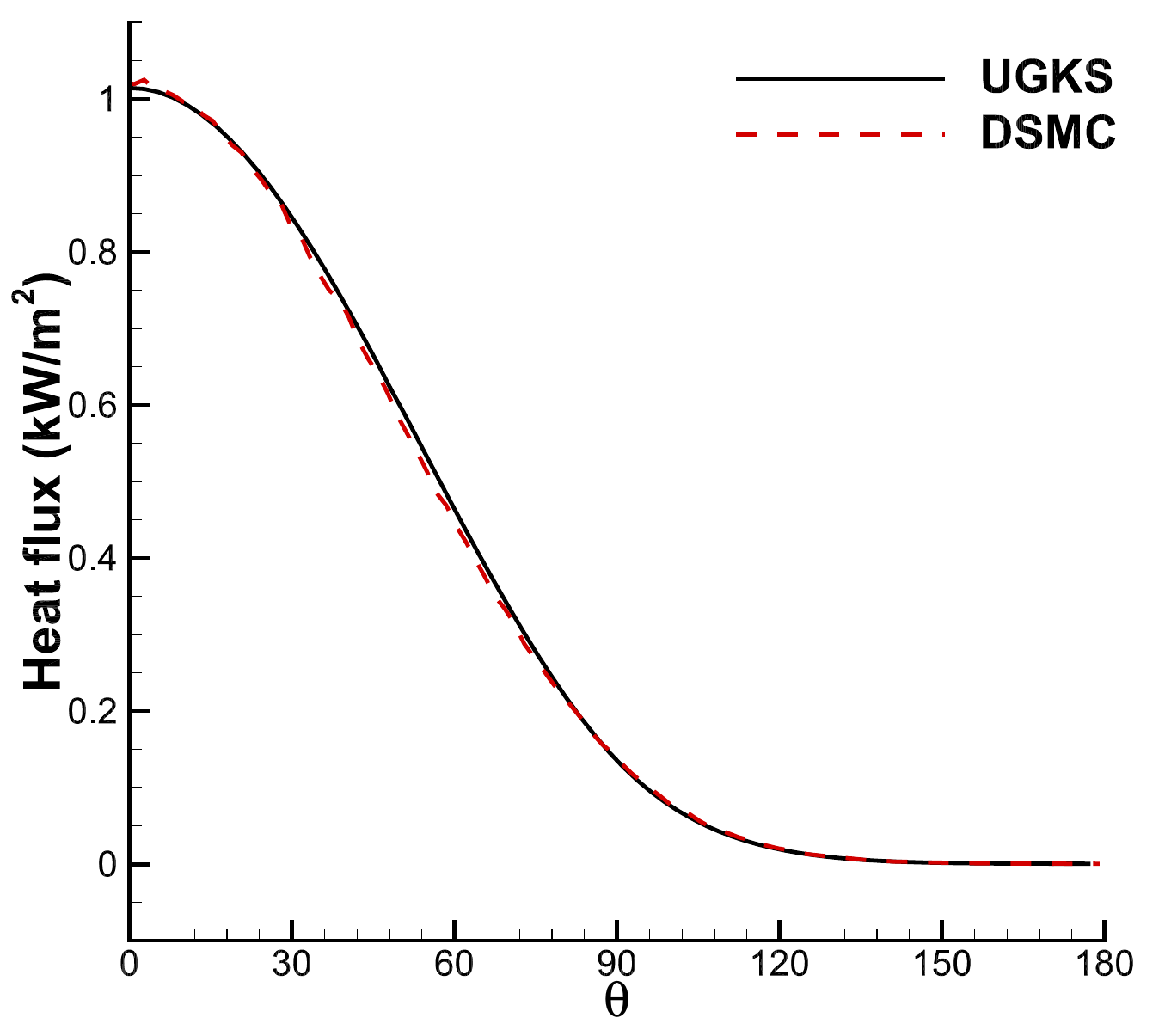}
		} 
	\caption{Hypersonic chemical-reaction flow around a circular cylinder at ${\rm Ma}_\infty=7.03$, ${\rm Kn}_\infty=0.488$. Distributions of (a) pressure, (b) shear stress, and (c) heat flux on the cylinder surface, compared with the DSMC method.}
	\label{fig:cylinder-0-488-wall}
\end{figure}

\begin{figure}[H]
	\centering
    \subfloat[]
    {
    	\includegraphics[width=0.45 \textwidth]{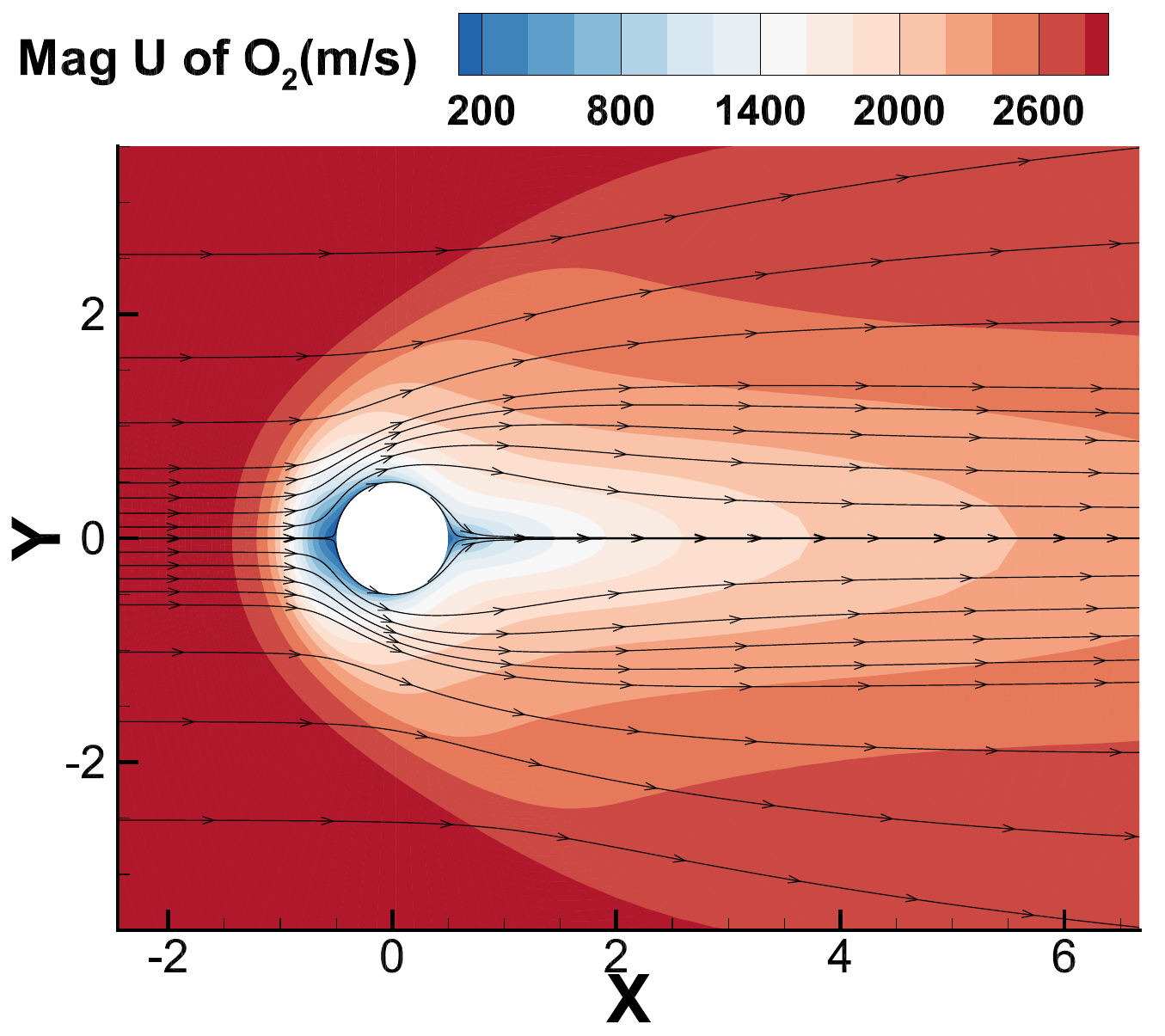}
    }
    \subfloat[]{
			\includegraphics[width=0.45 \textwidth]{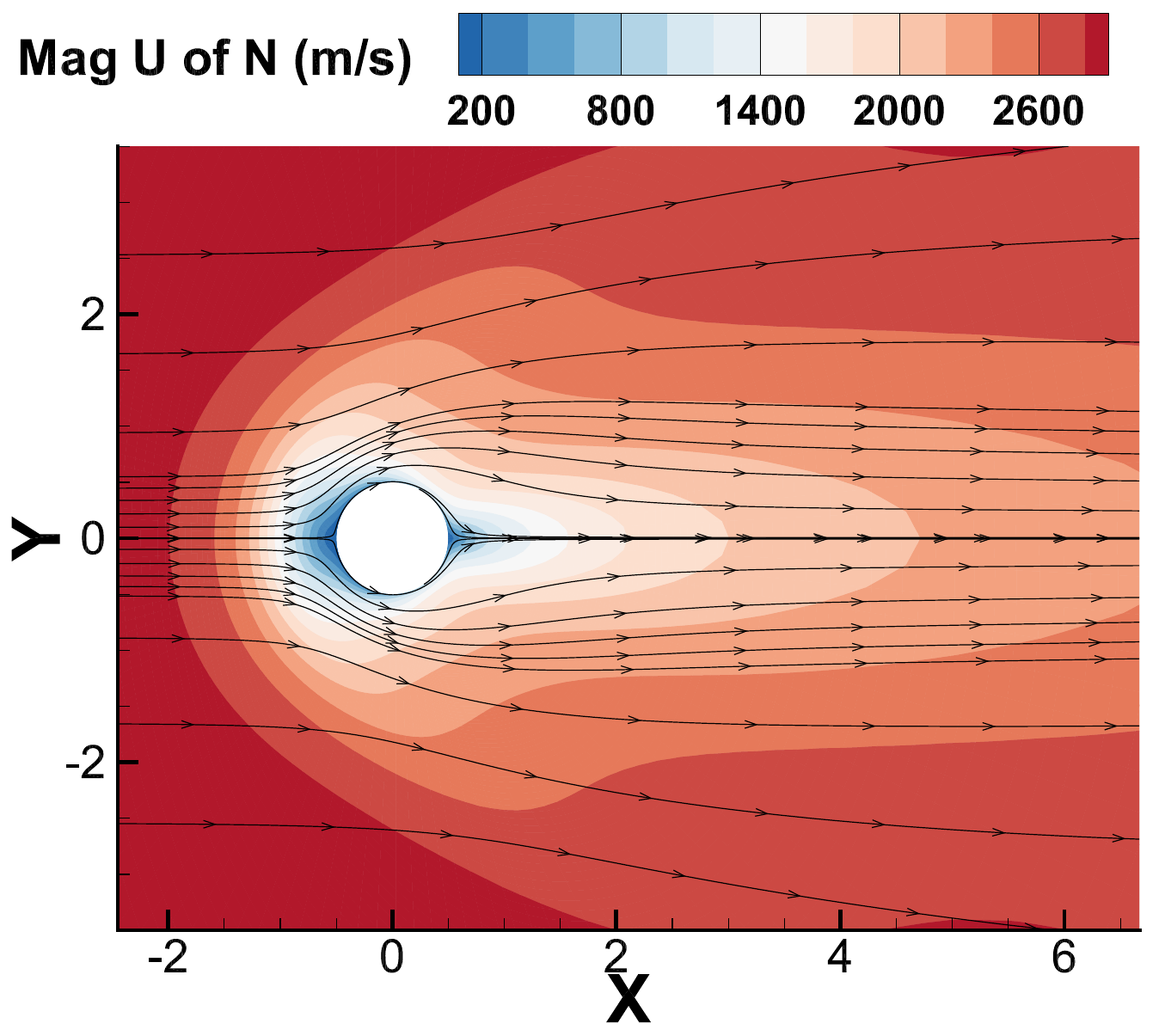}
	} \\ 
	\vspace{1.5mm}
    \subfloat[]{
    		\includegraphics[width=0.45 \textwidth]{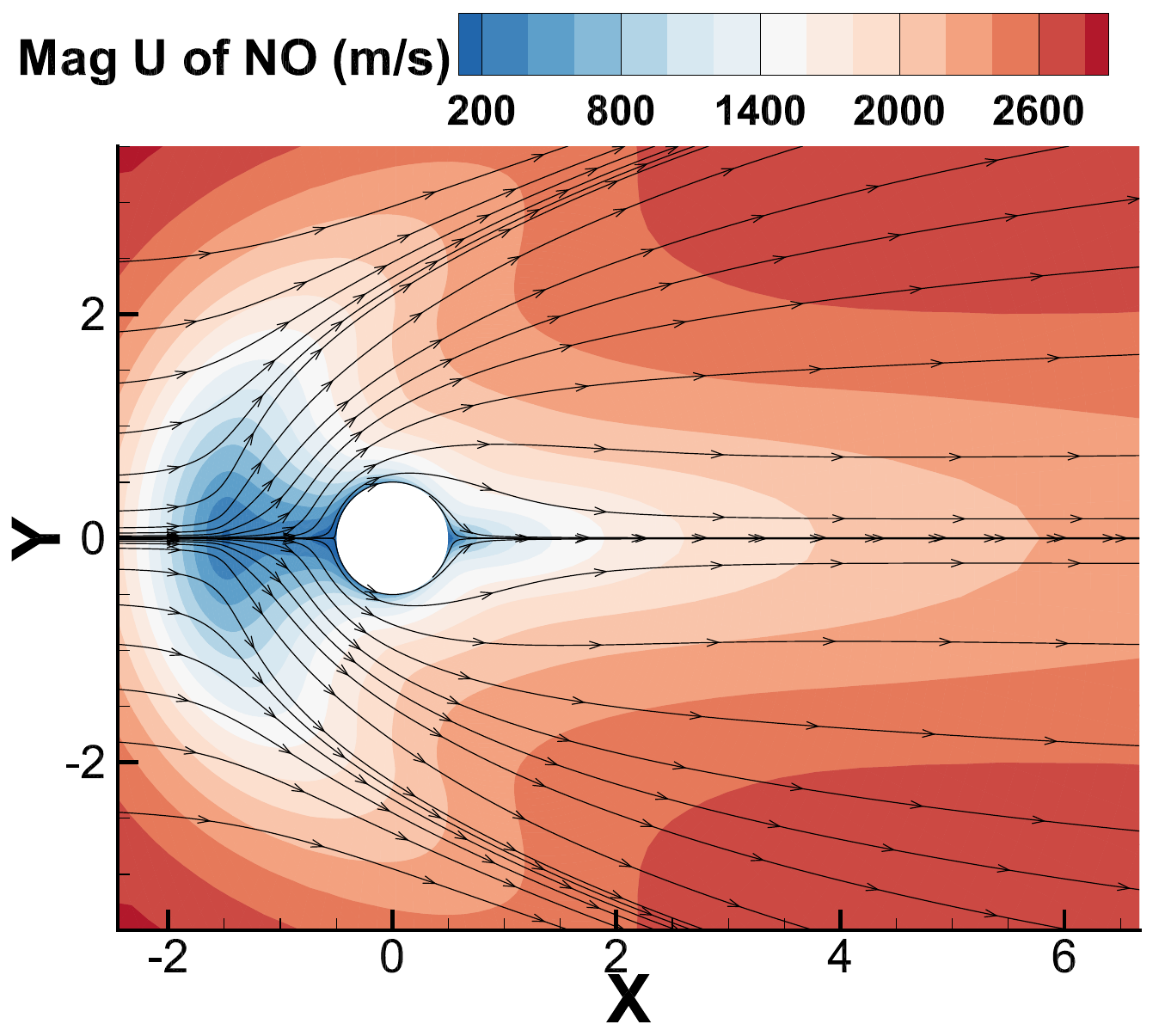}
    } 
    \subfloat[]{
			\includegraphics[width=0.45 \textwidth]{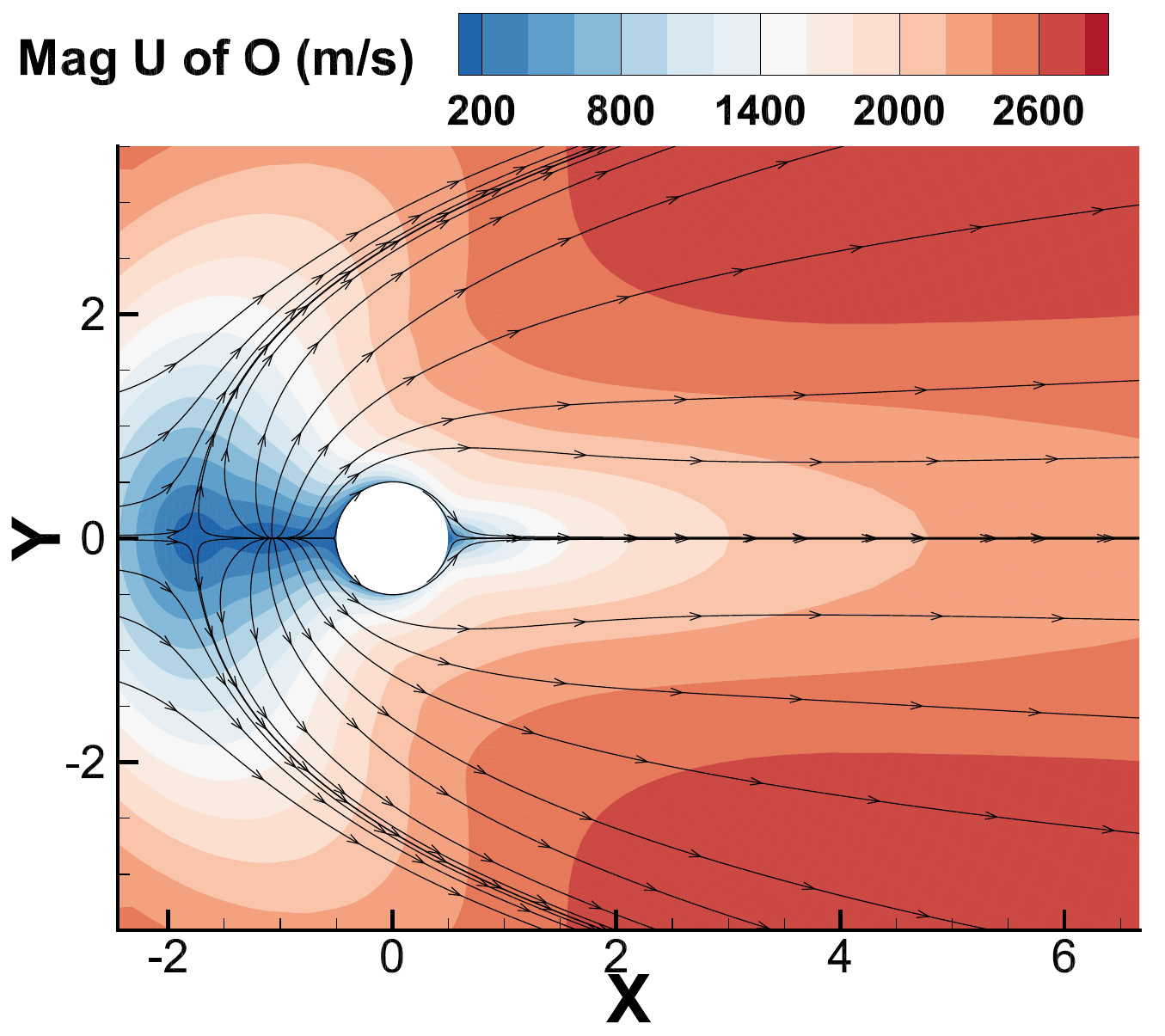}
	}
	\caption{Hypersonic chemical-reaction flow around a circular cylinder at ${\rm Ma}_\infty=7.03$, ${\rm Kn}_\infty=0.488$. The contours of magnitude of velocity and streamline of the species (a) ${\rm O}_2$, (b) ${\rm N}$, (c) ${\rm NO}$, and (d) ${\rm O}$.}
	\label{fig:cylinder-U-0-488}
\end{figure}

% Furthermore, to demonstrate the influence of rarefied effects on chemical reactions, the flow around the cylinder at a larger Kn number ${\rm Kn}_\infty=0.488$ is simulated, with the inflow states $n_{\rm O_2,\infty}=1.392\times 10^{18}{~\rm m}^{-3}$, $n_{\rm N,\infty}=2.784\times 10^{18}{~\rm m}^{-3}$. The simulation results are also compared with those of the DSMC method. 
The contours in Fig.~\ref{fig:cylinder-0-488-contour} illustrate that compared to ${\rm Kn}_\infty=0.0488$, the flow field at the larger Kn number becomes smoother, attributed to the rarefaction effects. On the other hand, by comparing with the results from the DSMC method, it shows that the current kinetic model and numerical scheme can still effectively characterize the non-equilibrium information of the flow field when rarefaction effects are more pronounced. Furthermore, from the distribution of pressure and mass fractions along the stagnation lines in the flow field shown in Fig.~\ref{fig:cylinder-0-488-stag}, it can be noted that in rarefied environments, chemical reactions are less intense compared to denser gas conditions, specifically reflected in smaller variations in mass fractions and pressure of the products. Concerning reactants, ${\rm O}_2$ diminishes near stagnation lines while ${\rm N}$ increases gradually, attributable to the different diffusion coefficients of the species. The phenomenon is also evident in other cases discussed later in the section. In the comparison of high-order quantities around the wall, shown in Fig.~\ref{fig:cylinder-0-488-wall}, we arrive at the same conclusion as in denser cases. The surface pressure and heat flux match well, while shear stress exhibits some error. Specifically, the error in shear forces is relatively smaller compared to denser conditions when contrasted with DSMC results, indicating that the single-relaxation multi-component model potentially outperforms chemical reaction models in predicting high-order quantities. It also expresses that future enhancements to the model could primarily focus on the collision term of chemical reactions. 
The occurrence of stagnation points far from the wall in the streamlines present in the denser gas condition also exists in rarefied environments, shown in Fig.~\ref{fig:cylinder-U-0-488}. However, the stagnation point for $\rm NO$ disappears. Combining this characteristic of weakened chemical reaction intensity in this rarefied condition, perhaps this phenomenon can be further attributed to chemical reactions leading to the generation of more reactant particles with zero initial velocity behind the shock.

\begin{figure}[H]
	\centering
    \subfloat[]{
    		\includegraphics[width=0.3 \textwidth]{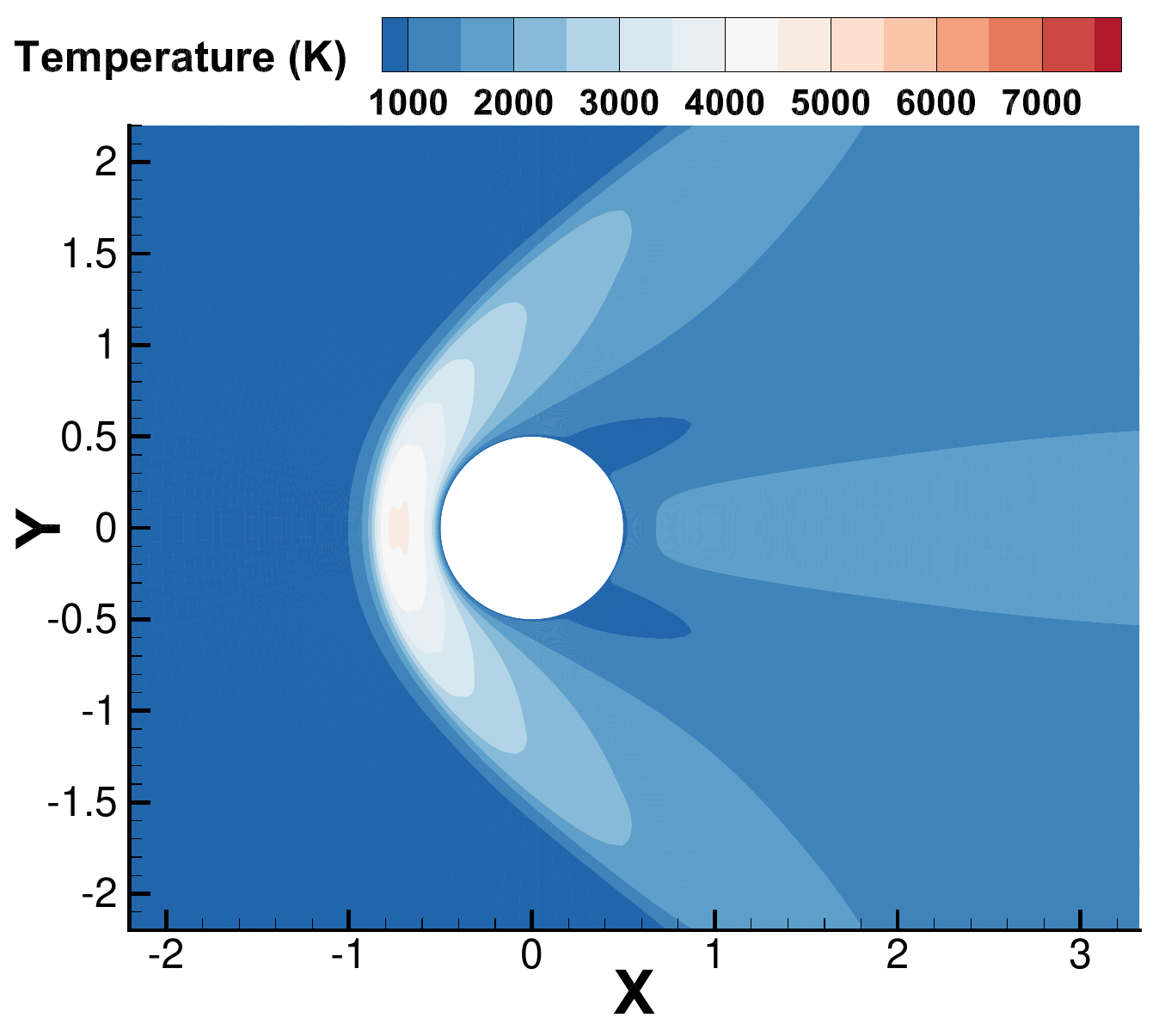}
    	}
    \subfloat[]{
			\includegraphics[width=0.3 \textwidth]{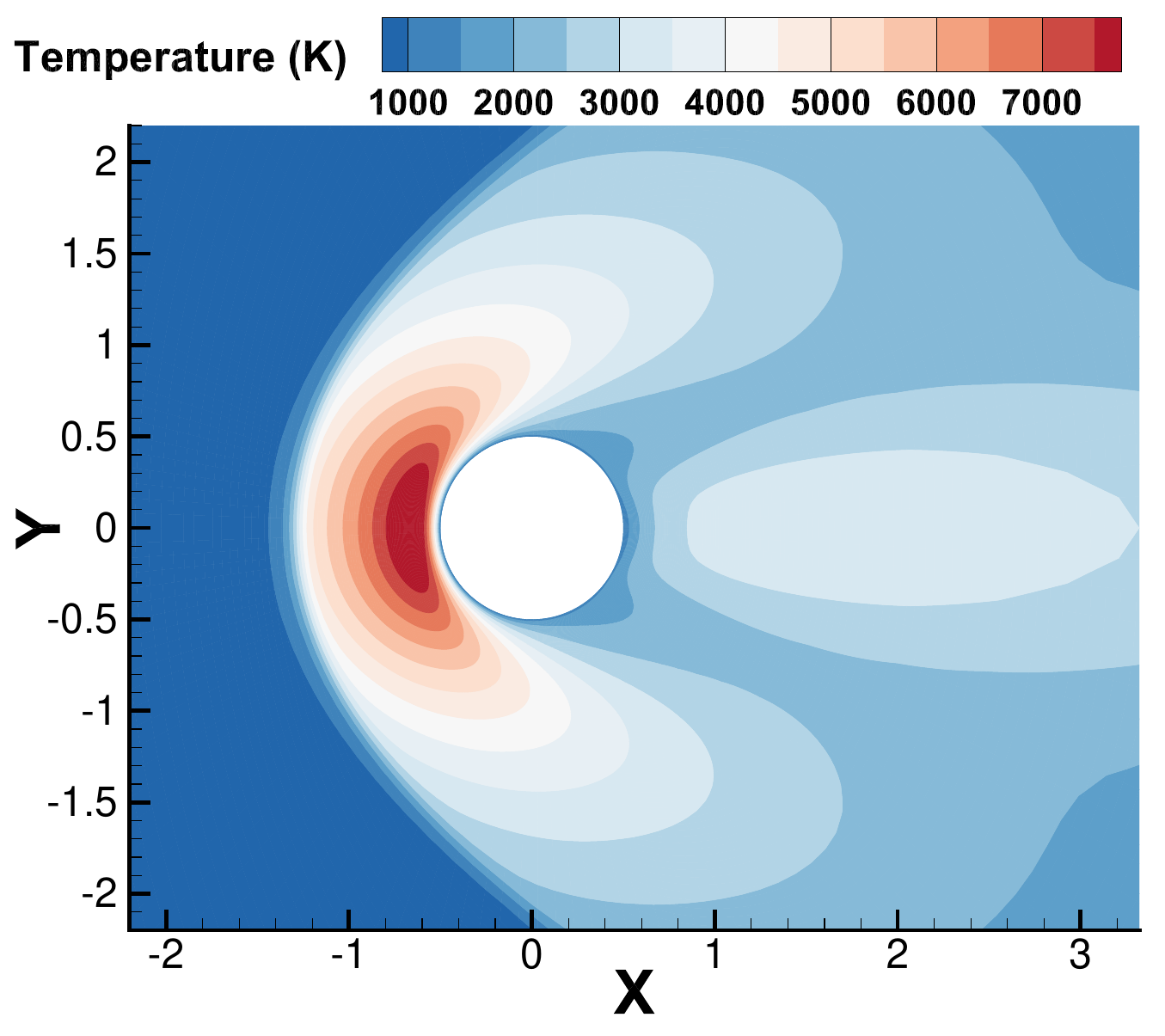}
		}
    \subfloat[]{
    		\includegraphics[width=0.3 \textwidth]{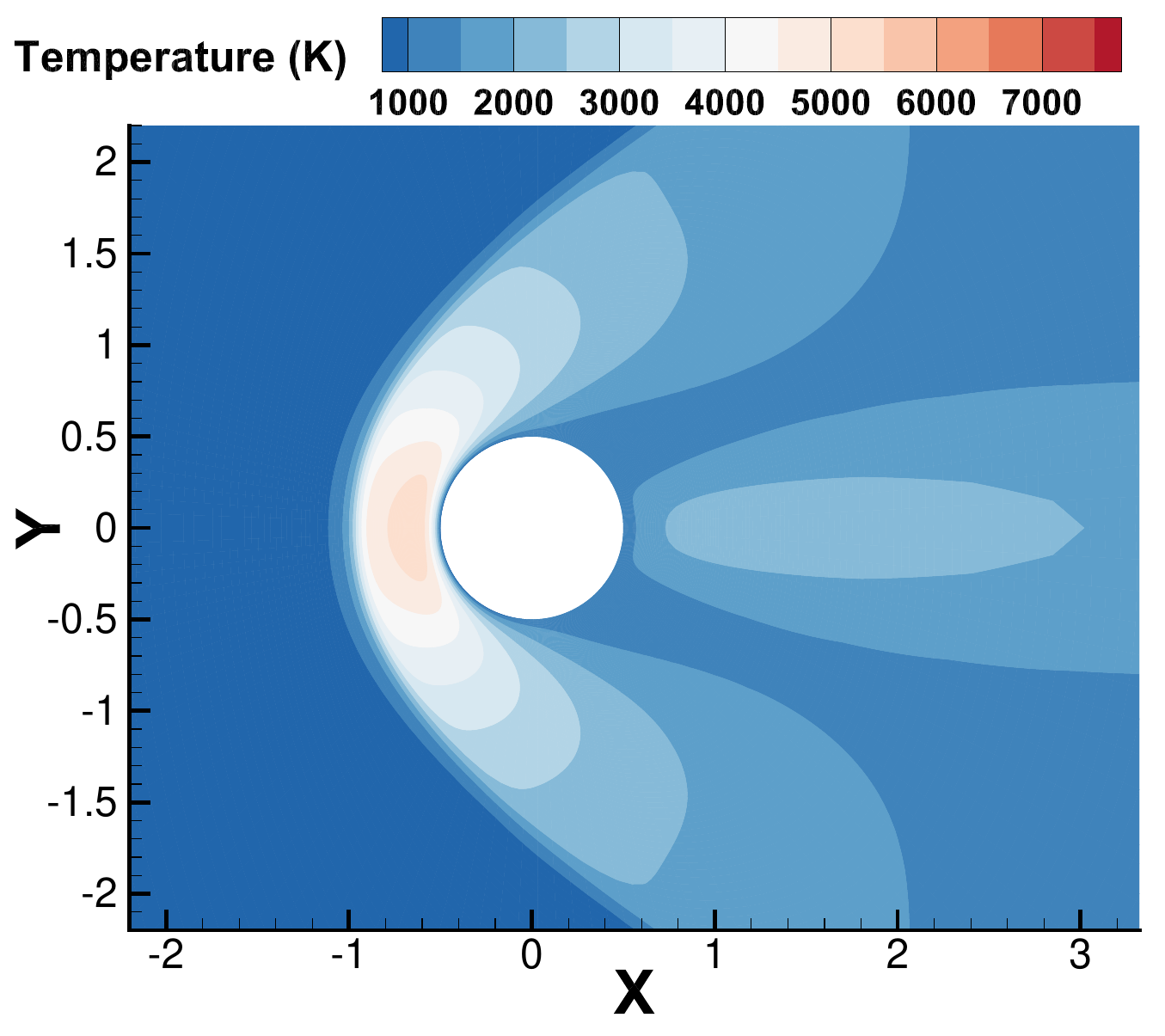}
    	}
	\caption{Hypersonic chemical-reaction flow around a circular cylinder at ${\rm Ma}_\infty=7.03$, ${\rm Kn}_\infty=0.0488$. Contours of global temperature of (a) forward endothermic reaction, (b) forward exothermic reaction, and (c) inert scenario with no reaction.}
	\label{fig:cylinder-3-compare-T}
\end{figure}

\begin{figure}[H]
	\centering
       \subfloat[]{
			\includegraphics[width=0.4 \textwidth]{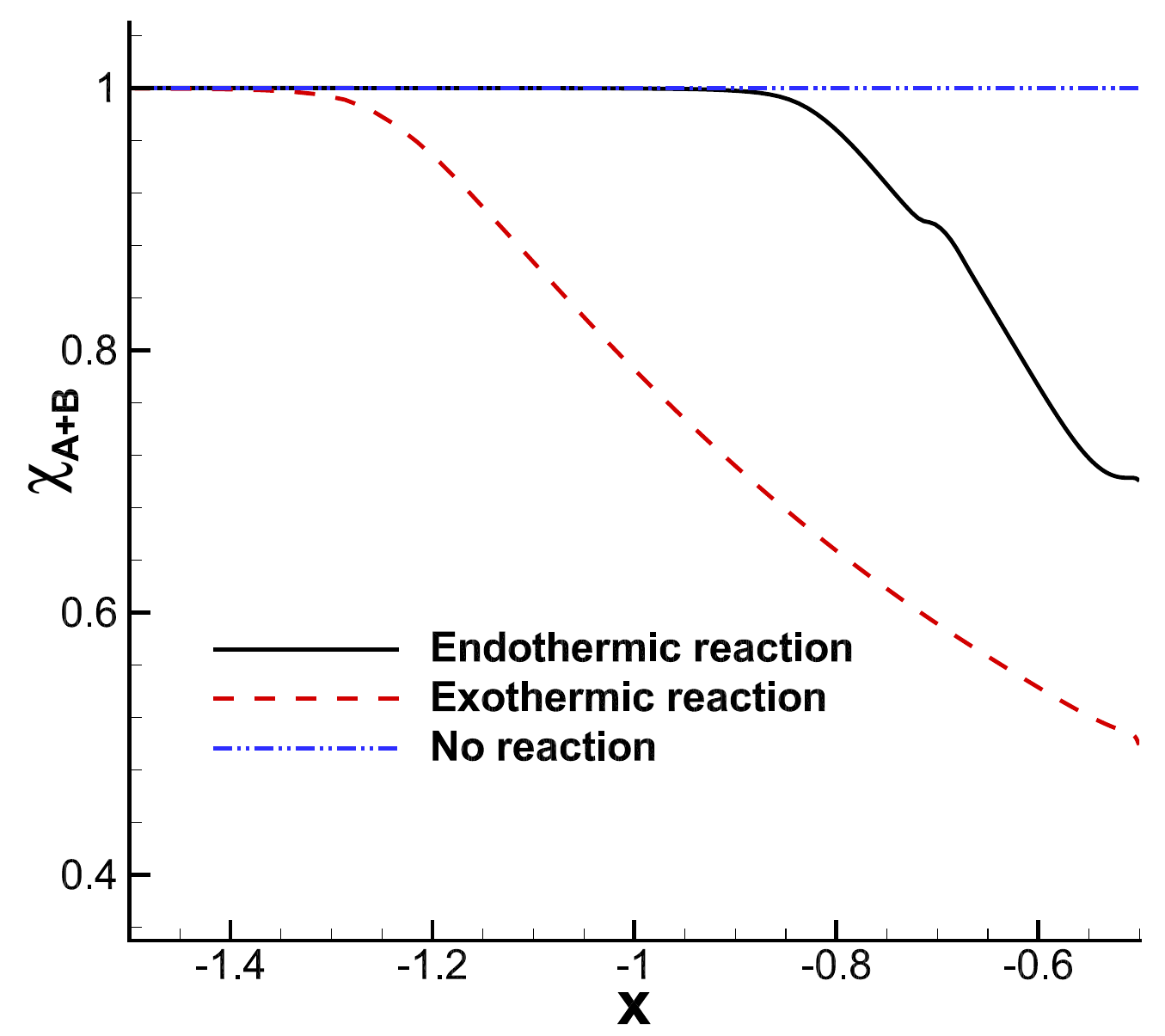}
		}
    \subfloat[]{
    		\includegraphics[width=0.4 \textwidth]{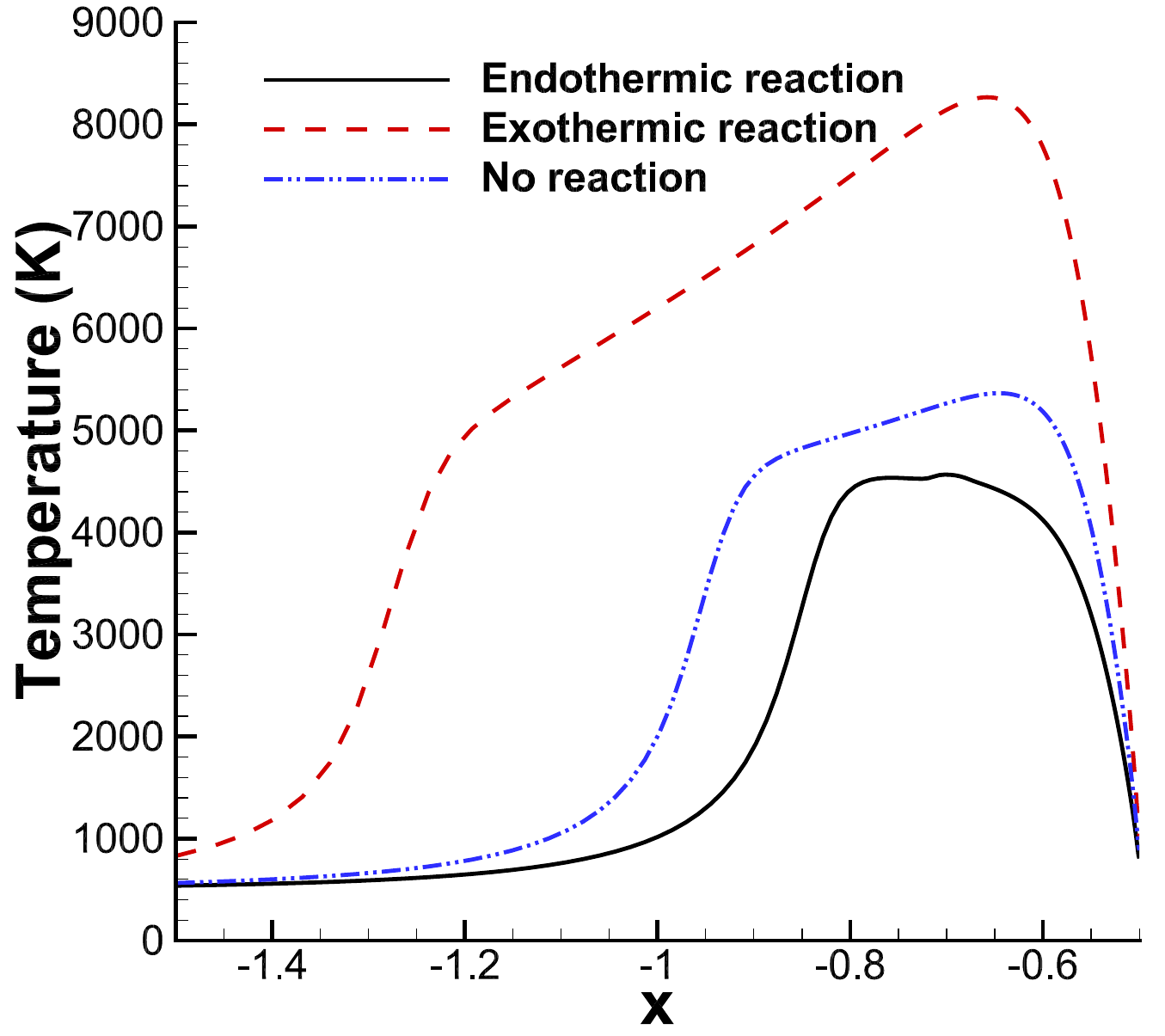}
    	}
	\caption{Hypersonic chemical-reaction flow around a circular cylinder at ${\rm Ma}_\infty=7.03$, ${\rm Kn}_\infty=0.0488$ of the forward endothermic reaction, forward exothermic reaction, and no reaction with (a) concentration fraction of ${\rm O}_2+{\rm N}$ at the stagnation line and (b) global temperature at the stagnation line.}
	\label{fig:cylinder-3-compare-T-stag}
\end{figure}

\begin{figure}[H]
	\centering
    \subfloat[]{
		\includegraphics[width=0.4 \textwidth]{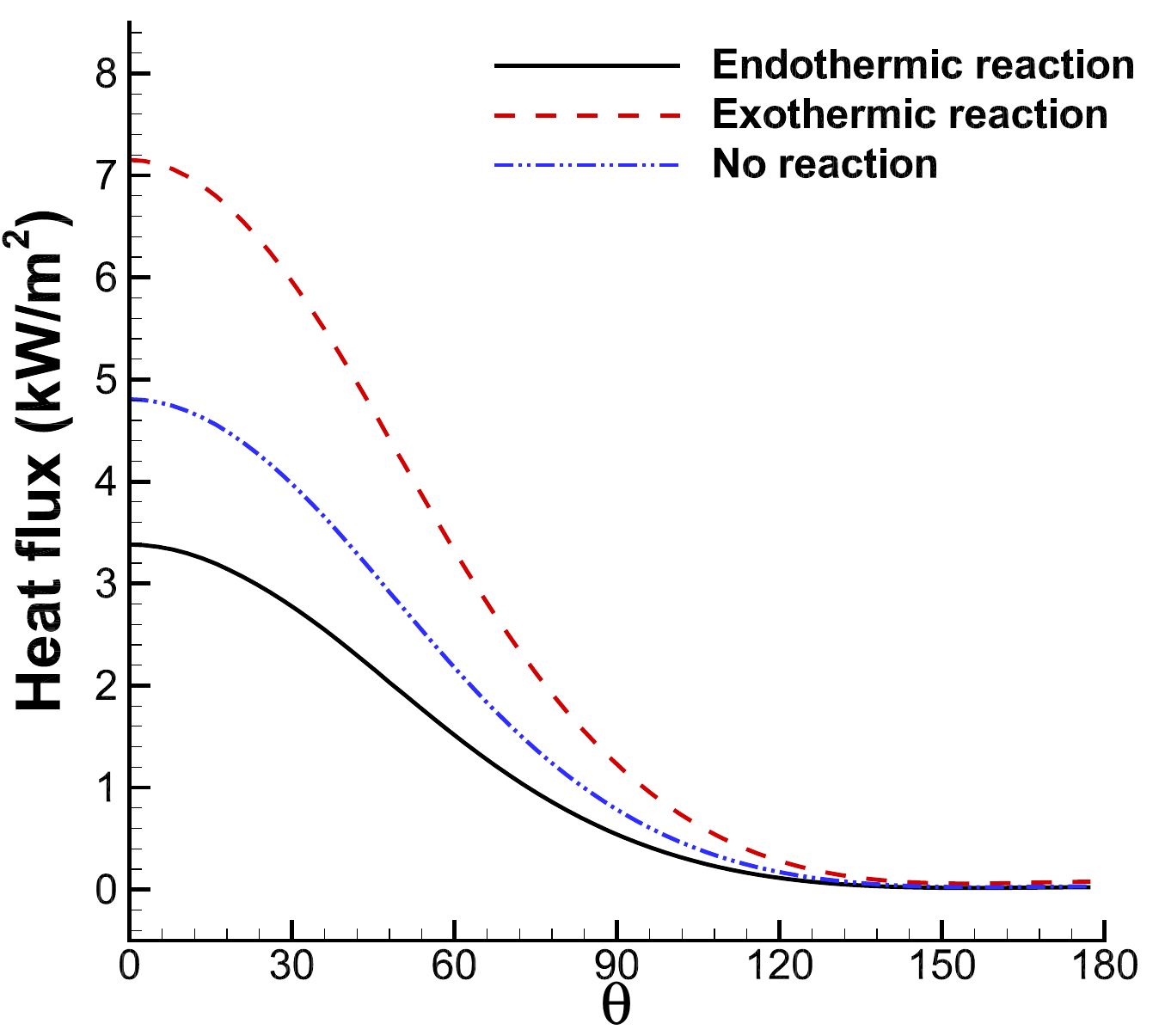}
    	}
	\subfloat[]{
			\includegraphics[width=0.4 \textwidth]{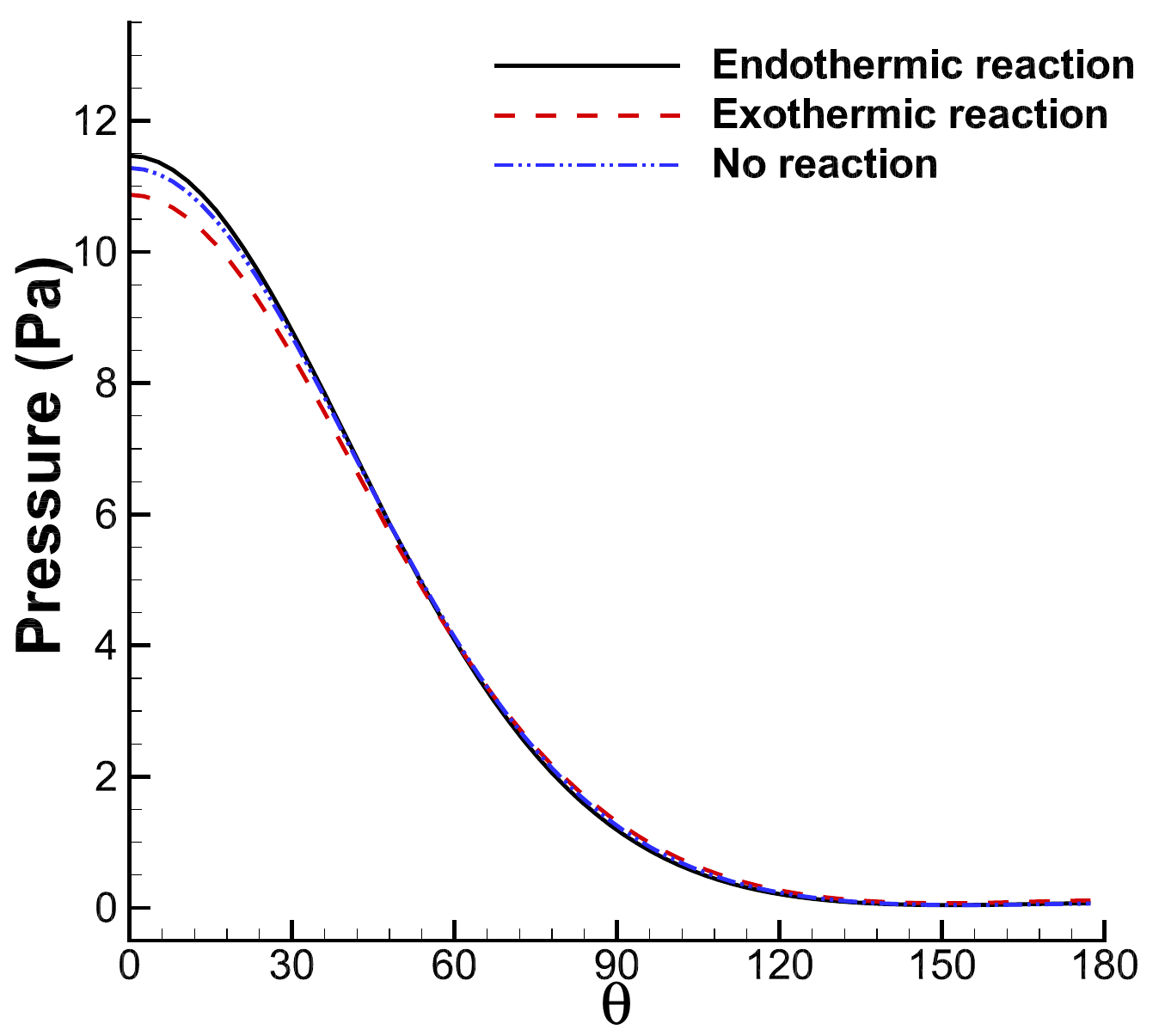}
		}
	\caption{Hypersonic chemical-reaction flow around a circular cylinder at ${\rm Ma}_\infty=7.03$, ${\rm Kn}_\infty=0.0488$ of the forward endothermic reaction, forward exothermic reaction, and no reaction with (a) heat flux and (b) pressure at the wall.}
	\label{fig:cylidner-3-compare-wall}
\end{figure}

At last, keeping the inflow state the same at ${\rm Kn}_{\infty}=0.0488$, the forward endothermic reaction is changed into a forward exothermic one ($\Delta E \rightarrow -\Delta E$) to show the influence of different energy release type of chemical reactions on the flow field, aerodynamic force and heat. In addition, the flow with no reaction is also simulated. As the global temperature contours for the three states, shown in Fig.~\ref{fig:cylinder-3-compare-T}, the peak value of global temperature and the location of shock wave vary significantly with different kinds of reactions. Quantificational results along the stagnation line in Fig.~\ref{fig:cylinder-3-compare-T-stag} illustrate that the peak of the temperature of the exothermic reaction is twice that of the endothermic reaction. 
Also, the higher global temperature caused by exothermic reactions will further strengthen the chemical collision. As a result, the reacted proportions of $\rm O_2$ and $\rm N$ are larger than the forward endothermic reaction, and the location of the shock wave is more forward, which is shown in Fig.~\ref{fig:cylinder-3-compare-T-stag}(a). From another perspective, as shown in Fig.~\ref{fig:cylinder-3-compare-T-stag}(b), when there is no reaction, the global temperature rises slowly at the post-shock because the velocity declines, and the kinetic energy is turned into internal energy. As to the forward exothermic reaction flow, the global temperature becomes higher because more energy is released. While as for the forward endothermic reaction flow, the equilibrium temperature even goes down, as the internal energy is absorbed with the reaction. On the other hand, as Fig.~\ref{fig:cylidner-3-compare-wall}(a) shows, the peak values of heat flux at the wall are quite different. The reason is also about the large quantity of energy released or absorbed. Nevertheless, in Fig.~\ref{fig:cylidner-3-compare-wall}(b), the deviations of pressure at the wall are relatively tiny, because even with large energy variation, the total moment keeps constant. To sum up, the energy release types have a greater influence on aerodynamic heating than forcing. The twofold difference can be observed in the peak of stagnation temperature and heat flux around the wall. 

\subsection{Hypersonic Flow around an Apollo Reentry}\label{sec:apollo}
The three-dimensional hypersonic flow around an Apollo Reentry is simulated. The inflow condition is set as $n_{\infty}=1.37\times 10^{19}{~\rm m}^{-3}$, corresponding to the atmospheric state at altitude $100$ km. The ratio of number density is $n_{\rm NO, \infty} : n_{\rm O,\infty} : n_{\rm O_2, \infty} : n_{\rm N, \infty} = 1:1:0:0$, velocity $U_{\infty}=10000 $ m/s, and temperature $T_{\infty} = 966.1$ K, which corresponds to ${\rm Ma}_{\infty}=13.15$ and ${\rm Kn}_{\infty}=0.0254$. The isothermal wall is applied with $T_w = 966.1$ K. The reference length is the base diameter of the Apollo $3.912$ m. In the computation, 48,693 unstructured meshes are used in the physical space, and the height of the first layer near the wall is set to be $0.008$ m. 
The un-DVS mesh is discretized into 29,988 cells in a sphere mesh with a radius of $\sqrt{5 k_B / m_\alpha T_s}$ for each species $\alpha$, where $T_s$ is the stagnation temperature in the primary flow calculated by the GKS simulation. The sphere center is located at $(0,0,0)$. The velocity space near the stagnation and free stream velocity points are refined. 

\begin{figure}[H]
	\centering
    	\includegraphics[width=0.42 \textwidth]
			{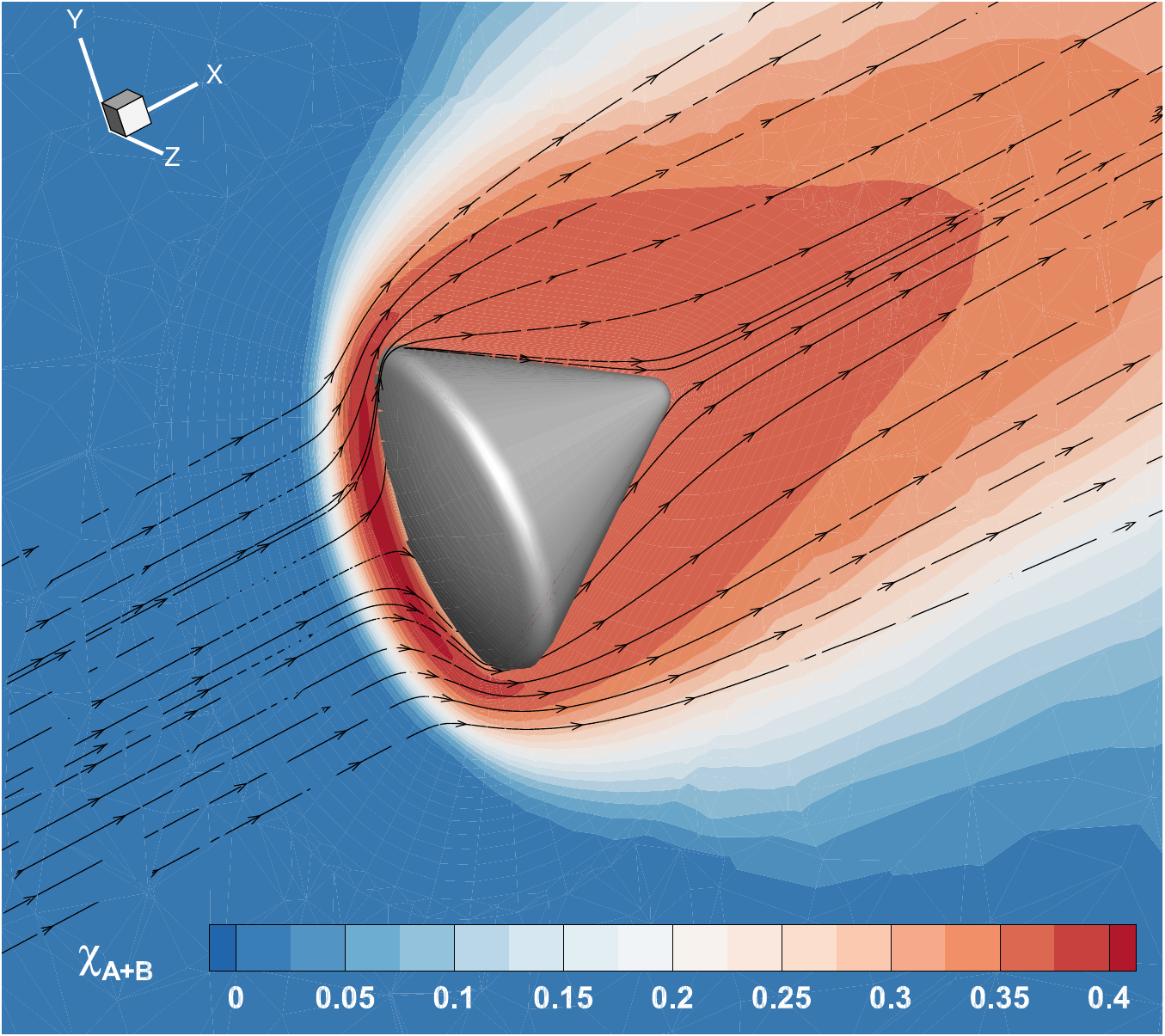}
	\caption{Hypersonic chemical reaction flow around an Apollo reentry at ${\rm Ma}_\infty=13.15$, ${\rm Kn}_\infty=0.0254$. The distribution of the ratio of number densities of reacting gases to the system.}
	\label{fig:apollo-3D}
\end{figure}

\begin{figure}[H]
	\centering
    \subfloat[]{
			\includegraphics[width=0.45 \textwidth]
			{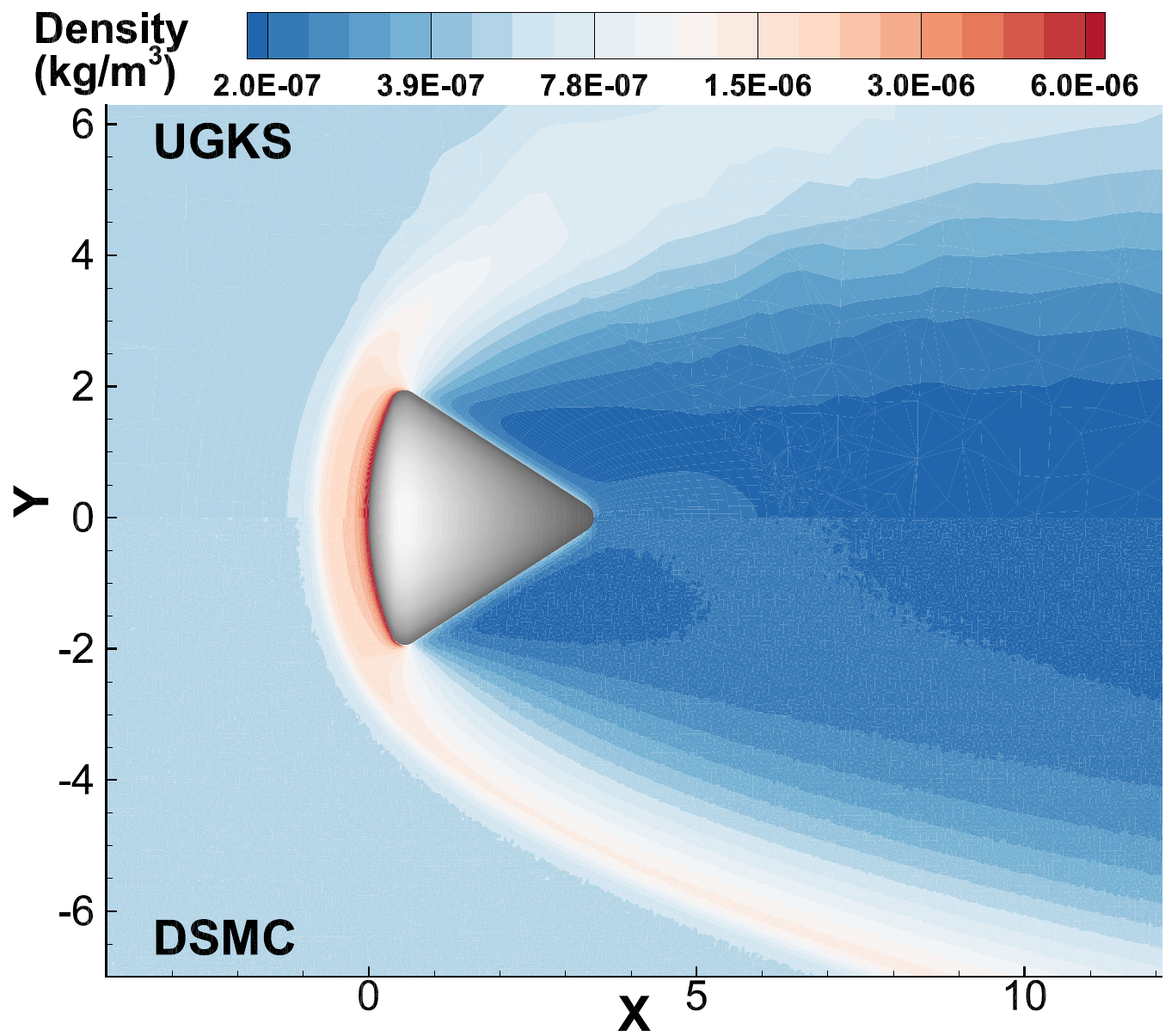}
	} 
	\hspace{1mm}
    \subfloat[]{
    	\includegraphics[width=0.45 \textwidth]{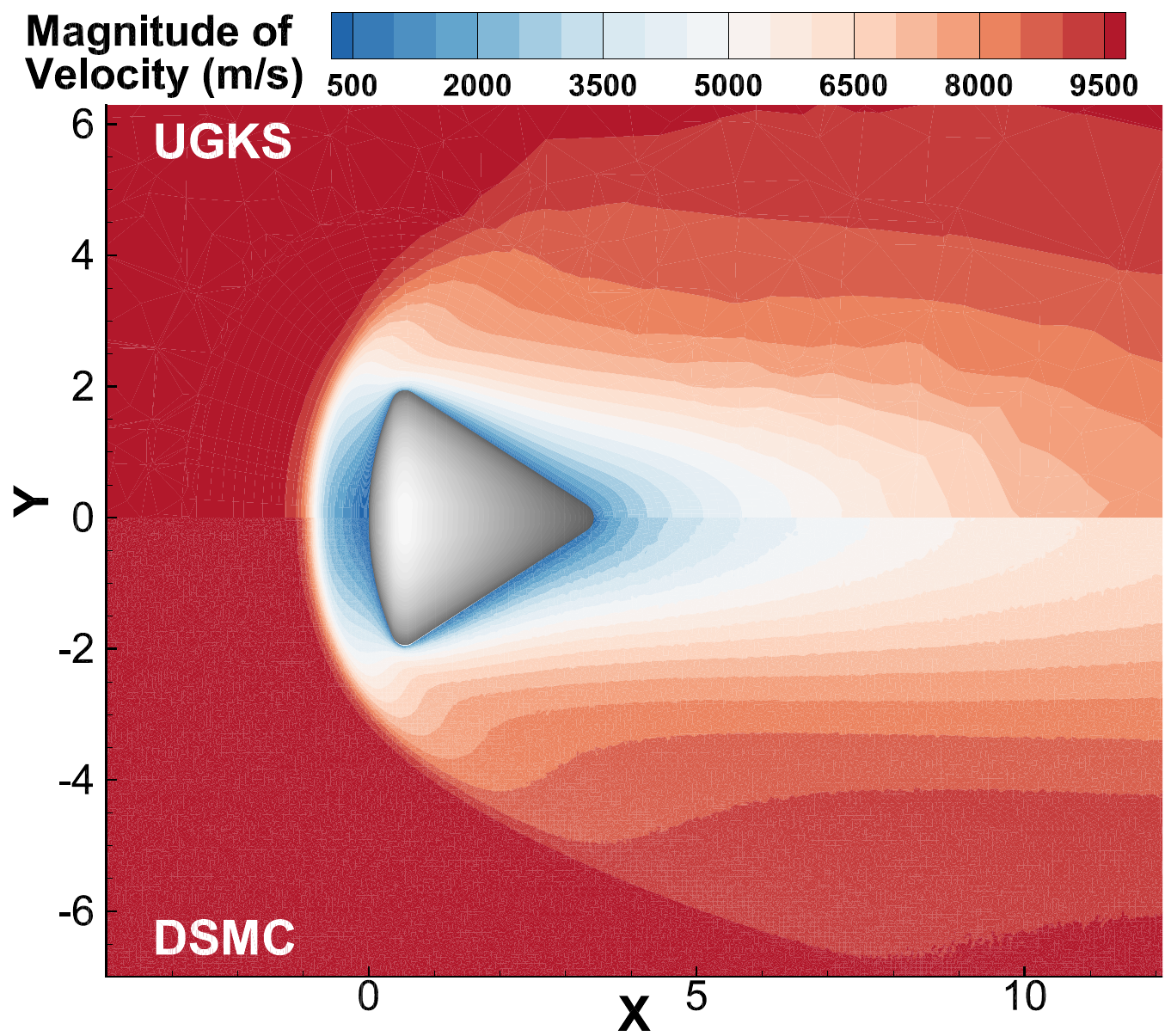}
    } \\
    \subfloat[]{
		\includegraphics[width=0.45 \textwidth]{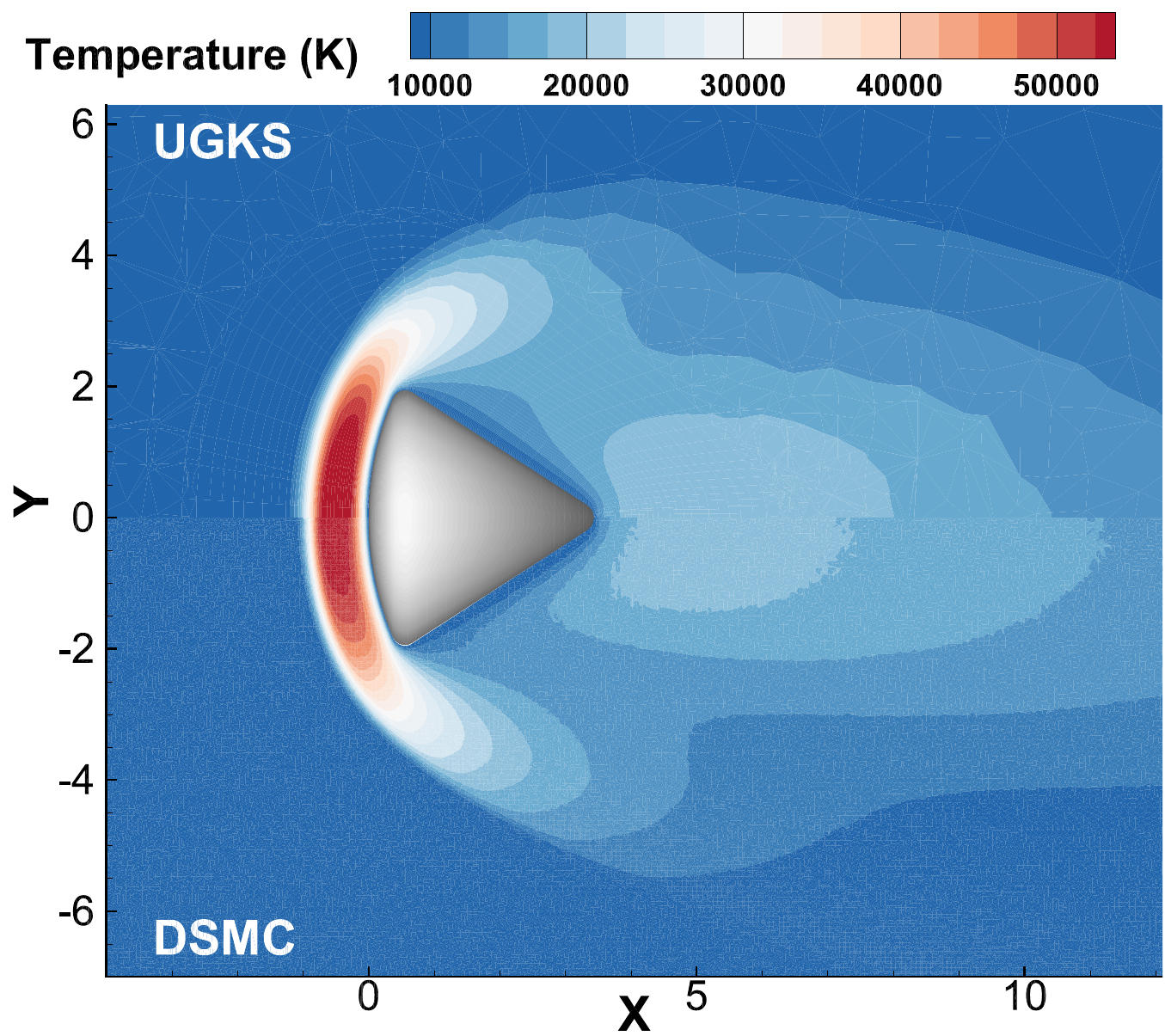}
	}
	\caption{Hypersonic chemical reaction flow around an Apollo reentry at ${\rm Ma}_\infty=13.15$ and ${\rm Kn}_\infty=0.0254$. Comparison of the UGKS with DSMC method with (a) density, (b) magnitude of velocity, and (c) temperature of the system.}
	\label{fig:apollo-contour}
\end{figure}

\begin{figure}[H]
	\centering
    \subfloat[]{
			\includegraphics[width=0.4 \textwidth]
			{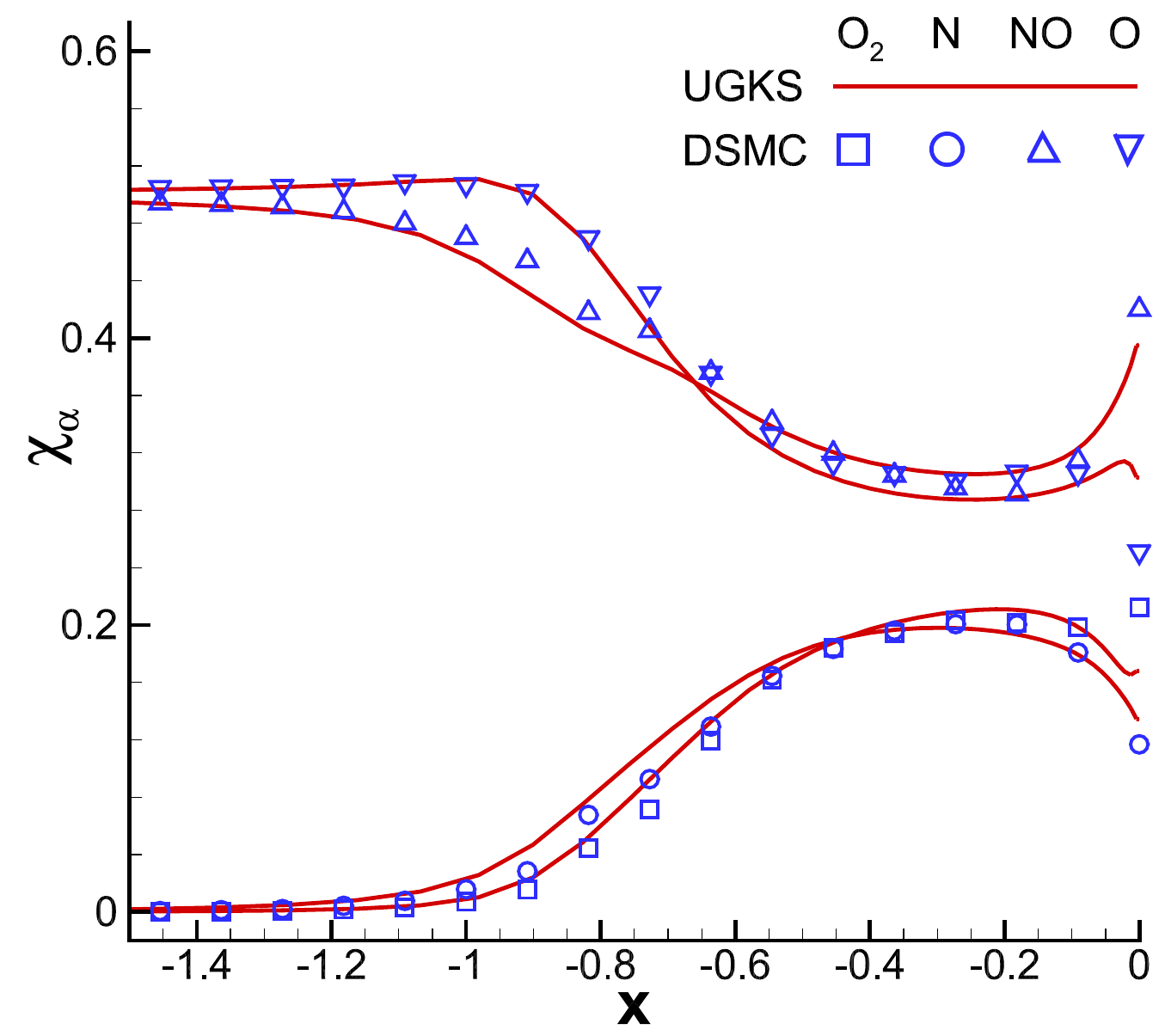}
	} 
    \subfloat[]{
    	\includegraphics[width=0.4 \textwidth]{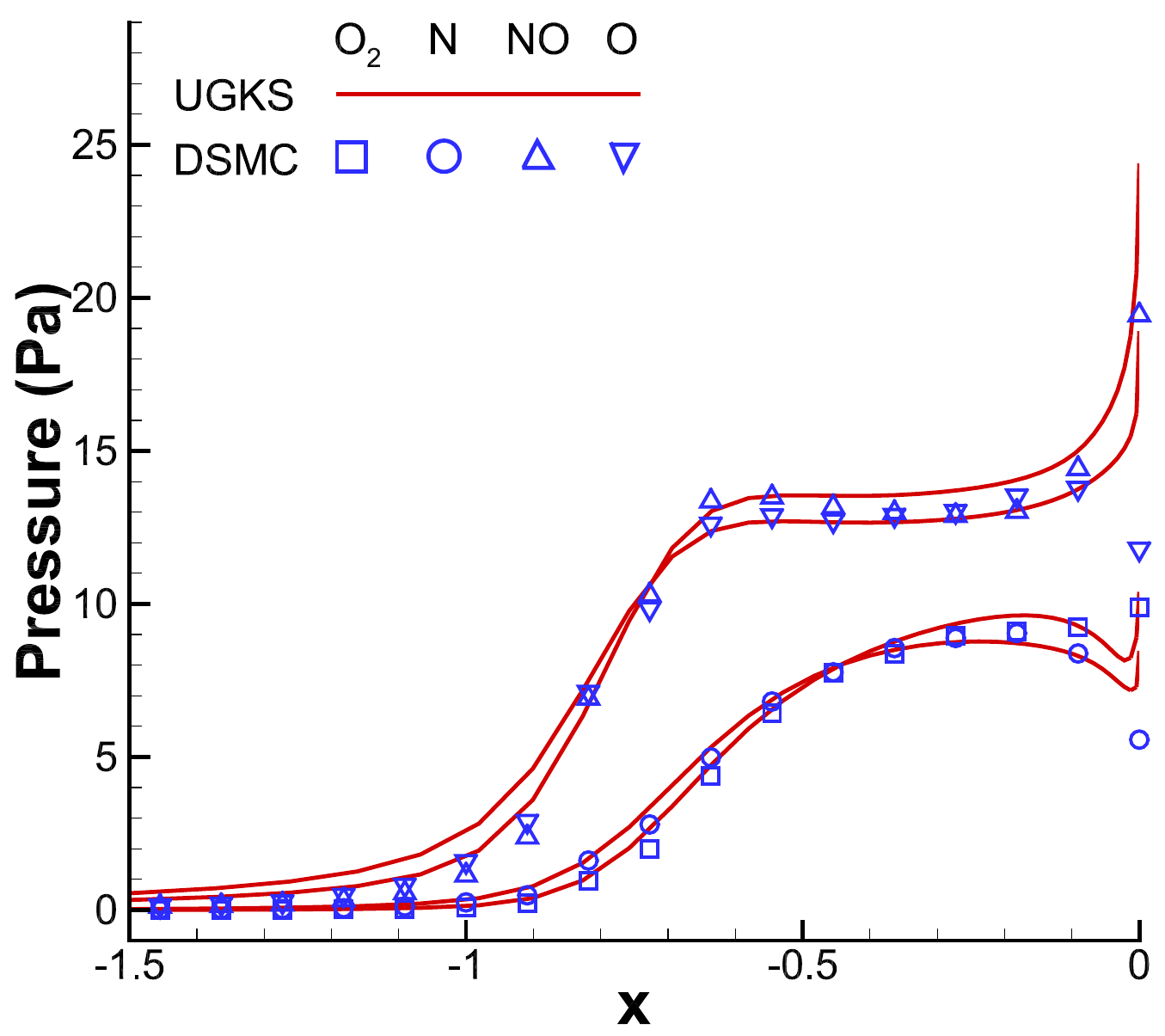}
    } 
	\caption{Hypersonic chemical reaction flow around an Apollo reentry at ${\rm Ma}_\infty=13.15$, ${\rm Kn}_\infty=0.0254$. The distributions of (a) ratio of number density of each species to the system and (b) pressure along the stagnation line.}
	\label{fig:apollo-stag}
\end{figure}

The distribution of concentration fractions of the products, as plotted in Fig.~\ref{fig:apollo-3D}, illustrates the intensity of the chemical reaction. It can be observed that the chemical reaction becomes more intense behind the shock wave. The products convect along streamlines into the leeward region. Fig.~\ref{fig:apollo-contour} displays contours of global density, velocity, and temperature, compared with the DSMC method. In three-dimensional computations, the current kinetic model and the UGKS also yield satisfactory simulation results. 
The UGKS can accurately determine the shock wave position in the windward region under the near-continuum flow regimes and the expansion distance in the leeward region in the near-rarefied flow regimes. This demonstrates the UGKS's capability in modeling multi-scale non-equilibrium flows. 
Additionally, the global temperature shown in Fig.~\ref{fig:apollo-contour}(c) sharply increases behind the shock wave, leading to a significant rise in the number of molecules exceeding the energy threshold, which intensifies the chemical reaction behind the shock wave. Furthermore, in the plot of stagnation lines in Fig.~\ref{fig:apollo-stag}, one can observe in more detail the spatial variations of concentration fractions of each component and pressure. These computational results also align well with the DSMC results. For more detailed wall high-order quantities, the current kinetic model and numerical scheme exhibit good agreement in surface pressure and shear stress. While they capture similar trends in surface heat flux, there are still some differences in peak values. These discrepancies stem from the inability of the single relaxation model to fully recover all transport coefficients, indicating a need for further research to enhance its capabilities.

\begin{figure}[H]
	\centering
    \subfloat[]{
		\includegraphics[width=0.3\textwidth]
		{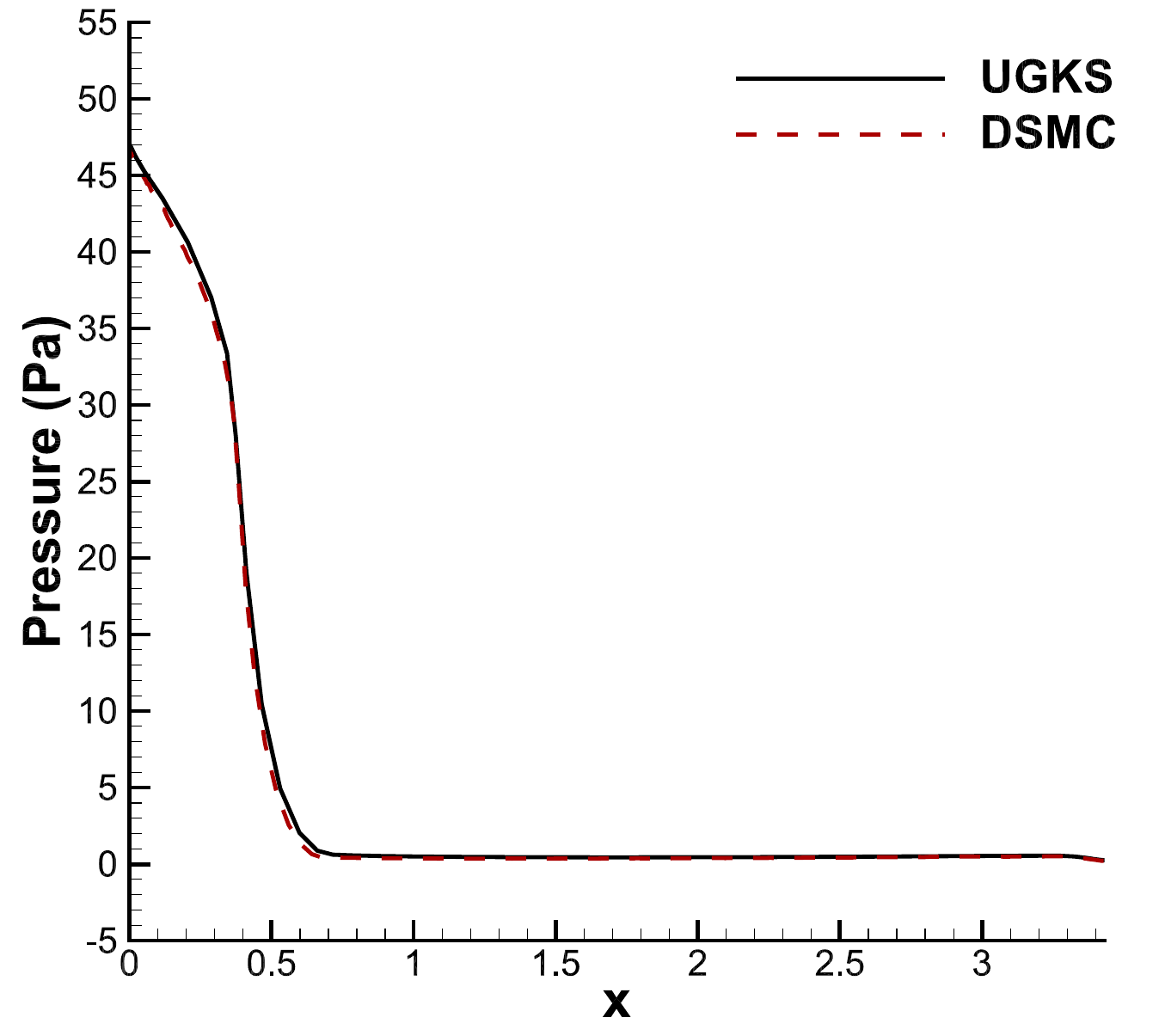}
	} 
    \subfloat[]{
    	\includegraphics[width=0.3\textwidth]{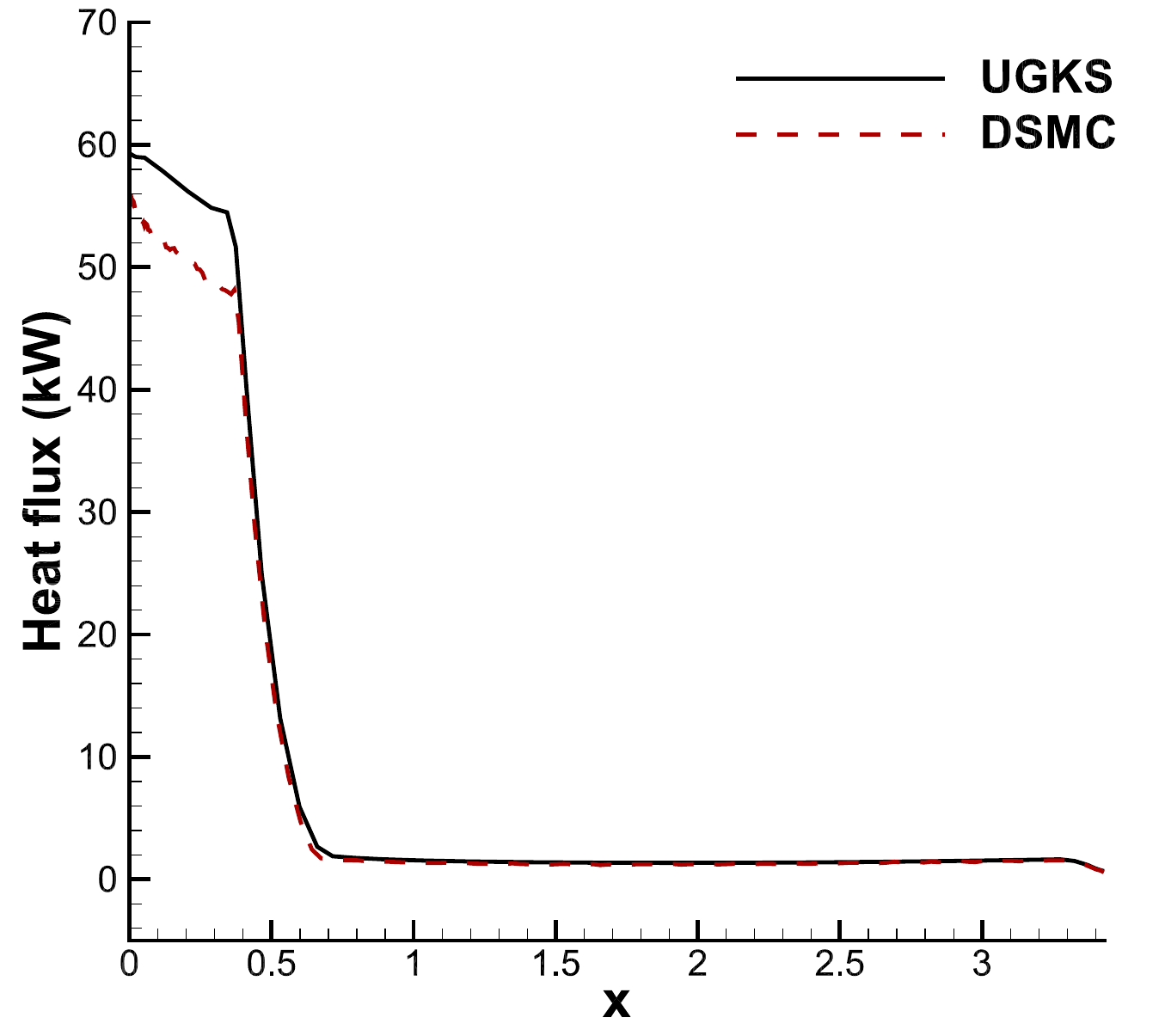}
    } 
    \subfloat[]{
    	\includegraphics[width=0.3\textwidth]{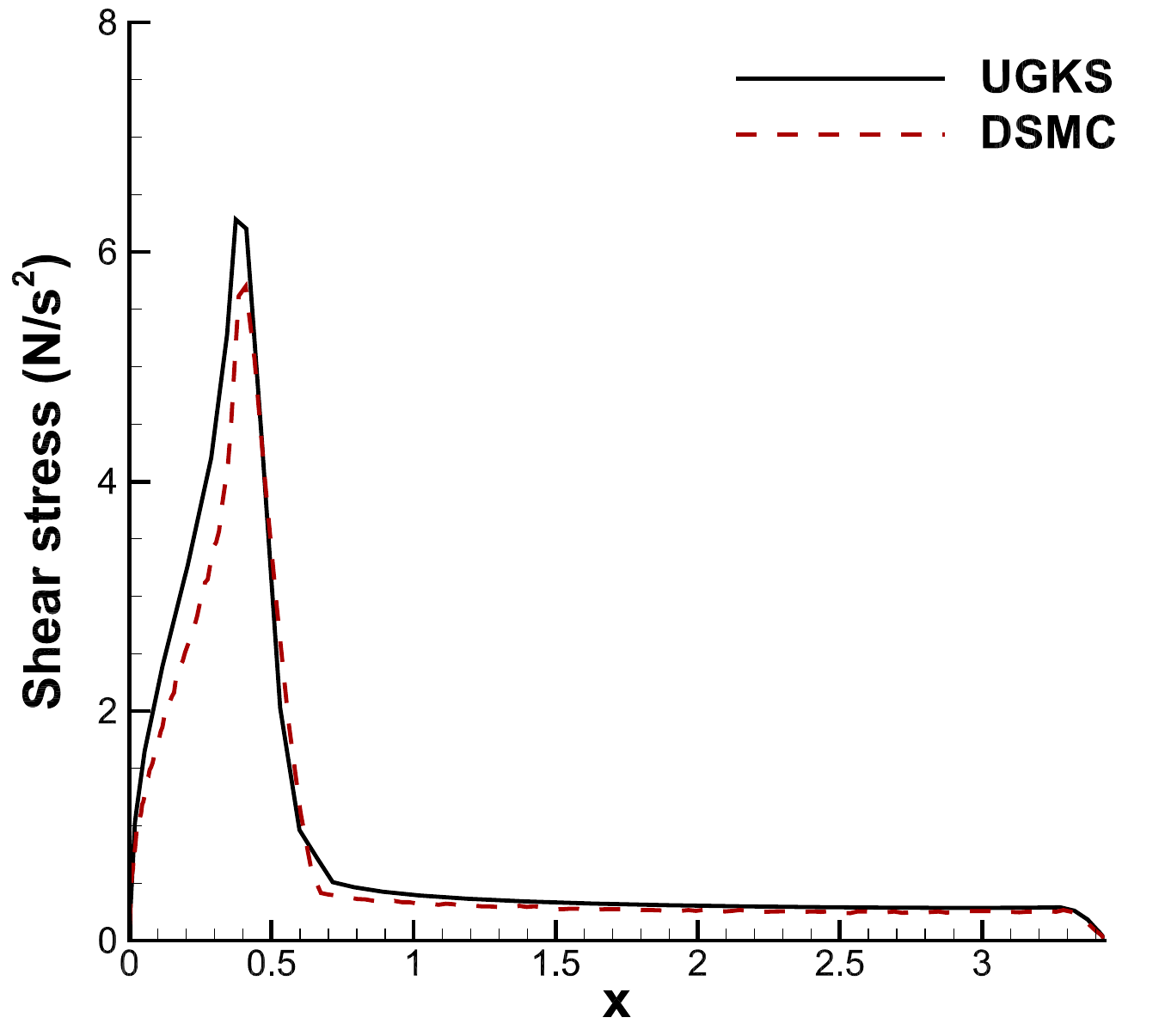}
    } 
	\caption{Hypersonic flow around a Apollo reentry with chemical reaction at ${\rm Ma}_\infty=13.15$, ${\rm Kn}_\infty=0.0254$. The distributions of (a) pressure, (b) heat flux, and (c) shear stress around the surface.}
\end{figure}

\section{Hypersonic Flow around an X38-Like Space Vehicle}

The hypersonic flow around an X38-like space vehicle is simulated at the Mach number ${\rm Ma}_\infty = 13.15$ at two inflow Knudsen numbers ${\rm Kn}_\infty$ is $0.001267$ and 0.03543. Those Knudsen numbers, with the reference length of the space vehicle 2.8 m, correspond to the atmospheric states at the altitudes of 80 km and 100 km, respectively. With the inflow temperature $T_\infty = 966.1$ K, the velocity is given by $U_\infty = 10000$ m. The isothermal wall is applied with $T_w = 966.1$ K. The total number density can be calculated by inflow Kn number as $n_\infty = 3.831\times 10^{20}{~\rm m}^{-3}$ for ${\rm Kn}_\infty = 0.001267$ and $n_\infty = 1.37\times 10^{19}{~\rm m}^{-3}$ for ${\rm Kn}_\infty = 0.03543$, with the ratio of the inflow number density $n_{\rm NO, \infty} : n_{\rm O,\infty} : n_{\rm O_2, \infty} : n_{\rm N, \infty} = 1:1:0:0$. In the computation, 277,004 unstructured meshes are used in the physical domain, and the height of the first layer near the wall is set to be 0.002 m. The CFL is taken as 0.5.The discretized velocity space applies $30\times30\times30$ points with a radius of $\sqrt{5 k_B / m_\alpha T_s}$

\begin{figure}[H]
	\centering
    \subfloat[]{
		\includegraphics[width=0.45\textwidth]{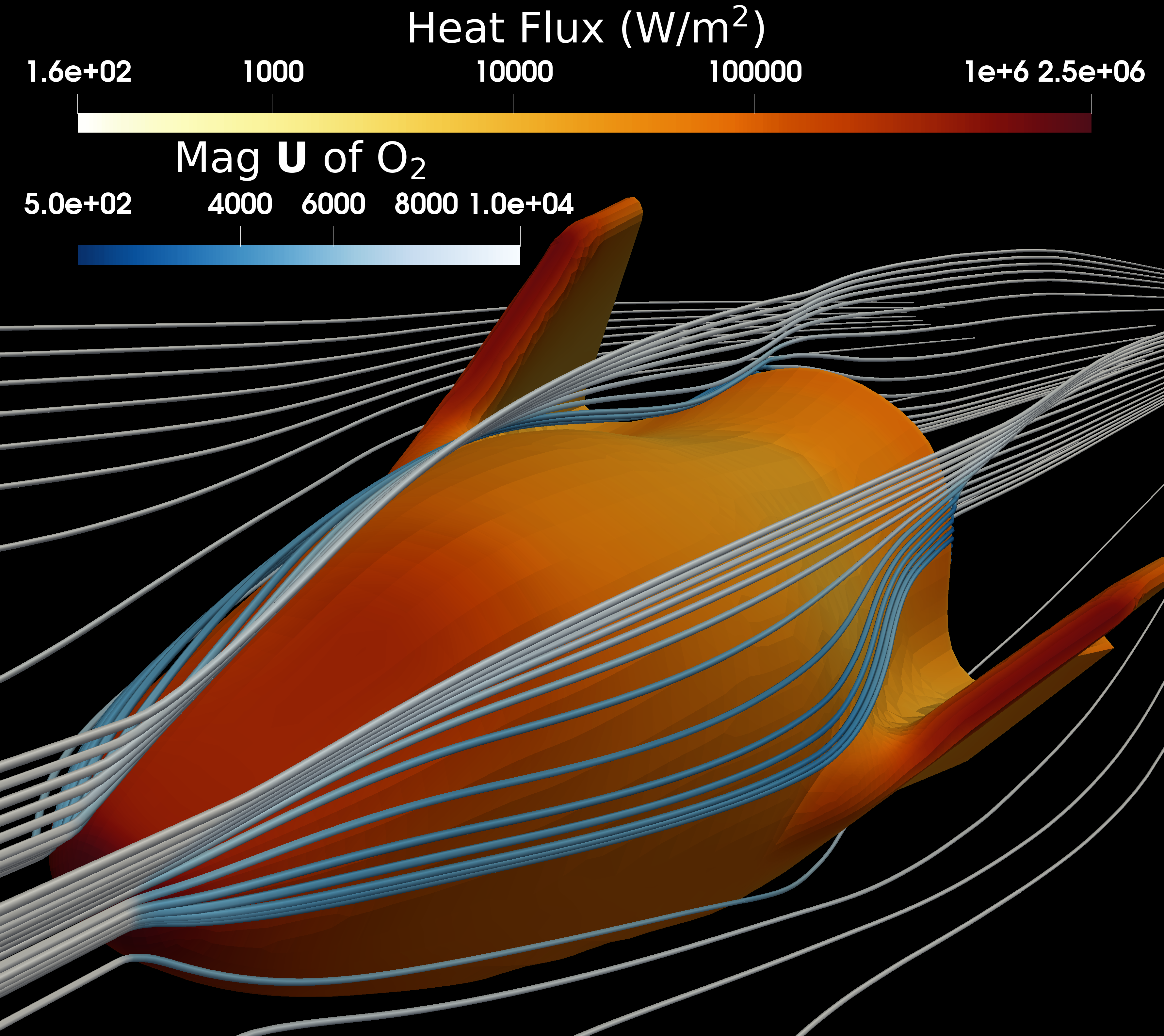}
	} 
	\hspace{0.5mm}
    \subfloat[]{
    	\includegraphics[width=0.45\textwidth]{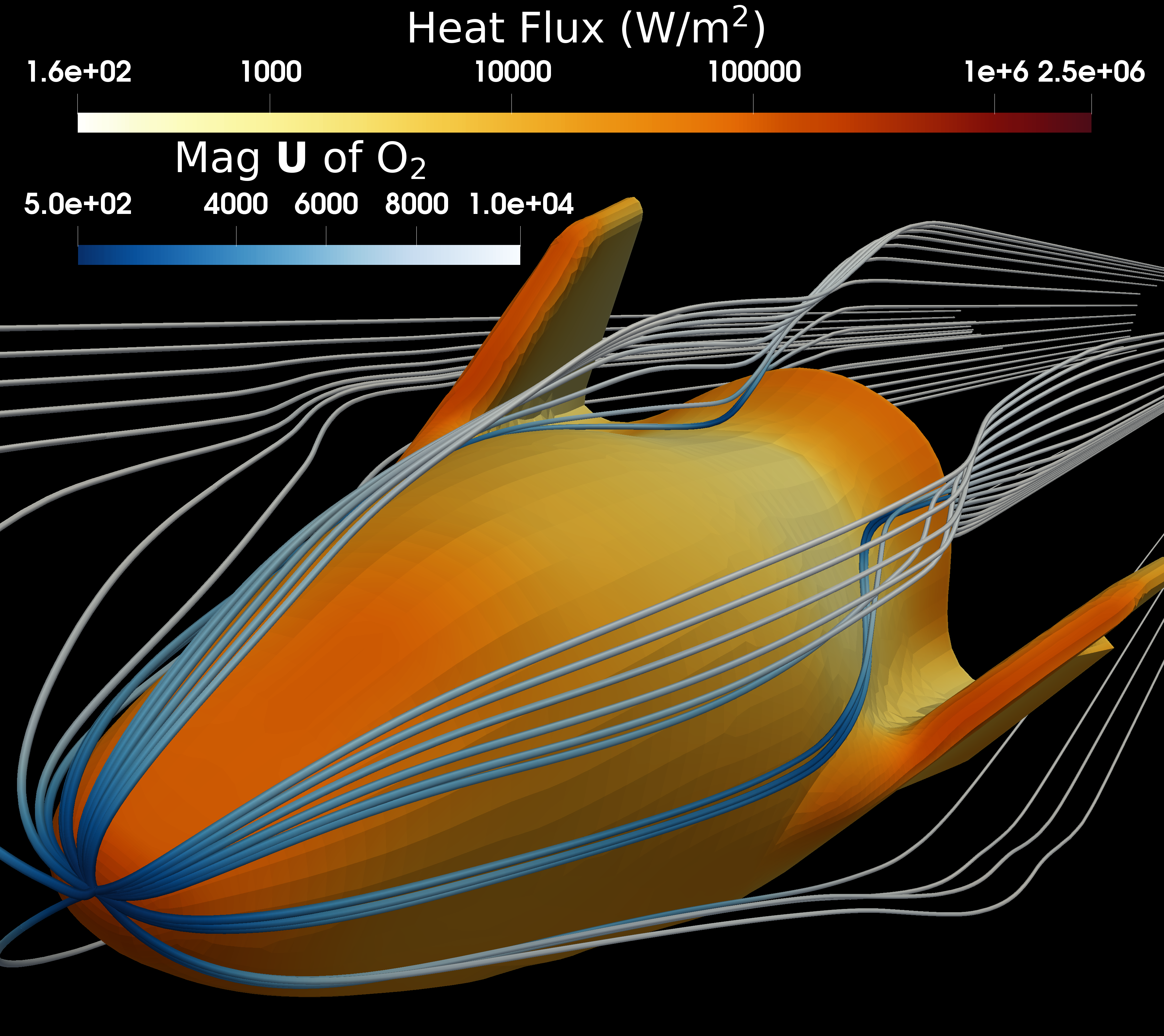}
    } 
	\caption{Hypersonic flow around an X38-like space vehicle with chemical reaction at ${\rm Ma}_\infty=13.15$. The distributions of heat flux on the surface and the magnitude of velocity of the product ${\rm O}_2$ along the streamline at (a) ${\rm Kn}_\infty = 0.001267$ and (b) ${\rm Kn}_\infty = 0.03543$, corresponding to atmospheric states at altitudes 80 km and 100 km, respectively.}
	\label{fig:x38-3D}
\end{figure}

\begin{figure}[H]
	\centering
    \subfloat[]{
		\includegraphics[width=0.9\textwidth]{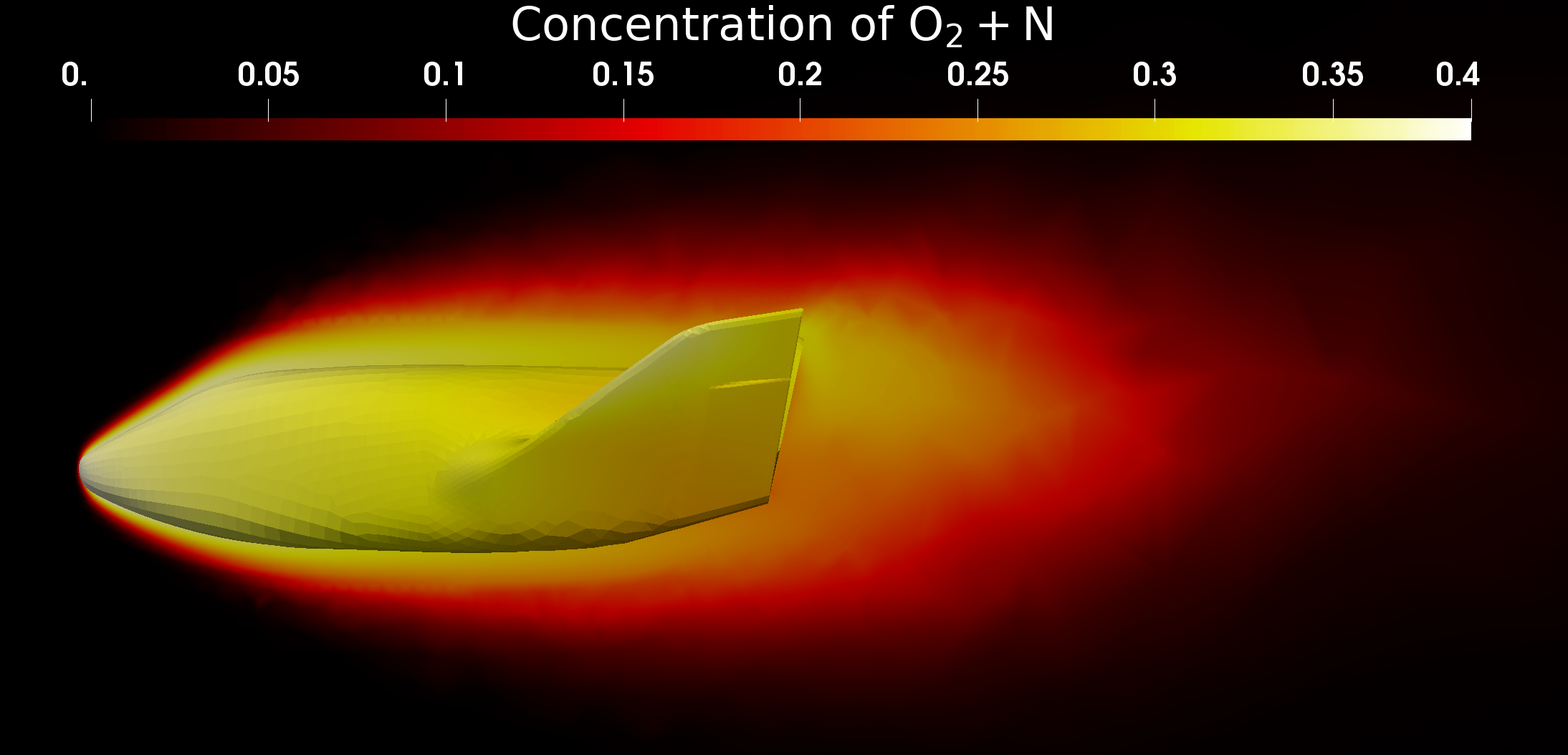}
	}  \\
    \subfloat[]{
    	\includegraphics[width=0.9\textwidth]{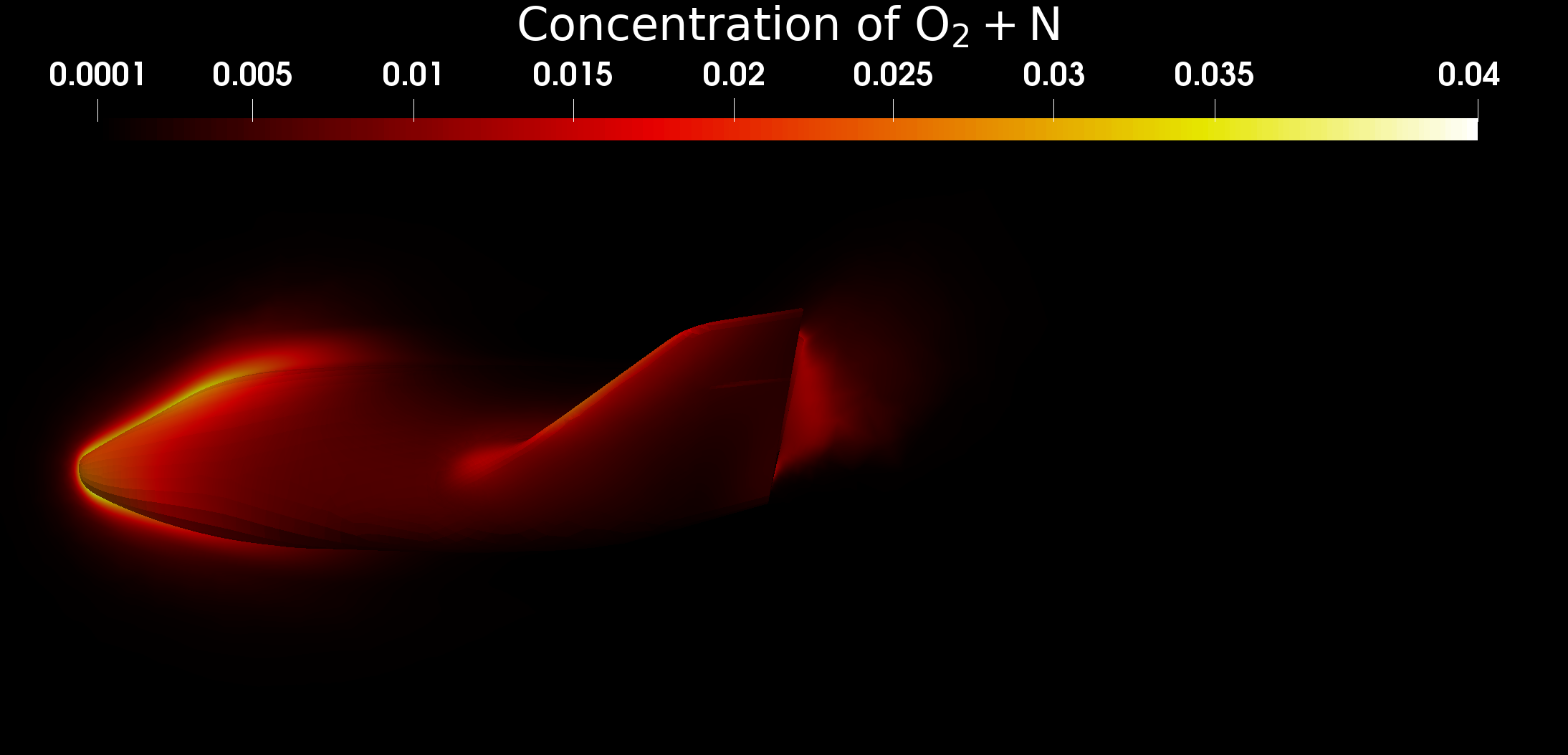}
    } 
	\caption{Hypersonic flow around an X38-like space vehicle with chemical reaction at ${\rm Ma}_\infty=13.15$. The distributions of the concentration of products ${\rm O}_2 + {\rm N}$ at (a) ${\rm Kn}_\infty = 0.001267$ and (b) ${\rm Kn}_\infty = 0.03543$, corresponding to atmospheric states at altitudes 80 km and 100 km, respectively.}
	\label{fig:x38-ratio}
\end{figure}

\begin{figure}[H]
	\centering
	\subfloat[]{
		\includegraphics[width=0.45\textwidth]{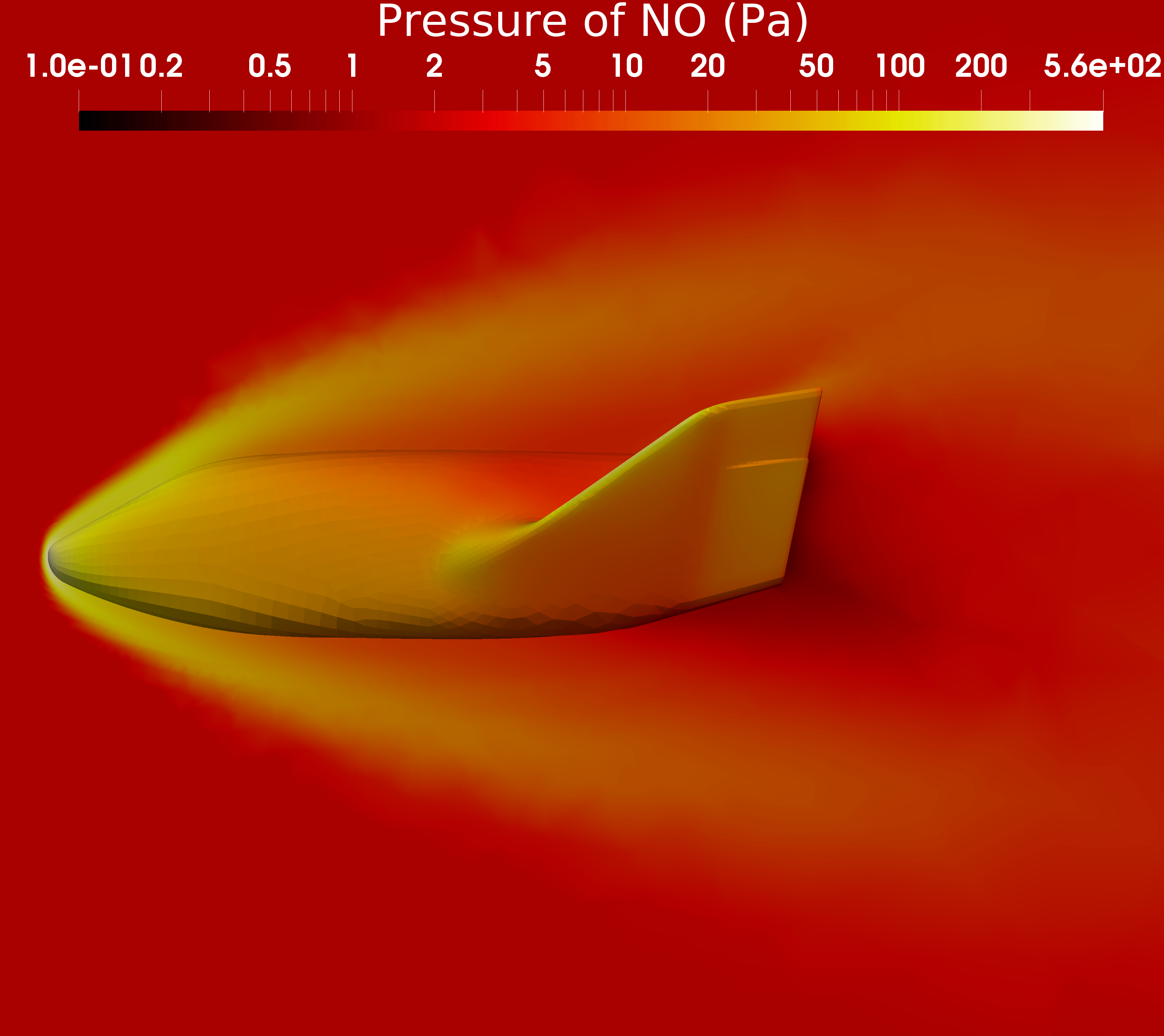}
	} 
	\hspace{0.5mm}
    \subfloat[]{
    	\includegraphics[width=0.45\textwidth]{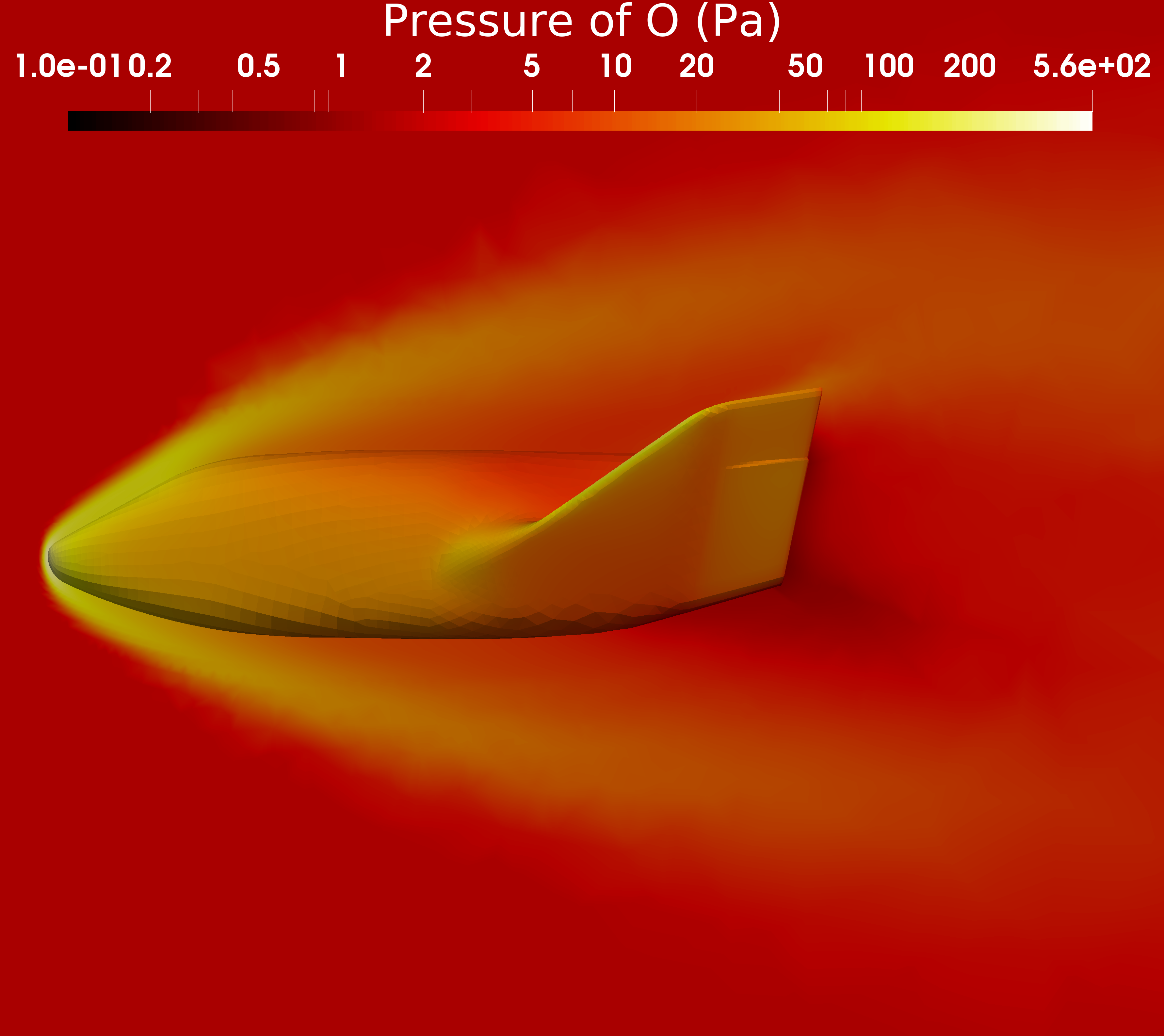}
    } \\
    \subfloat[]{
		\includegraphics[width=0.45\textwidth]{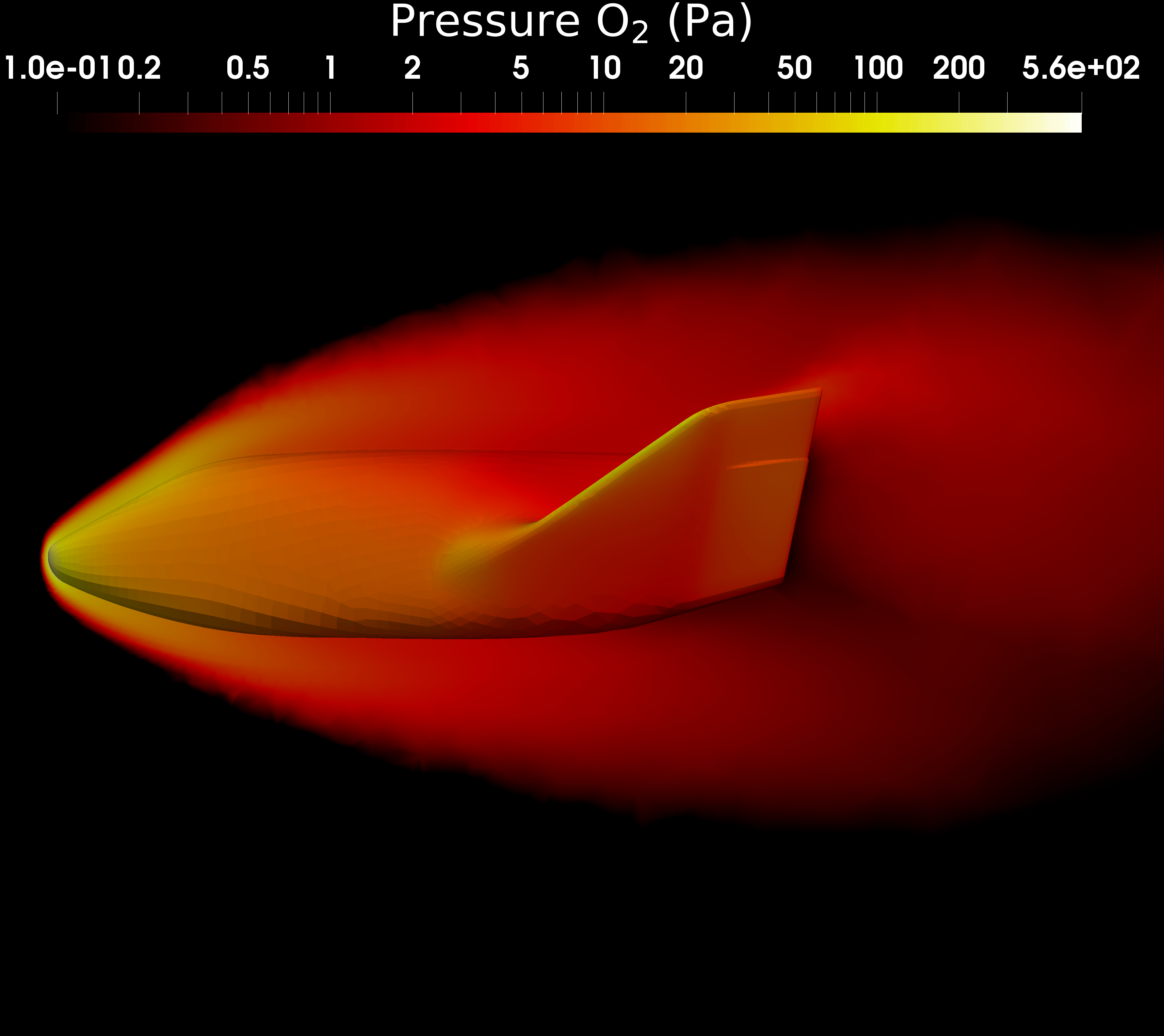}
	} 
	\hspace{0.5mm}
    \subfloat[]{
    	\includegraphics[width=0.45\textwidth]{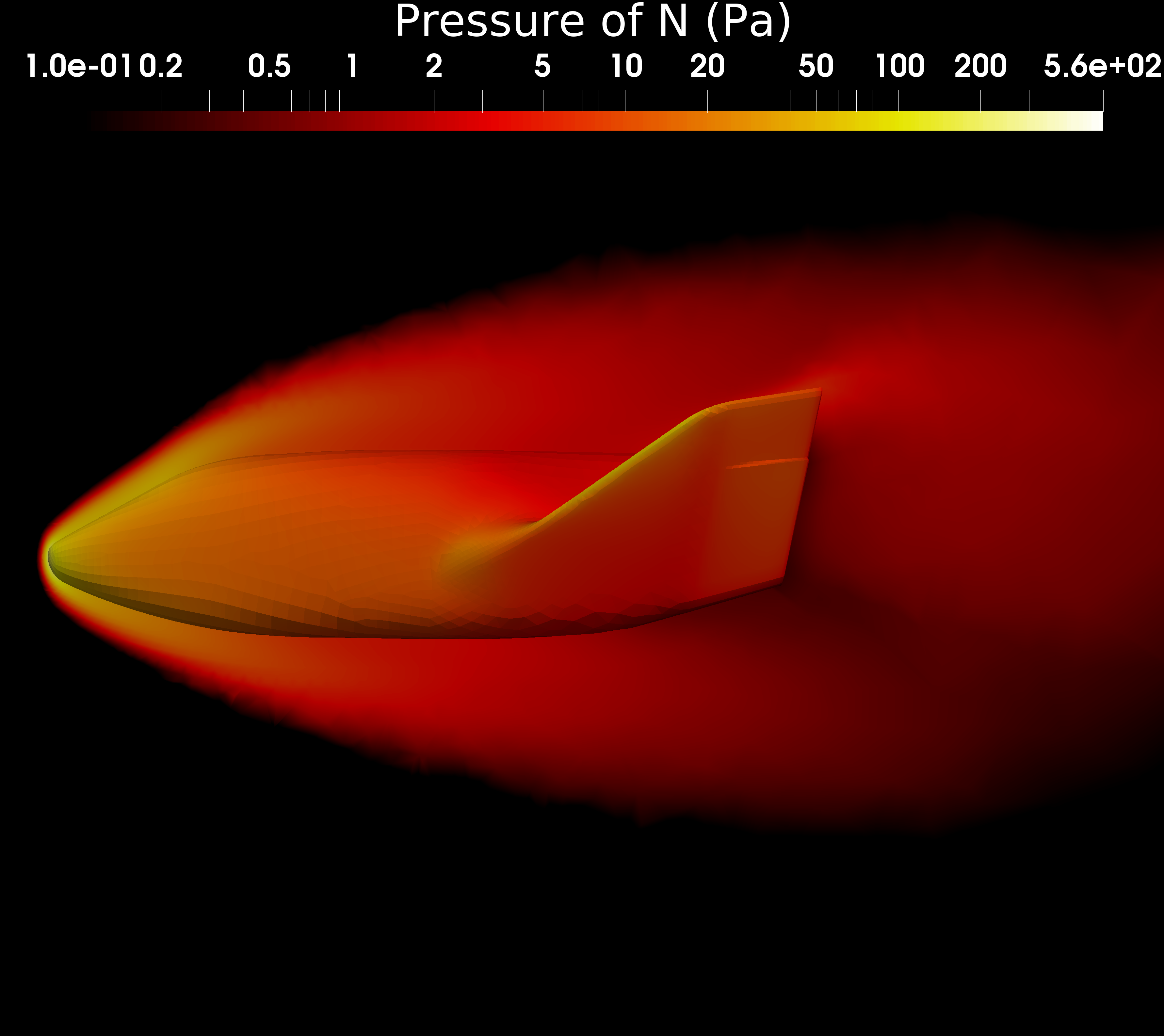}
    }  
	\caption{Hypersonic flow around an X38-like space vehicle with chemical reaction at ${\rm Ma}_\infty=13.15$. The distributions of pressure of the reactants (a) ${\rm NO}$ and (b) ${\rm O}$, and the products (c) ${\rm O}_2$ and (d) ${\rm N}$ at the ${\rm Kn}_\infty = 0.001267$ corresponding to atmospheric state at altitude 80 km.}
	\label{fig:x38-pressure-80} 
\end{figure}

\begin{figure}[H]
	\centering
	\subfloat[]{
		\includegraphics[width=0.45\textwidth]{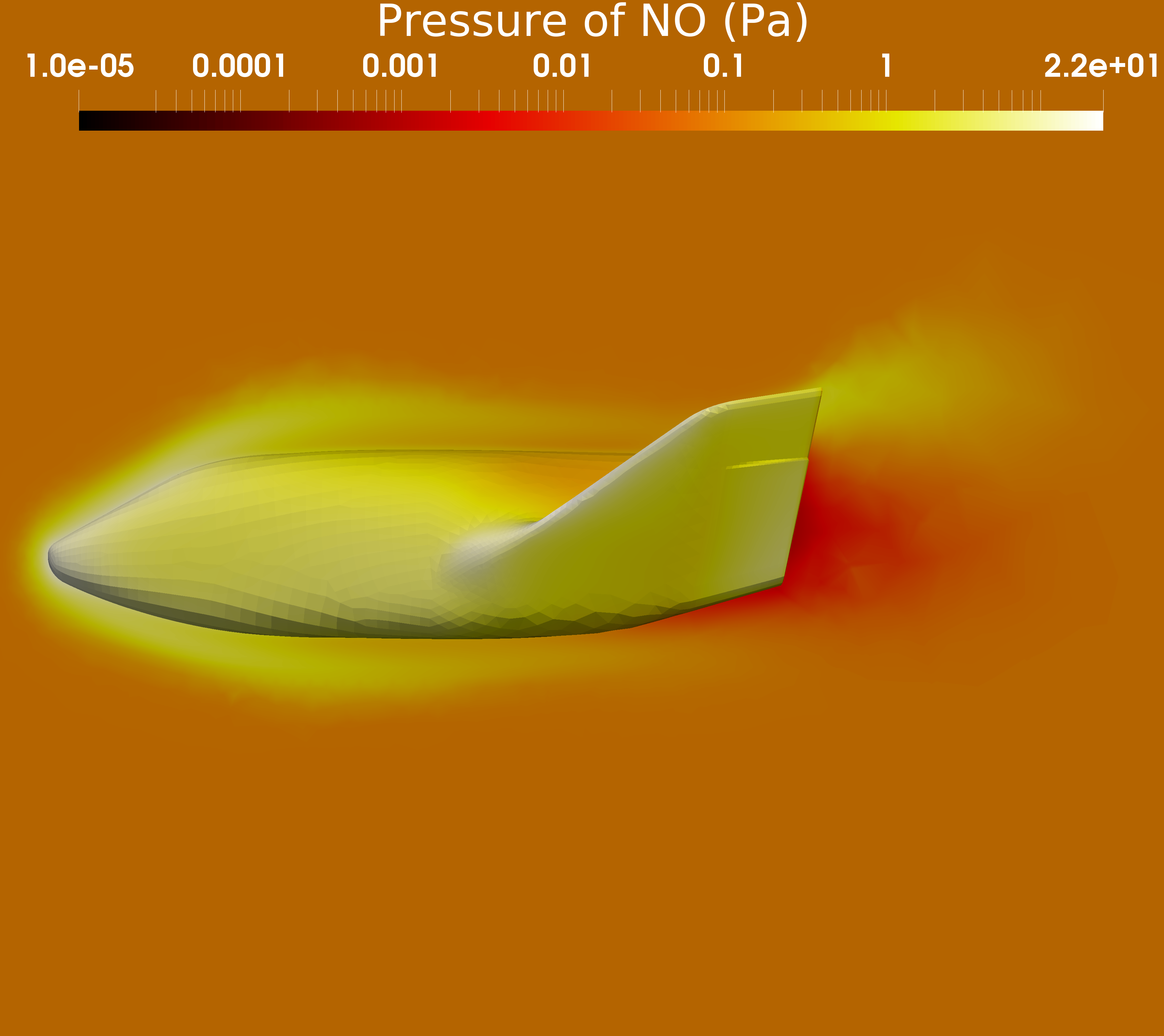}
	} 
	\hspace{0.5mm}
    \subfloat[]{
    	\includegraphics[width=0.45\textwidth]{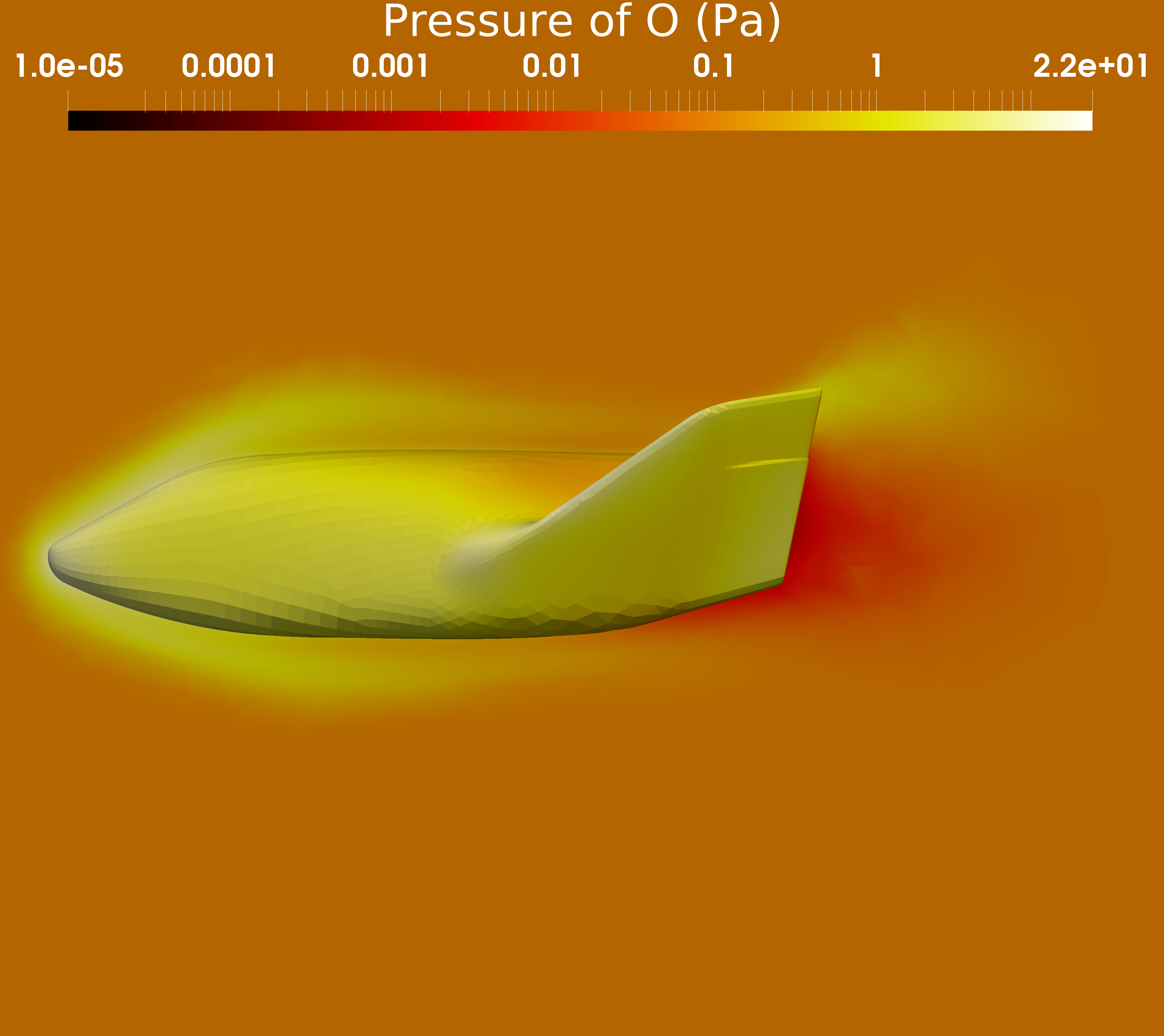}
	} \\
    \subfloat[]{
		\includegraphics[width=0.45\textwidth]{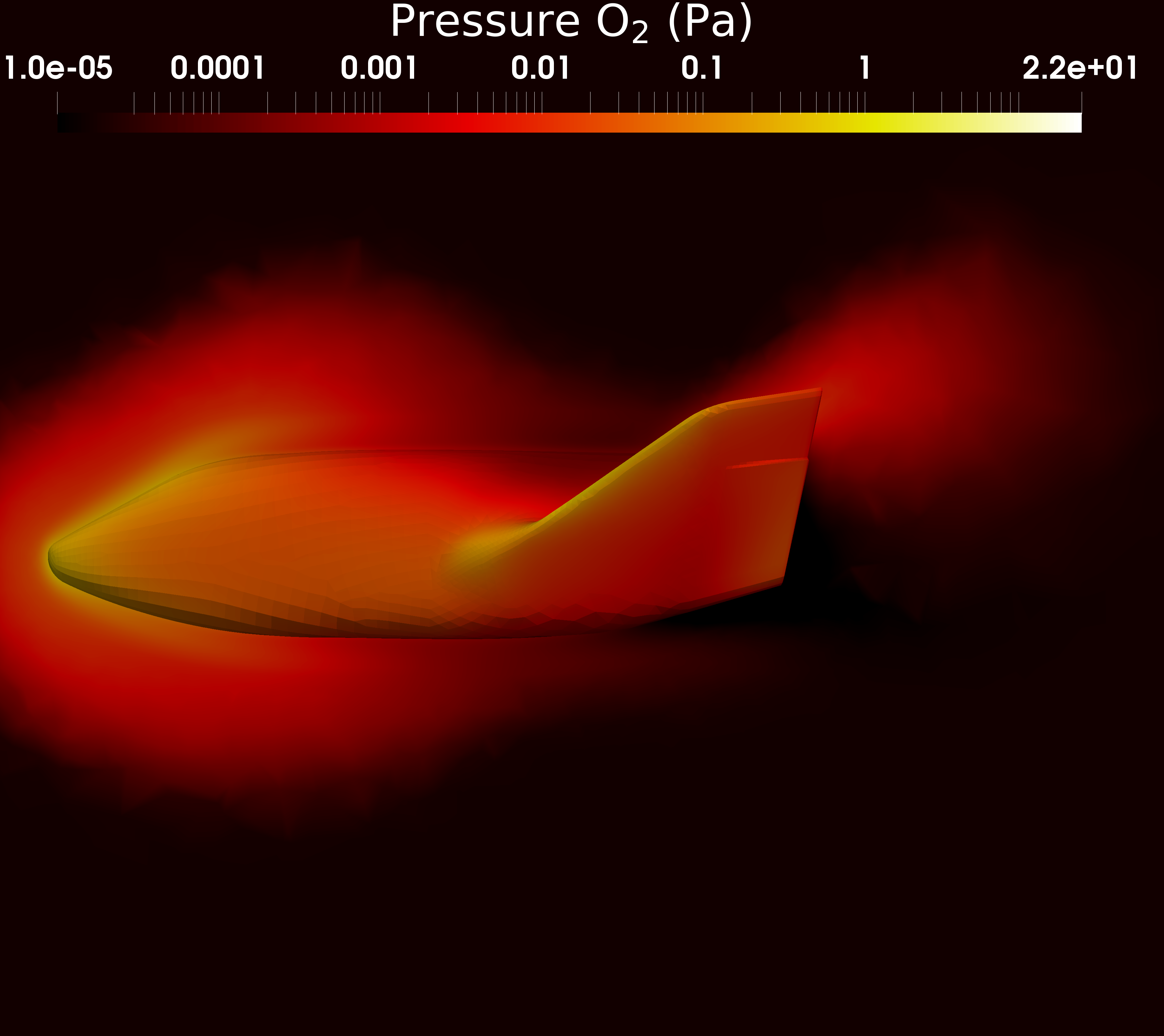}
	} 
	\hspace{0.5mm}
    \subfloat[]{
    	\includegraphics[width=0.45\textwidth]{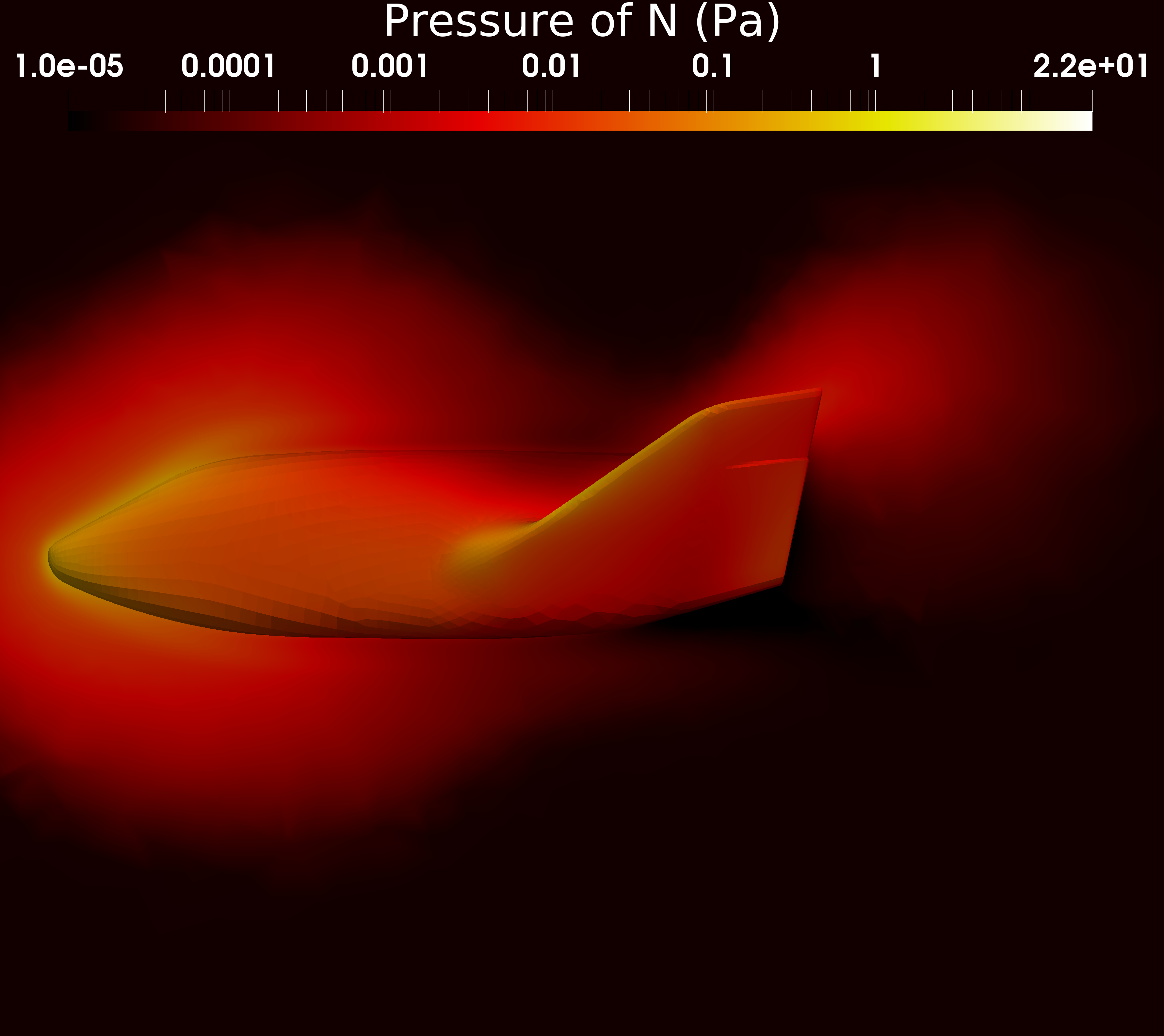}
    } 
	\caption{Hypersonic flow around an X38-like space vehicle with chemical reaction at ${\rm Ma}_\infty=13.15$. The distributions of pressure of  the reactants (a) ${\rm NO}$ and (b) ${\rm O}$, and the products (c) ${\rm O}_2$ and (d) ${\rm N}$ at the ${\rm Kn}_\infty = 0.03543$ corresponding to atmospheric state at altitude 100 km.}
	\label{fig:x38-pressure-100} 
\end{figure}

Figure.~\ref{fig:x38-3D} shows the velocity streamline plots of surface heat flux and the product ${\rm O_2}$ at altitudes of 80 km and 100 km, respectively. It can be observed that the surface heat flux at 80 km is higher than that at 100 km. Furthermore, the velocity stagnation point away from the wall observed in the flow around a cylinder is also presented in this case for the product streamline plot at 100 km. The specific reasons for this can be seen in the concentration fraction of products in Fig.~\ref{fig:x38-ratio} which describes the intensity of the chemical reaction. The concentration fraction of products at 80 km is around 0.4, whereas it decreases to 0.04 at 100 km. Combined with the distributions of pressures of each species shown in Figs.~\ref{fig:x38-pressure-80}-\ref{fig:x38-pressure-100}, especially considering the especially the differences in shock wave thickness of reactants at 80 km and 100 km, it can be observed that the rarefied environment has a certain attenuating effect on the pressure distribution and shock wave in the compressed region. Such reduced compressibility with the decreased pressure and the gradually smoothed shock wave in the rarefied environments is a key factor leading to the reduction in chemical reaction intensity. Moreover, it is worth noting that the simulation at 80 km presents a multi-scale transport of particles' collision and free streaming, multi-species interaction, and chemical non-equilibrium, which is a comprehensive reflection of rarefaction effects and real gas effects, making the flow highly complex. The multi-scale characteristics of the UGKS can effectively couple various non-equilibrium physics mentioned above, achieving a unified solution from continuum to rarefied flow regimes. This typical case clearly demonstrates the significant potential of the UGKS in simulating multi-scale flow with complex non-equilibrium physics.

\subsection{Nozzle plume into a background vacuum}

The last case in this section involves the nozzle plume flow into a background vacuum. Unlike the multi-scale flow caused by compressed and expanded air in high-speed flows, the multi-scale issue in nozzle flow arises from differences in the internal and external environments \cite{george1999simulation}. Single-scale methods struggle to accurately tackle such problems, and the presence of multi-species and chemical reactions further complicates this issue. The UGKS is applied to this case, showcasing its potential in handling external flow dynamics. In this case, the impact of different energy release types on nozzle flow is illustrated, considering cases of endothermic reaction, exothermic reaction, reaction with no energy release, and no chemical reactions. At the nozzle inlet, the total density is $1.64 \times 10^{-4}{\rm kg/m^3}$, with a pressure of 77 Pa, temperature of 2000 K, and velocity of 600 ${\rm m/s}$. The wall temperature of the nozzle is $500$ K. The pressure in near vacuum conditions is $10^{-13}$ Pa, with a temperature of 500 K. The concentration fractions of species are $n_{\rm NO, \infty} : n_{\rm O,\infty} : n_{\rm O_2, \infty} : n_{\rm N, \infty} = 1:1:0:0$. The case is set with $5025$ physical meshes and $89\times 89$ discretized velocity points. The CFL number is 0.8. 

\begin{figure}[H]
	\centering
    \subfloat[]{
			\includegraphics[width=0.48 \textwidth]
			{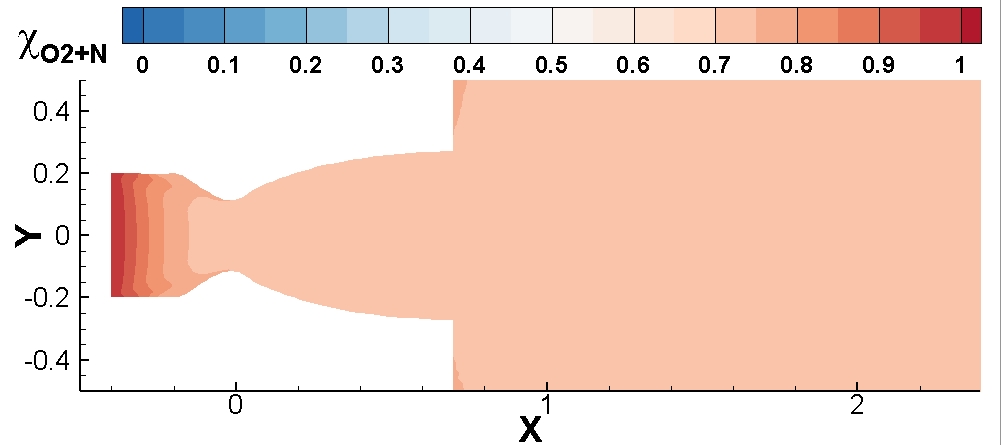}} 
    \subfloat[]{
			\includegraphics[width=0.48 \textwidth]
			{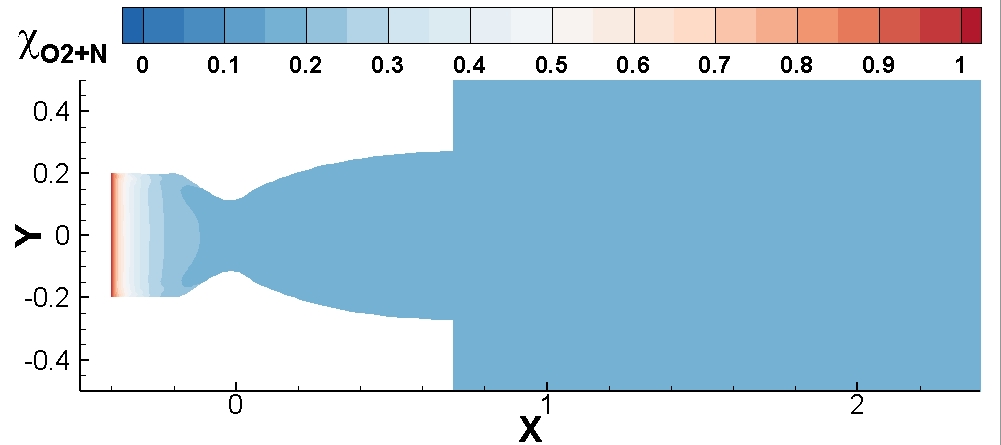}}  \\
	\subfloat[]{
			\includegraphics[width=0.48 \textwidth]
			{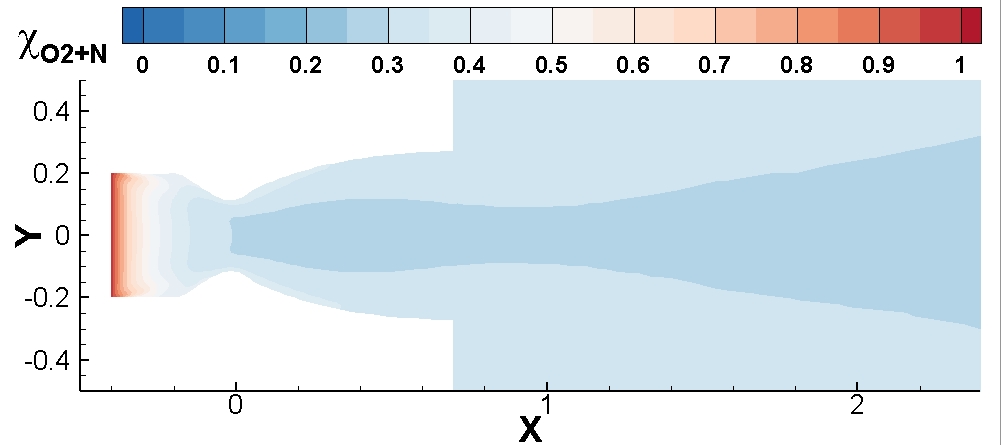}} 
	\subfloat[]{
			\includegraphics[width=0.48 \textwidth]
			{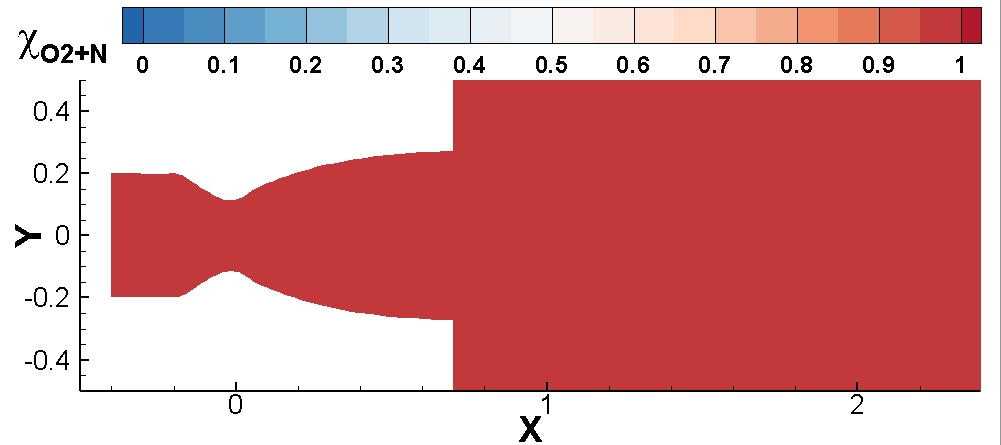}}  
	\caption{Nozzle plume flow to a background vacuum. Distributions of concentration fractions of the reactions under the energy release type (a) forward endothermic reaction, (b) forward exothermic reaction, (c) reaction with no energy release $\Delta E = 0$, and (d) no reactions with pure mechanical collisions.}
	\label{fig:nozzle-chi}
\end{figure}

\begin{figure}[H]
	\centering
    \subfloat[]{
			\includegraphics[width=0.48 \textwidth]
			{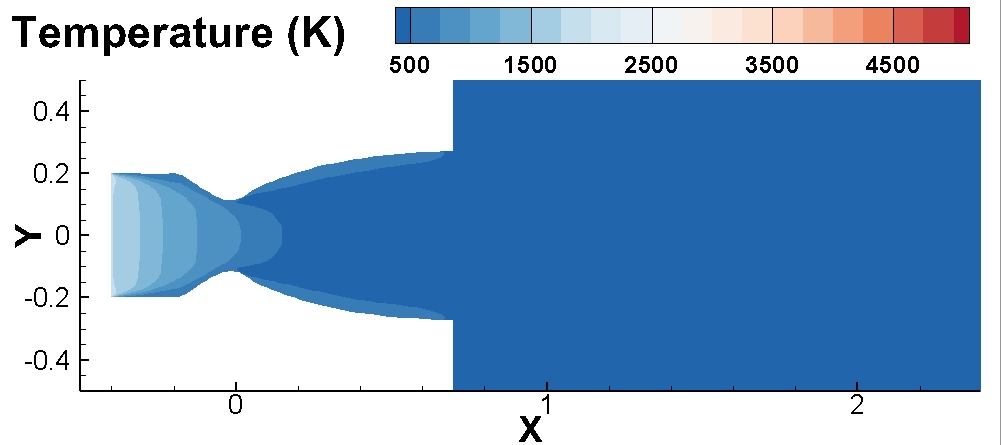}} 
    \subfloat[]{
			\includegraphics[width=0.48 \textwidth]
			{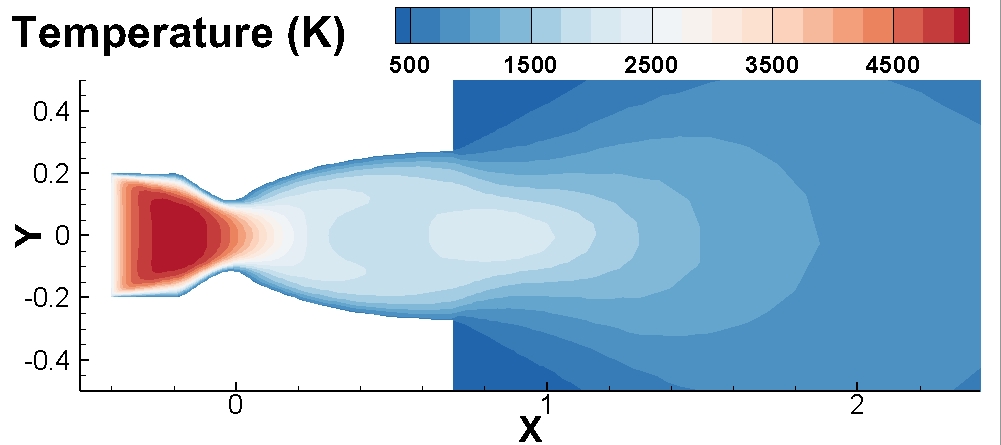}}  \\
	\subfloat[]{
			\includegraphics[width=0.48 \textwidth]
			{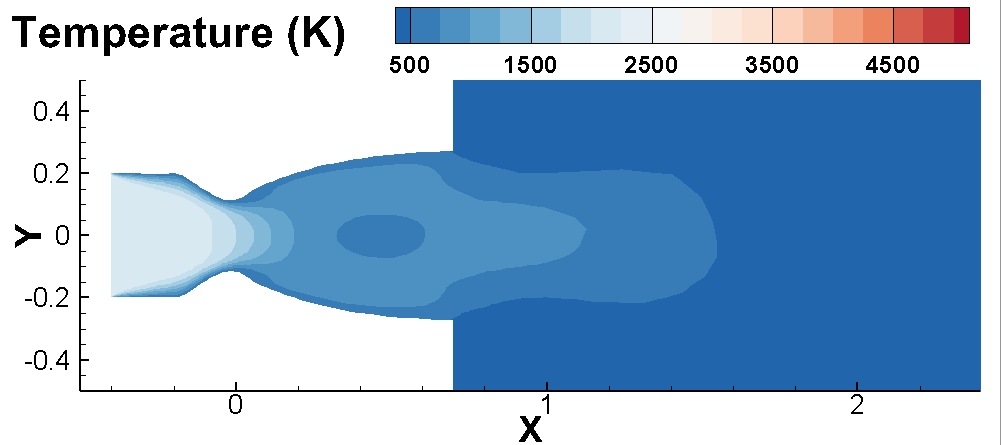}} 
	\subfloat[]{
			\includegraphics[width=0.48 \textwidth]
			{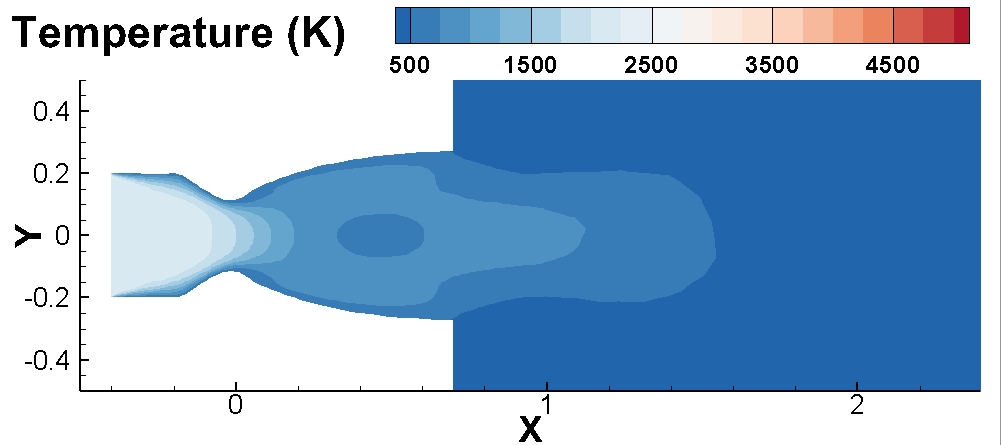}}	
	\caption{Nozzle plume flow to a background vacuum. Distributions of global temperature under the energy release type (a) forward endothermic reaction, (b) forward exothermic reaction, (c) reaction with no energy release $\Delta E = 0$, and (d) no reactions with pure mechanical collisions.}
	\label{fig:nozzle-T}
\end{figure}

\begin{figure}[H]
	\centering
    \subfloat[]{
			\includegraphics[width=0.48 \textwidth]
			{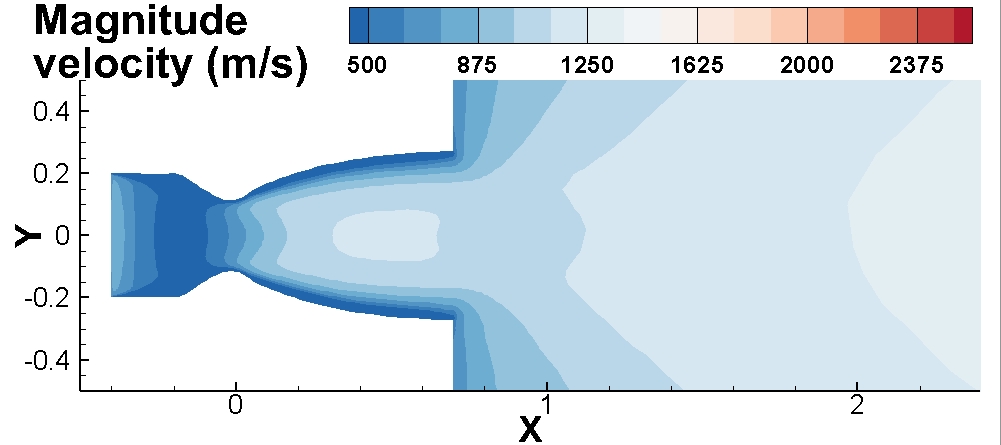}} 
    \subfloat[]{
			\includegraphics[width=0.48 \textwidth]
			{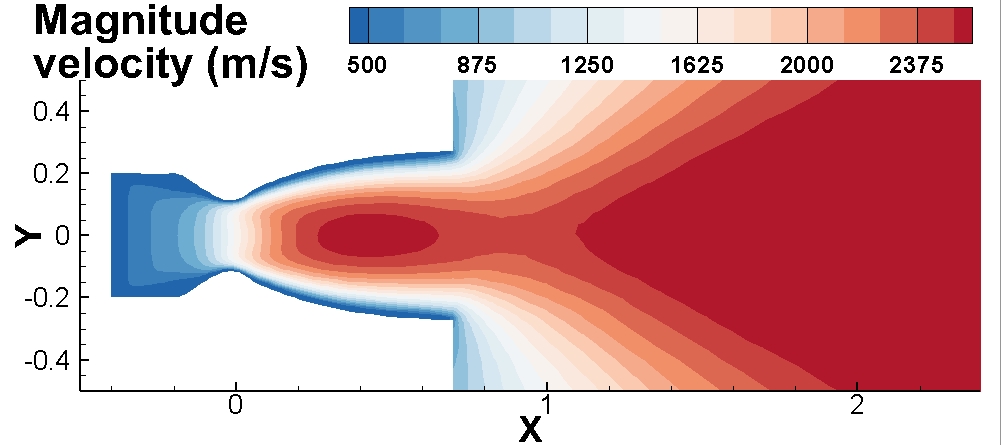}}  \\
    \subfloat[]{
			\includegraphics[width=0.48 \textwidth]
			{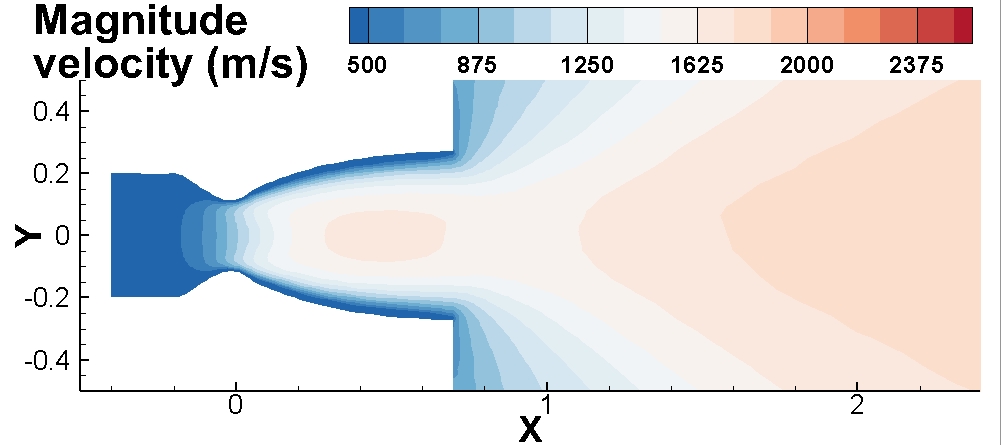}} 
    \subfloat[]{
			\includegraphics[width=0.48 \textwidth]
			{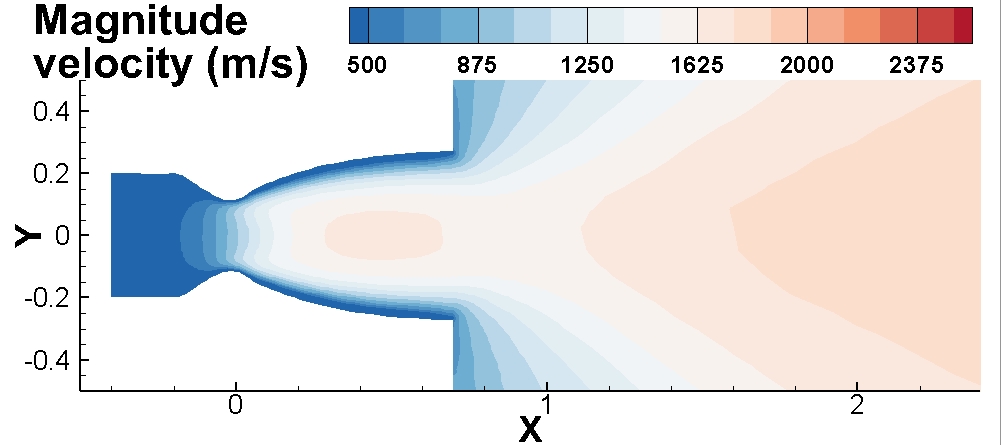}} 
	\caption{Nozzle plume flow to a background vacuum. Distributions of the magnitude of global velocity under the energy release type (a) forward endothermic reaction, (b) forward exothermic reaction, (c) reaction with no energy release $\Delta E = 0$, and (d) no reactions with pure mechanical collisions.}
	\label{fig:nozzle-U}
\end{figure}

\begin{figure}[H]
	\centering
    \subfloat[]{
			\includegraphics[width=0.48 \textwidth]
			{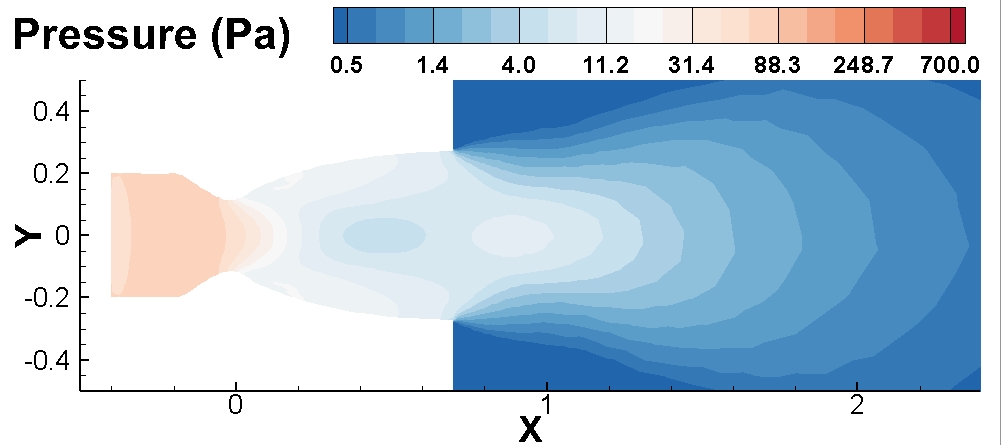}} 
    \subfloat[]{
			\includegraphics[width=0.48 \textwidth]
			{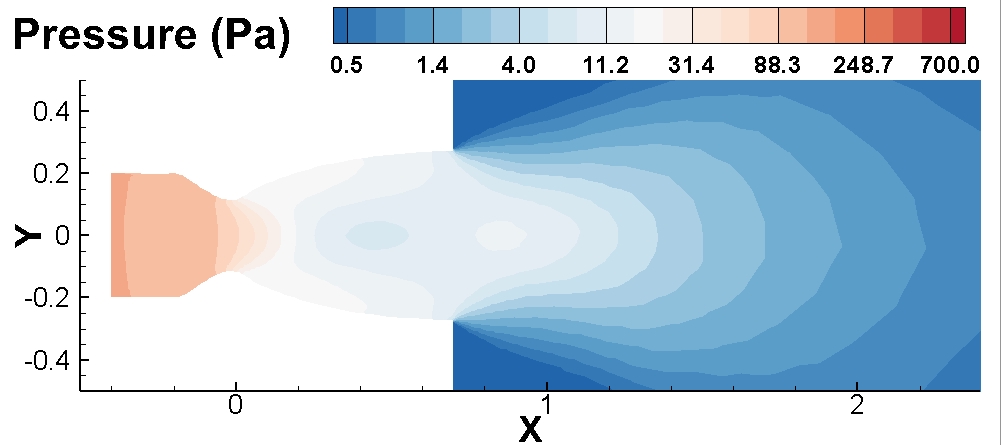}}  \\
    \subfloat[]{
			\includegraphics[width=0.48 \textwidth]
			{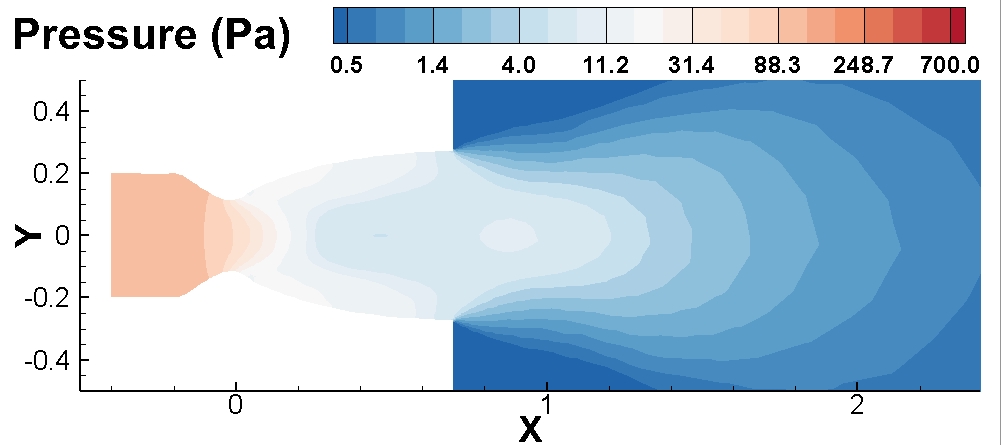}} 
    \subfloat[]{
			\includegraphics[width=0.48 \textwidth]
			{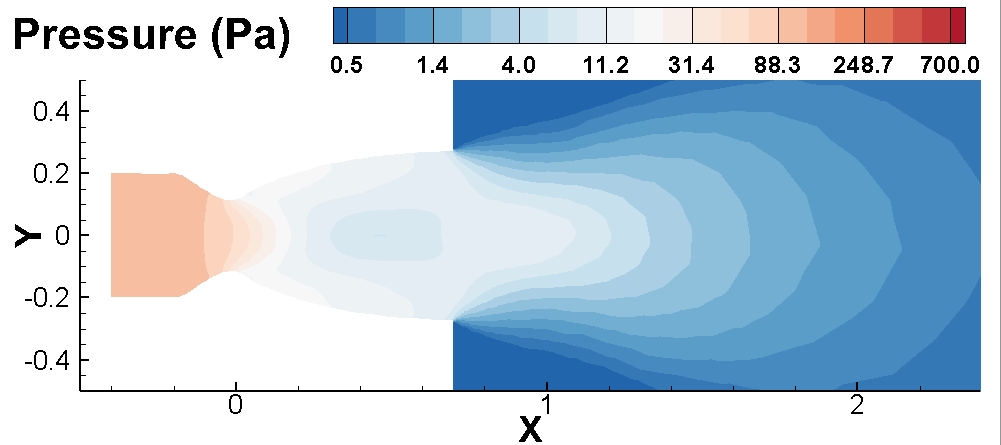}} 
	\caption{Nozzle plume flow to a background vacuum. Distributions of global pressure under the energy release type (a) forward endothermic reaction, (b) forward exothermic reaction, (c) reaction with no energy release $\Delta E = 0$, and (d) no reactions with pure mechanical collisions.}
	\label{fig:nozzle-P}
\end{figure}

\begin{figure}[H]
	\centering
    \subfloat[]{
			\includegraphics[width=0.48 \textwidth]
			{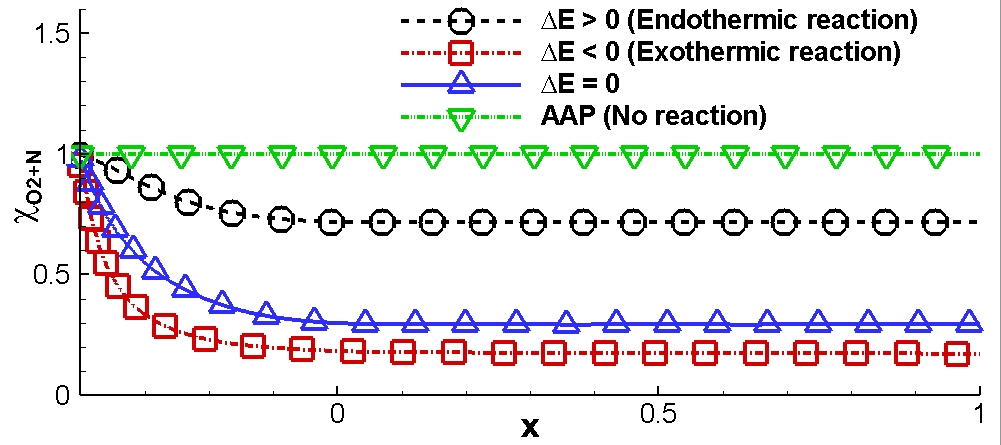}} 
    \subfloat[]{
			\includegraphics[width=0.48 \textwidth]
			{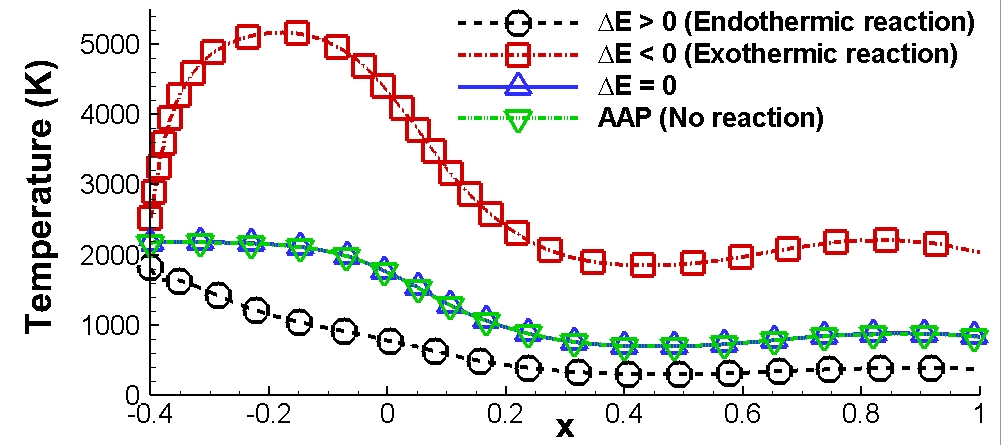}}  \\
	\subfloat[]{
			\includegraphics[width=0.48 \textwidth]
			{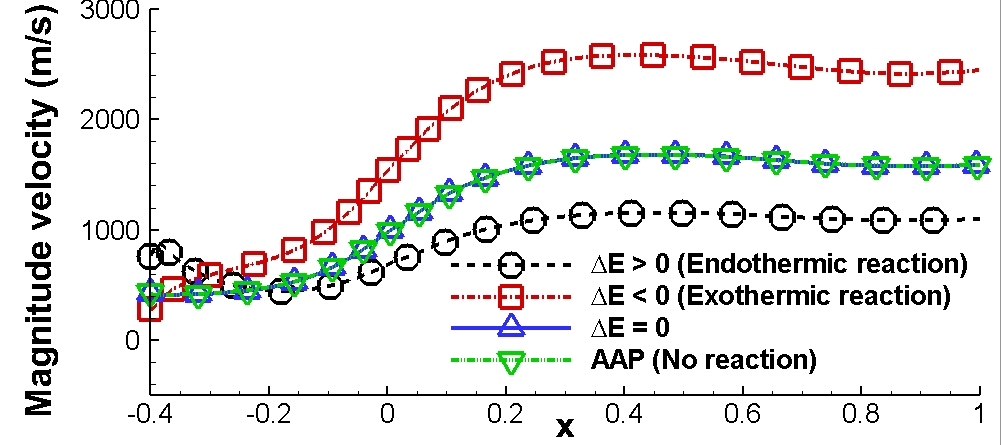}} 
	\subfloat[]{
			\includegraphics[width=0.48 \textwidth]
			{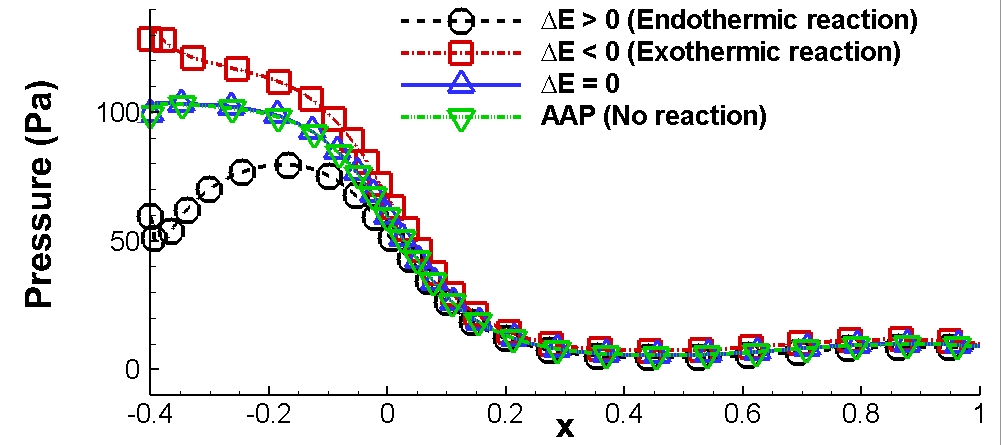}} 
	\caption{Nozzle plume flow to a background vacuum. Distributions of (a) concentration fractions of reactions, (b) global temperatures (c) magnitude of velocities, and (d) pressures of species along the axis of the nozzle, under the energy release type, such as forward endothermic reaction, forward exothermic reaction, reaction with no energy release $\Delta E = 0$, and no reactions with pure mechanical collisions.}
	\label{fig:nozzle-line}
\end{figure}

This case study mainly focuses on physical quantities such as global pressure, concentration fraction of reactants, global temperature, and magnitude of global velocities. 
Figures.~\ref{fig:nozzle-chi}-\ref{fig:nozzle-T} respectively show the contours of these quantities under endothermic, exothermic, no energy release, and no chemical reactions cases. 
Firstly, Fig~\ref{fig:nozzle-P} illustrates the distribution of pressure during the expansion process of the nozzle, revealing that the pressure difference between the internal and external environments can vary by orders of magnitude, which requires a multi-scale scheme for simulation. Figure~\ref{fig:nozzle-line} shows the distribution of various physical quantities along the axis of the nozzle under four energy release types. The concentration fractions shown in Fig.~\ref{fig:nozzle-chi} demonstrate that exothermic reactions exhibit the most intense chemical reactions. As shown in Fig.~\ref{fig:nozzle-chi}(b) and Fig.~\ref{fig:nozzle-line}(a), the reactant concentration is 1.0 at the inlet, decreasing to around 0.2 at the throat after the contraction section due to chemical reactions. The second intense one is the reaction with no energy release, which decreases to around 0.3 at the throat. Endothermic reactions are more moderate, reducing to around 0.7. Combined with the temperature contours in Fig.~\ref{fig:nozzle-T}, it can be observed again that the intensity of chemical reactions is closely related to global temperature. The gas in the contraction section receives heat from compression, promoting chemical reactions. Exothermic reactions further release energy, enhancing the intensity of chemical reactions, while endothermic reactions lower the temperature in the contraction section, slowing down the chemical reaction process. These series of physical processes ultimately impact the velocity of the nozzle in the expansion section. Figure~\ref{fig:nozzle-U} shows that the velocities under endothermic and exothermic reactions can differ by approximately a factor of two. In summary, this case study showcases the significant impact of different exothermic types on the flow field and the capability of the UGKS with a chemical reaction model in handling multi-scale internal to external flow.

\section{Conclusion}
Reactive flow is a highly complex issue involving multi-species interaction and reactions. In rarefied environments, the complexity increases due to the involvement of multi-scale transport with particles' scattering and collisions. The unified gas-kinetic scheme, coupling the free transport and collisions of particles through the integral solution of the kinetic model, can capture the multi-scale transport without time step and mesh size limitations posed by splitting methods. This study extends the UGKS to the chemical non-equilibrium flow, validated against the benchmark results from the DSMC method. Moreover, the hypersonic flows around an X38-like space vehicle at 80 km and 100 km and nozzle plume flows into a vacuum are simulated to showcase the extended UGKS's proficiency in multi-scale and non-equilibrium simulations. With the methodology of direct modeling, the UGKS shows great potential for simulating multi-scale flows with complex non-equilibrium physics. Furthermore, the current scheme provides a basic method for the future development of multi-scale methods, such as the implicit adaptive unified gas-kinetic scheme and the unified gas-kinetic wave-particle method.

\section*{Author's contributions}
All authors contributed equally to this work.

\section*{Acknowledgments}

This work was supported by National Key R$\&$D Program of China (Grant Nos. 2022YFA1004500), National Natural Science Foundation of China (12172316), Hong Kong research grant council (16208021,16301222).

\section*{Data Availability}

The data that support the findings of this study are available from the corresponding author upon reasonable request.

\appendix

\section{Upstream and Downstream Condition of a Shock Structure with Chemical Non-equilibrium}
\label{ch:reacting-shock}
It is more difficult to calculate the state at both sides of the shock structure in the chemical reaction flow. Because the internal energy is uncertain and more energy is released or absorbed by the reaction, the Rankine--Hugoniot condition cannot be given directly, but should be derived by iterations, as shown below. 
At first, the upstream (-) and downstream (+) states should be at the chemical equilibrium states, respectively, with the rates of chemical reaction equal to zero
\begin{equation}\label{eq:upstream}
	- K_f n_\mathcal{A}^{-} n_\mathcal{B}^{-}
			    +   K_b n_\mathcal{C}^{-} n_\mathcal{D}^{-} = 0,
\end{equation}
and
\begin{equation}\label{eq:downstream}
	- K_f n_\mathcal{A}^{+} n_\mathcal{B}^{+}
			    +   K_b n_\mathcal{C}^{+} n_\mathcal{D}^{+} = 0.
\end{equation}
The downstream state can be calculated by
\begin{equation}
\label{eq:chemical-RH}
\begin{aligned}
\frac{n^{+}}{n^{-}} &= 2\left(1-\frac{T^{-}}{T^{+}}\right)
                     - \frac{\Delta E}{k_B T^{+}}\Delta \chi + \sqrt{\left[2\left(1-\frac{T^{-}}{T^{+}}\right)-\frac{\Delta E}{k_B T^{+}}\Delta \chi\right]^2+\frac{T^{-}}{T^{+}}},\\
U^{-} &= \sqrt{\frac{n^{-}k_B T^{-}}{\rho^{-}}}\sqrt{\frac{n^{+}}{n^{-}}\frac{1-(n^{+}/n^{-})(T^{+}/T^{-})}{1-(n^{+}/n^{-})}},\\
U^{+} &= U^{-}\frac{n^{-}}{n^{+}},
\end{aligned}
\end{equation}
where $\chi_\alpha^{+} = \chi_\alpha^{-} + \Lambda_\alpha \Delta \chi$. Once $\Delta \chi$ is given, the downstream temperature $T^{+}$ can be calculated through equilibrium reacting state Eq.~\eqref{eq:downstream} by iteration. Then other state parameters can be calculated by Rankine--Hugoniot Eq.~\eqref{eq:chemical-RH}. With the sound speed $a$ for the considered reactive mixture \cite{rossani_travelling_2004,conforto_steady_2006}
\begin{equation*}
	a = \sqrt{
		\frac{5}{3}\frac{n k_B T}{\rho}
		\left[\sum_\alpha \chi_\alpha^{-1}
		+\frac{2}{5}\left(\frac{\Delta E}{k_B T}\right)^2
		\right]
		\left[
			\sum_\alpha \chi_\alpha^{-1}+\frac{2}{3}\left(\frac{\Delta E}{k_B T}\right)^2
		\right]^{-1}
		}, 
\end{equation*}
the upstream Mach number ${\rm Ma}^{-} = U^{-} / a^{-}$ can be obtained.
	
%% References


% \end{CJK*}

\bibliographystyle{elsarticle-num}
% \section*{\refname}
\bibliography{ugks-reacting.bib}






% \end{thebibliography}

\end{document}